\documentclass[twocolumn]{aastex62}

\usepackage{graphicx}	
\usepackage{amsmath}	
\usepackage{amssymb}	
\usepackage{mathrsfs}

\def\figg#1#2#3#4#5{\hfill\vbox{\parskip=0pt\hsize=#2
\includegraphics[scale=#5,angle=#4]{#1}\vskip2pt\vtop{\centering
\footnotesize
\hsize=#2
#3\vskip1pt
}}\hfill}

\newcommand{\reffiglambdasPwA}{{\color{blue}a}}
\newcommand{\reffiglambdasPmA}{{\color{blue}b}}
\newcommand{\reffiglambdasPsA}{{\color{blue}c}}
\newcommand{\reffigemPwA}{{\color{blue}d}}

\newcommand{\reffigemPsA}{{\color{blue}f}}
\newcommand{\reffigAmFTPwA}{{\color{blue}a}}
\newcommand{\reffigAmFTPmA}{{\color{blue}b}}
\newcommand{\reffigAmFTPsA}{{\color{blue}c}}
\newcommand{\reffigAmFTwPwA}{{\color{blue}a}}
\newcommand{\reffigAmFTwPmA}{{\color{blue}b}}
\newcommand{\reffigAmFTwPsA}{{\color{blue}c}}
\newcommand{\reffigAmFTwPwAd}{{\color{blue}d}}
\shorttitle{Disk Stability Parameters}
\shortauthors{D. Valencia-Enr\'{i}quez, I. Puerari and I. Rodrigues.}

\begin{document}

\title{Assessing Disk Galaxy Stability through Time}

\author[0000-0002-0786-7307]{D. Valencia-Enr\'{i}quez}
\affil{Instituto Nacional de Astrof\'{i}sica, \'{O}ptica y Electr\'{o}nica,  Calle Luis Enrique Erro 1, Santa Mar\'{i}a Tonantzintla, 72840 Puebla, Mexico\\
}
\affiliation{Corporaci\'on Universitaria Aut\'onoma de Nari\~{n}o, Carrera 28 No. 19-24, 52001	Pasto, Colombia}

\author{I. Puerari}
\affiliation{Instituto Nacional de Astrof\'{i}sica, \'{O}ptica y Electr\'{o}nica,  Calle Luis Enrique Erro 1, Santa Mar\'{i}a Tonantzintla, 72840 Puebla, Mexico\\ 
}

\author{I. Rodrigues}
\affiliation{Instituto de Pesquisa e Desenvolvimento, Universidade do Vale do Para\'{i}ba, Av. Shishima Hifumi, 2911 - Urbanova, S\~{a}o Jos\'{e} dos Campos - SP, 12244-000, Brasil\\
Revised 2019 February 21; accepted 2019 March 13; published 2019 April 16; to AJ}






\begin{abstract}
N-body simulations have shown that the bar can be triggered by
two processes: (1)  by own instabilities in the disk, or (2)
by interactions with other galaxies. Both mechanisms have been
widely studied. However, the literature has not shown measurements
of the critical limits of the Disk Stability Parameters, DSP.
We showed measurements of those parameters 
through whole evolution in isolated disk models finding that the 
initial rotation configuration of those models is saved in the stable
or unstable regimen from the initial to the final evolution. Then, 
we perturbed such isolated models to study the evolution 
of DSP under perturbation. We find that the critical limits of DSP are 
not much affected in barred models, however, when the bar is triggered 
by the perturbation, the disk fall in the unstable regimen.
We show in our models that the bar triggered by a light perturbation grows 
into two phases: first, the bar appears as slow rotator, then it evolves 
toward fast rotator; second, when the perturbation is far from the target 
galaxy, the bar evolves from fast to slow rotator. Nevertheless, when the 
bar is triggered by a heavy perturbation, it appears as fast rotator 
and evolves toward slow rotator similar to classical bar models. 
\end{abstract}



\section{Introduction} 

Evaluate the stability of a disk galaxy as well as to
disentangle the formation and evolution of a barred galaxy
from the observations is a difficult task. Analytical works and  
N-body simulations are alternatives to study its dynamics and bar evolution. The analyzes of the stabilities
are based on the determination of the dispersion relations
and then investigate the unstable modes. If the size of the
perturbation is much smaller than the size of the disk, 
the perturbation is cataloged as local stability. The Toomre
stability criterion $Q$ play an important role in the formation
of spiral arm fragmentation to local scale. But,
if the size of the perturbation is comparable with
the size of the disk, the perturbation is classified
as global stability; however, it is very difficult to
write down a universal dispersion relation stability
criterion. In the cosmological context, where the
halo is much larger than the disk, the bar perturbation
can not  classified as a global perturbation rather is a 
central characteristic of full galaxy.

On the other hand, the bar formation in disk galaxies has been
studied for several decades. Basically,
the bar-like structure is triggered by
two processes: (1) those which stem from internal causes, 
e.g., dynamical instabilities within individual galaxies
and (2) those which are produced by external (e.g., tidal)
influences. 

\cite{1996LNP...474....7L} and \cite{2004ARep...48..877P}
reviewed five different mechanisms for the bar formation
in isolated galaxies:
(a) Since real bars have internal stream motions,
Freeman's picture of bar formation 
tried to associate these motions with incompressible 
Jacoby and Rieman fluids.
(b) Two density waves are reflected in the center of the galaxy
and amplified via Toomre's swing mechanism
	when the disk has not Inner Lindblad Resonace(ILR).
(c) The Contopoulos's picture shows that the bar is a result
of the distortion of circular orbits into eccentric orbits
 when the bar potential is already formed.
(d) \cite{1971ApJ...166..275K,1977ApJ...212..637K}
made a full stability
analysis which led to an eigenvalue problem for normal modes
of an axisymmetric stellar disk. 
The self-gravity of
an ensemble of orbits (lobes) cooperate with one other to
generate a disk instability radial orbits, which results
in the formation of slow Lynden-Bell bars.
In practice, it has been proved to be very difficult to
find eigenvalues \citep{1993RPPh...56..173S}.
\cite{1979MNRAS.187..101L} 
suggested that bars may grow slowly
through the gradual alignment of eccentric orbits. 
In addition, \cite{2004ARep...48..877P}
  discuss the formation of galactic structures
  viewed as low-frequency normal modes in disk
  consisting of precessing stellar orbits.
  Studying the properties of an integral equation
  via the Lynden-Bell derivate of the distribution
  function, which depends on the variation
  	of angular momentum. They found that if such derivate is positive,
  the bar mode can form;  otherwise, the derivate is
  negative, then spiral modes growth. The mechanism of
  bar formation constraints on the angular velocity
  of the mode, which should be larger than the angular
  velocity of the orbital precession for fast bars or
  should be similar for slow bars \citep{2003AstL...29..447P}.
  Then, the bar mode develops as a result of azimuthal
  tunning orbits in a massive disk. However, 
  the wave decay in a light disk due to the wave mode
  meets the ILR.
(e) A statistical focus that examines the bar formation
from the rotating initial
configurations when its spin parameter is lower than
some critical limit.

These channels of bar formation are all connected by
the change of angular momentum.
It has shown that the 
formation and evolution of bars in isolated disk galaxies
depends on the angular momentum exchange between their resonances and components (halo,
disk, or bulge) \citep[and references therein]{2002ApJ...569L..83A}. 
Therefore, a study of the
spin parameter $\lambda_d$ \added{(eq. \ref{eq:lm})}  which is a
measure of specific angular momentum, and its
Critical Limit $\lambda_{crit}$ \added{(eq. \ref{eq:lcrit})} can assess the stability {\it in situ}
of a disk model, and diagnosing the growth of a bar
instability as well. These parameters can be related
to an empirical stability parameter of 
\cite{1982MNRAS.199.1069E} (hereafter EF82), 
which only depends on
the rotation curve, disk mass, and the disk radius
scale, and it is relatively easy to obtain
in real observed galaxies \added{(eq. \ref{eq:em})}. 
\deleted{ This parameter is a simple
comparison of the rotation curve and the circular velocity
of a hypothetical particle subject to a point-mass potential
which has a mass equal to that of the disk. Thus,}
\replaced{this}{This} parameter can evaluate roughly
the stability of a disk because
it is indirectly related to the exchange of angular
momentum. In other words, the angular
momentum exchange can be more efficient when the velocity
dispersion is low and affects the mass distribution 
of the disk, halo or bulge, which in turns affect
the rotation curve and the scale radius of a disk.
Therefore, our goal is to
figure out the critical limit of the spin parameter
which relates the energy, the mass distribution 
and the angular momentum of
the a disk galaxy by using N-body simulations to show the utility
of $\lambda_{crit}$ for diagnosing the disk stability
in N-body simulations. 
Furthermore, we claim the possibility of use $\epsilon_m$
to give a rough estimation of the stability in a
real galaxy, since getting the angular momentum or spin parameter from
a real galaxy is very difficult task.

The \cite{1998MNRAS.295..319M} (hereafter MO98) models
let us manipulate the spin parameter directly and get a
maximum and sub-maximum disk with the same $M_D/M_H$ ratio. 
These models can be stable or
unstable to the bar formation, depending on
the disk stability parameters
$(\lambda_d,\lambda_{crit})$ (hereafter DSP).
Therefore, to accomplish our goal, we generate models 
with the same $M_D/M_H$ ratio, but with different initial disk
stability parameters to be stable or unstable to the bar
formation, and then we follow these parameters through the
time, since a follow-up of $\lambda_{crit}$ over time
has not been done. Likewise, we obtain measures of
$\epsilon_m$ parameter through the time to show
and connect its behavior to the bar instability.

We know that the galaxies in the
universe are not isolated; in fact, they are interacting.
\cite{1981ApJ...244L..43T} showed that there are a large
fraction of barred galaxies in the core of the Coma
cluster indicating that tidal interactions can trigger
bar-formation. \cite{1990ApJ...364..415E} 
studied binary galaxy samples to search for possible
correlations between the bar and the Hubble type founding 
that binary systems have a factor of $\sim$2 excess 
of barred galaxies. \cite{1996AJ....111.1805A} made
a study of the velocity distribution of disk galaxies in the
Virgo cluster, with the result that only the barred spiral 
galaxies in the core of the Virgo Cluster may have been
triggered by interactions. 
\cite{2011MSAIS..18...61M} investigated the properties of 
bright barred and unbarred galaxies in the Abell 901/902 cluster
explaining that high-velocity dispersion in the 
core regions benefits flyby interactions, 
which may increase the bar fraction, while preserving 
intact the galaxy disk.

N-body simulations have shown that a bar in a disk galaxy
can be triggered by interactions
\citep[and reference therein]{1987MNRAS.228..635N,1990A&A...230...37G,
	1993A&A...280..105S,1998ApJ...499..149M}.
\cite{1990A&A...230...37G}, and \cite{1993A&A...280..105S} showed
that tidal effects can increase or decrease 
the strength of the bar, which depends on 
the mass implicated, the pericentre distance,
and the relative phase between the bar and
the companion. 
\cite{1998ApJ...499..149M} simulated close encounters showing
that bar generated by these encounters  are confined to
the Inner Lindblad Resonance, producing slow bars; it
depends on the mass of the perturber. 
Recently, \cite{2017arXiv170306002M} showed how encounters
with low-mass satellite galaxies may cause a delay
in the bar formation compared to the isolated case.
Instead, they can cause an advance in the
bar formation after a small bar 
is already formed in the center, but its amplitude is still
insignificant. They explain that the spiral wave
created by the perturbation can
interfere positively or negatively via
Swing Amplification to form a bar.
Likewise, \cite{2017MNRAS.464.1502M}
showed that the evolution of the bar parameters (strength,
length, and pattern speed) in disk
galaxies, which forms a bar-like structure in isolation,
are not much affected, while such parameters triggered
by a perturbation show some difference with its counterpart.
The angular velocity
of the bar which was triggered by a flyby is slower than
such structure formed by a self-instability of the disk.
Besides, they showed that a slow flyby has a greater
effect on the target galaxy.

Moreover, cosmological simulations show that bars form, 
and then destroy in a response of asymmetric halos and
interactions with substructure \citep{2008ApJ...687L..13R}.
\cite{2014ApJ...790L..33L} used N-body simulations
to investigate the ability of galaxy flyby interactions
to form bars founding that the mass ratio between
the main galaxy and the perturbation determines
some properties of the bar in the target galaxy.
This type of encounters can be as strong as minor merger
\citep{2000ApJ...534..598V} and therefore change
the properties of the disk as well as the
Disk Stabilities, transforming the galaxy in a 
permanent way.

Therefore, flyby encounters appear in both simulations and
observations and such interactions may change the properties 
of the model and in turn the DSP; then the disk might fall in
the instability regimen so it triggers the bar growth. Hence,
the another goal is to study the behavior of the DSP and the 
experimental parameter $\epsilon_m$  as the bar or spirals 
structures evolve in our models, now under perturbation, and 
prove they keep in the stability or instability regimens.

The growth of a barred galaxy has
three main phases \citep{2006ApJ...637..214M},
which are characterized 
by three main observational parameters: length, strength
and pattern speed. \added{In this paper, we also measure these parameters to study the growth of the bar.}
The first phase corresponds to the bar formation
and extends for $\sim2$Gyr; the bar strength and the
bar length grow quickly.  
The second phase is the buckling of the bar where
the vertical symmetry in the bar is broken
weakening the bar. In this phase, the amplitude m=2 of the Fourier
Transform $A_2$  reach a maximum saturating the bar
\citep{2006ApJ...637..214M}.

The final phase of the bar is the secular evolution. 
\cite{1981A&A....99..362S} showed that the bar grows
slowly by increasing its strength and length,
\cite{1981A&A....96..164C}, on the other hand,
reported that bars tend to weaken in the long term.
The rate at which bar parameters change depends on
the properties of the model. \cite{1998ApJ...493L...5D}
showed that the bar slows by dynamical friction in a dense
dark matter halo, while \cite{2013MNRAS.429.1949A} found
that higher is the gas content on the disk, slower is the
growth of the bar. They also found that the halo triaxiality
triggers the bar formation earlier, and leads to considerably
less increase of the bar strength. On the other hand,
the pattern speed of the bar slows down during all these phases 
\citep{1985MNRAS.213..451W, 1991MNRAS.250..161L,2003MNRAS.341.1179A}.

This paper is organized as follows: section \ref{theory} presents
a theoretical view, section \ref{methodology} shows a description of the
N-body simulations and methods, sections \ref{resultsI} and \ref{results} show
the results for isolated and perturbed models, respectively, section 
\ref{discussion} the discussion and section \ref{conclusions} the conclusions.

\section{THEORETICAL INPUT}\label{theory}

We have performed collisionless N-body simulations with
Gadget-2 code \citep{2001NewA....6...79S,2005MNRAS.364.1105S}. 
We present here fifteen simulations of fully
self-consistent models, all of them with a live exponential
disk and live dark matter (DM) halo.  The live halo ensures
disk-halo angular momentum exchange, which plays an important
role in the formation and evolution of bars as discussed by
\cite{2002ApJ...569L..83A}. We simulated
barred and unbarred models aiming to
monitor the disk stability parameters of the disk. 

\subsection{Models of disk galaxies}\label{sectio:mdg1}

The initial conditions were set down following the methodology
delineated by \cite{1999MNRAS.307..162S} (hereafter SW99) and
\cite{2005MNRAS.361..776S} which is based on the analytic model of
\cite{1998MNRAS.295..319M} (hereafter MO98). 

The dark matter mass distribution was modeled with a
\cite{1990ApJ...356..359H} profile, 

\begin{equation}
\rho_{dm} = \frac{M_{dm}}{2\pi}\frac{a}{r(r+a)^3},
\label{eq:Hernquist}
\end{equation}

\noindent with cumulative mass profile $M(<r) = M_{dm} r^2/(r+a)^2$.
This profile has the same dark matter to NFW profile
\citep{1996ApJ...462..563N, 1997ApJ...490..493N}
within the $r_{200}$ radius ($r_{200}$ is the radius of which the mean enclosed dark
matter density is 200 times the critical density, i.e., it 
contains the virial mass). The NFW profile is
often given in terms of the concentration index $c$, defined as
$c=r_{200}/r_s$, where $r_s$ is the scale length of the NFW halo.
We then have the relation

\begin{equation}
a=\frac{r_{200}}{c}\sqrt{2[ln(1+c)-c/(1+c)]}
\label{eq:ah}
\end{equation}

Furthermore, let's define

\begin{equation}
V_{200}^2 = \frac{GM_{200}}{r_{200}}
\label{eq:v200}
\end{equation}

\noindent to be the circular velocity at the virial radius.

The stellar component is modeled with an exponential
surface density profile of scale length $r_d$, i.e.

\begin{equation}
\Sigma(r)=\Sigma_0 e^{-r/r_d}
\label{eq:pd}
\end{equation}

\noindent where $\Sigma_0=M_d/(2\pi r_d)$. The vertical mass
distribution is given by an isothermal sheet with a radially
constant vertical scale length $z_0$. Therefore, three-dimensional stellar
density in the disk is

\begin{equation}
\rho_d(r,z) = \Sigma(r)\left[\frac{1}{2z_0}sech^2\left(\frac{z}{2z_0} \right) \right].
\end{equation}

A self-gravitating model is the one in which
the initial kinetic energy of the spherically symmetric halo may
be computed by assuming that all particles move around the center
on circular orbits, with speed equal to the circular velocity
(SW99 and MO98), so that $E_{kin} = (GM_{200}^2/(2r_{200}))f_c$, where 

\begin{equation}
f_c = \frac{c\left[1-1/(1+c)^2-2ln(1+c)/(1+c)\right]}{2\left[ln(1+c)-c/(1+c)\right]^2},
\end{equation}

\noindent which comes from the change in the total energy resulting from
the different density profile.

The total angular momentum of the halo $\bf J_h$ with total 
energy $E_h$ is often characterized by the dimensionless spin
parameter

\begin{equation}
\lambda_h = \frac{J_h|E_h|^{1/2}}{GM_h^{5/2}}.
\label{eq:lh}
\end{equation}

The disk has a structure of a thin exponential disk, and it
is cold and centrifugally supported. The mass disk $M_d$ is a fraction $m_d$ of $M_{200}$ 

\begin{equation}
M_d = m_dM_{200}
\label{eq:Md}
\end{equation}

In a similar way, the angular momentum of the disk $J_d$ is a fraction $j_d$ of $J_h$ 

\begin{equation}
J_d = j_d J_h
\label{eq:Jd}
\end{equation}

Consequently, the spin parameter of the disk is

\begin{equation}
\lambda_d = \left(\frac{j_d}{m_d}\right)\lambda_h.
\label{eq:lm}
\end{equation}

From this equation, we then determine the scale radius of the 
disk (MO98), given by 

\begin{equation}
r_d = \frac{1}{\sqrt{2}}\lambda_d r_{200} f_c^{-1/2}f_r
\label{eq:rd}
\end{equation}

\noindent where

\begin{equation}\label{eq:fr}
f_r = 2\left[\int_0^{r_{200}}e^{-r/r_d}\frac{r^2}{r_d^3}\frac{v_c(r)}{V_{200}}dr\right]^{-1}.
\end{equation}

Note that in practice the scale length $r_d$ in the initial 
disk is determined iteratively in order to
satisfy equations (13) and (17) from \cite{1999MNRAS.307..162S}.

\subsection{Criterion of Instability}\label{sectio:mdg2}

Instabilities play a very important role in transforming
and regulating the properties of disk galaxies. Local
stabilities are affected by perturbations with lengths
much smaller than the size of the disk. They can be
transient and can regulate the evolution of a disk
by driven features as transient spiral structures and
star formation due to the fragmentation and collapse
of gas clouds. On the other hand,  large disk instabilities (LDI),
which are comparable to the size of the disk,
can cause a significant transformation of the overall disk.
Whenever a disk galaxy has LDI, it will evolve
towards a new stable configuration, erasing information about
the initial conditions under which the system was formed
\citep{2010gfe..book.....M}.

We first focus on disk instabilities that can trigger a bar
in an isolated disk galaxy. 
For that purpose, the most relevant studies are those of 
EF82 and MO98. EF82 used N-body techniques to investigate
disk instabilities of exponential disks embedded in a
variety of halos and found that the bar instability for a
stellar disk is characterized by the parameter $\epsilon_m$:
\begin{equation}
\epsilon_m = \frac{V_{max}}{\sqrt{G M_d/r_d}}.
\label{eq:em}
\end{equation}

They found that if $0.7 \leq \epsilon_m \leq 1.2$
then the disk is unstable to bar formation. Else, if
$\epsilon_m>1.2$, then the disk is stable,
but this parameter seems not to work well
\citep{2013MNRAS.434.1287S,2008MNRAS.390L..69A}.
However, MO98 and SW99 show that disk stability
is characterized by a lower limit of disk spin parameter
$\lambda_d$, then they obtain a relation of this
critical limit $\lambda_{crit}$ with the help of
$\epsilon_m$, but they use an approximation of the
$V_{max}$ in their models; we use the $V_{max}$ 
obtained directly from the N-body simulation and
we deduce a similar relation to MO98
of $\lambda_{crit}$.
Thus, following the same methodology as MO98,
we use the $\epsilon_m$ formula together with
equation \ref{eq:rd} and establishing
$\epsilon_{m,crit}\approx 1$ for the disk stability; then,
we obtain a relation of lower limit of disk spin parameter,
as follow:
\begin{equation}
\lambda_{crit} = \left(\frac{\epsilon_{m,crit}}{V_{max}}\right)^2 G M_d \frac{\sqrt{2f_c}}{r_{200}f_r}=\frac{GM_d}{V_{max}^2}\frac{\sqrt{2f_c}}{r_{200}f_r},
\label{eq:lcrit}
\end{equation}

where $V_{max}$ is the maximum circular velocity, $M_d$ 
is the mass disk, and $f_r$ is an integral which
depends on $r_d$, $v_c(r)$ is the circular velocity, 
$V_{200}$ is the circular velocity given by equation \ref{eq:v200},
and $f_c$ is a function which depends
of the concentration of the halo. Thus, we can find
a disk stable against bar formation if 
$\lambda_d>\lambda_{crit}$, otherwise it is unstable.
It is:

\begin{equation}
\begin{cases}\lambda_d > \lambda_{crit}, & \textrm{ stable against bar formation}\\ \lambda_d \leqslant \lambda_{crit}, & \textrm{ unstable to bar formation}\end{cases}
\end{equation}

This parameter has the advantage that depends on the angular
momentum exchange, and the mass distribution of the 
components, concentration and scale radius. In addition,
it also depends on the velocity dispersion $\sigma$ 
due that for higher $\sigma$ the particles
have less time to exchange energy and angular momentum,
and therefore, $\lambda_d$ varies fewer.
To prove this criterion, we set down four isolated disk models
and these model were subjected to other eleven interactions.

\section{METHODOLOGY}\label{methodology}

This section includes a description of the initial conditions
of the models, which consist of a disk and a Dark Mater 
halo, the interactions, and the tools we develop to analyze
the models.

\subsection{Isolated models setup }\label{isolated_setup}

\begin{figure*}
	\includegraphics[scale=0.35,angle=-90]{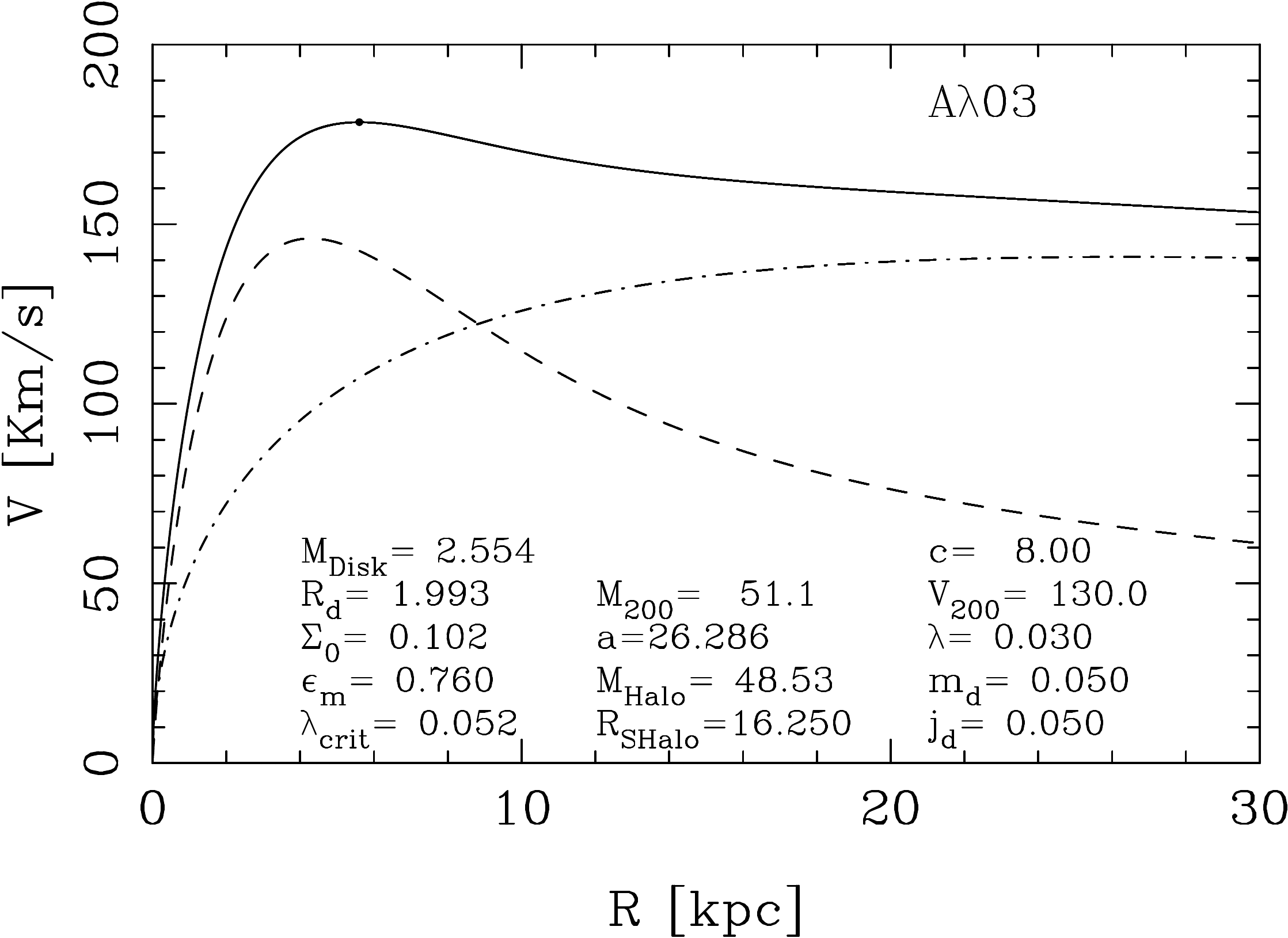}
	\includegraphics[scale=0.35,angle=-90]{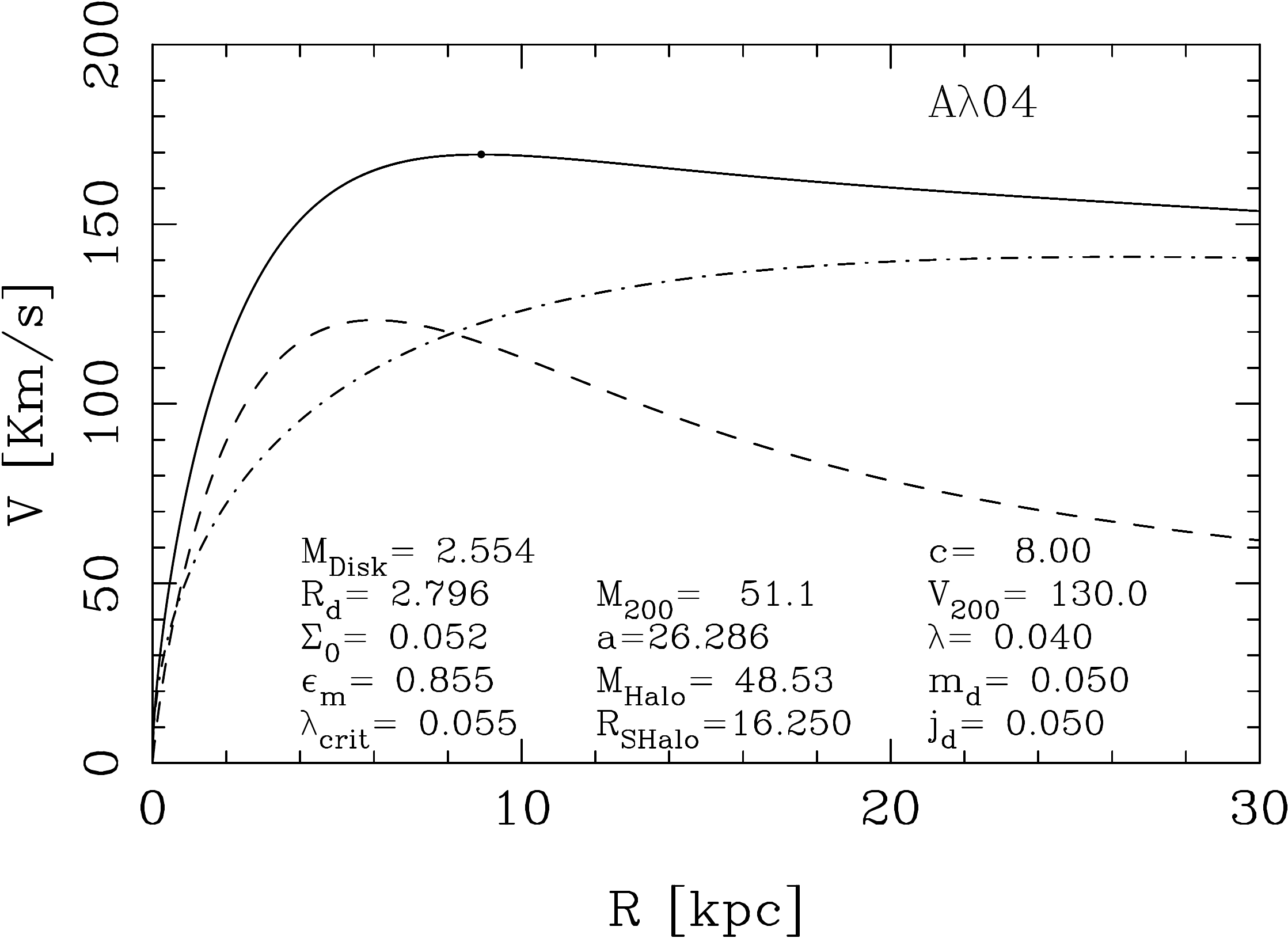}\\
	\includegraphics[scale=0.35,angle=-90]{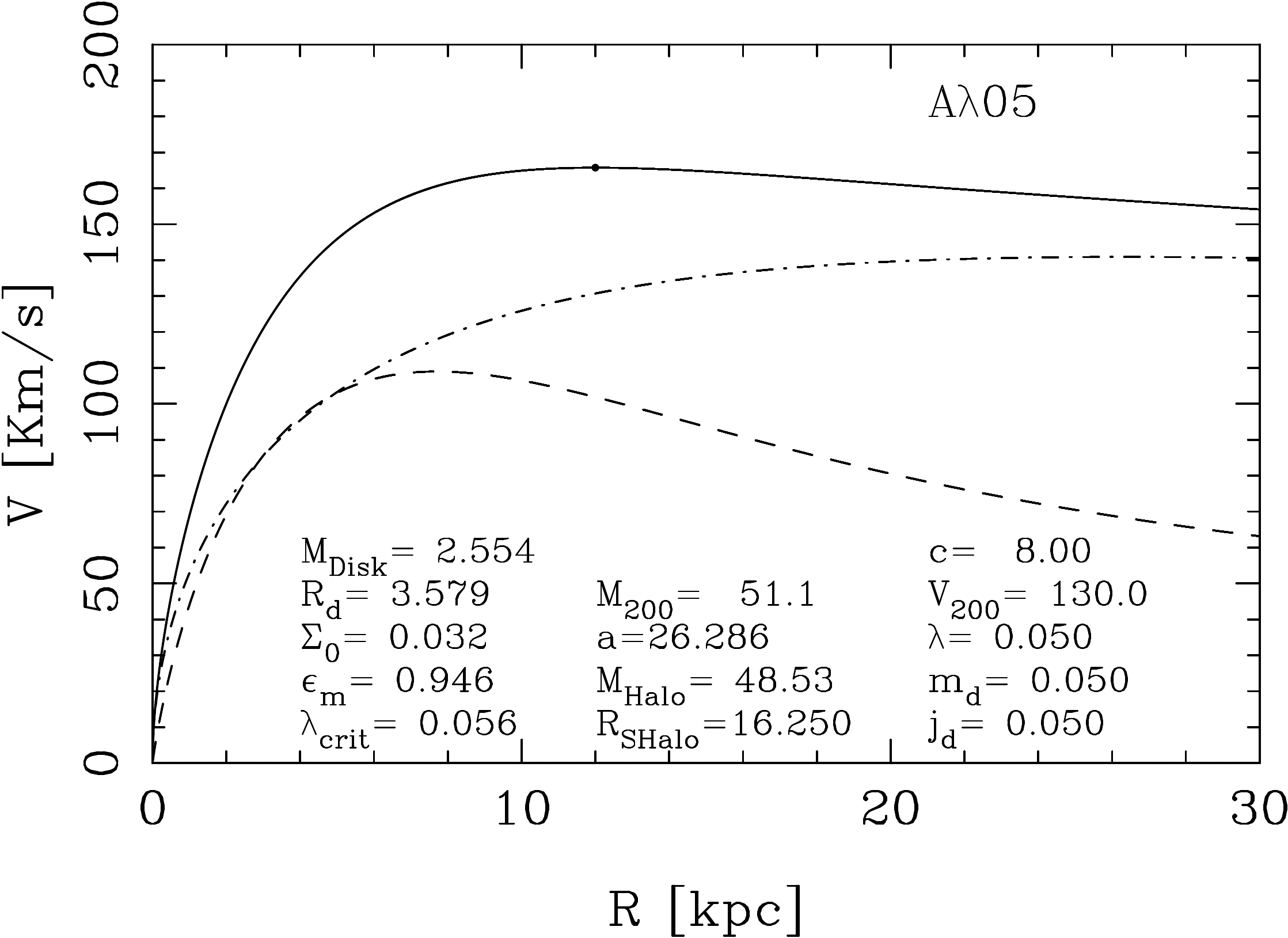}
	\includegraphics[scale=0.35,angle=-90]{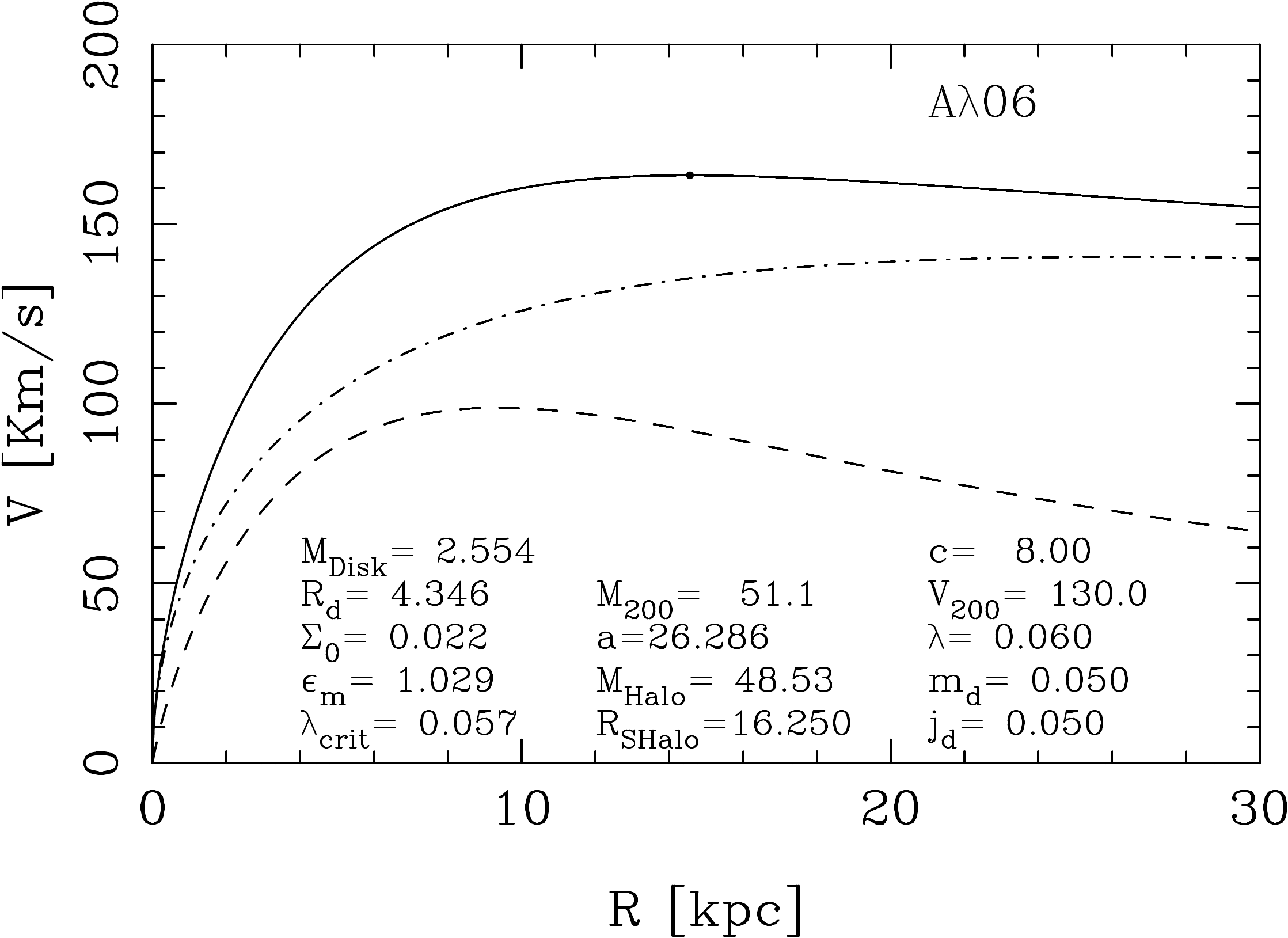}	
	\caption{Initial rotation curves for isolated
	models. Their structural parameters are shown in
	each plot.\label{fig:rotationcurve}}
\end{figure*}

\begin{figure}
	\includegraphics[scale=0.35,angle=-90]{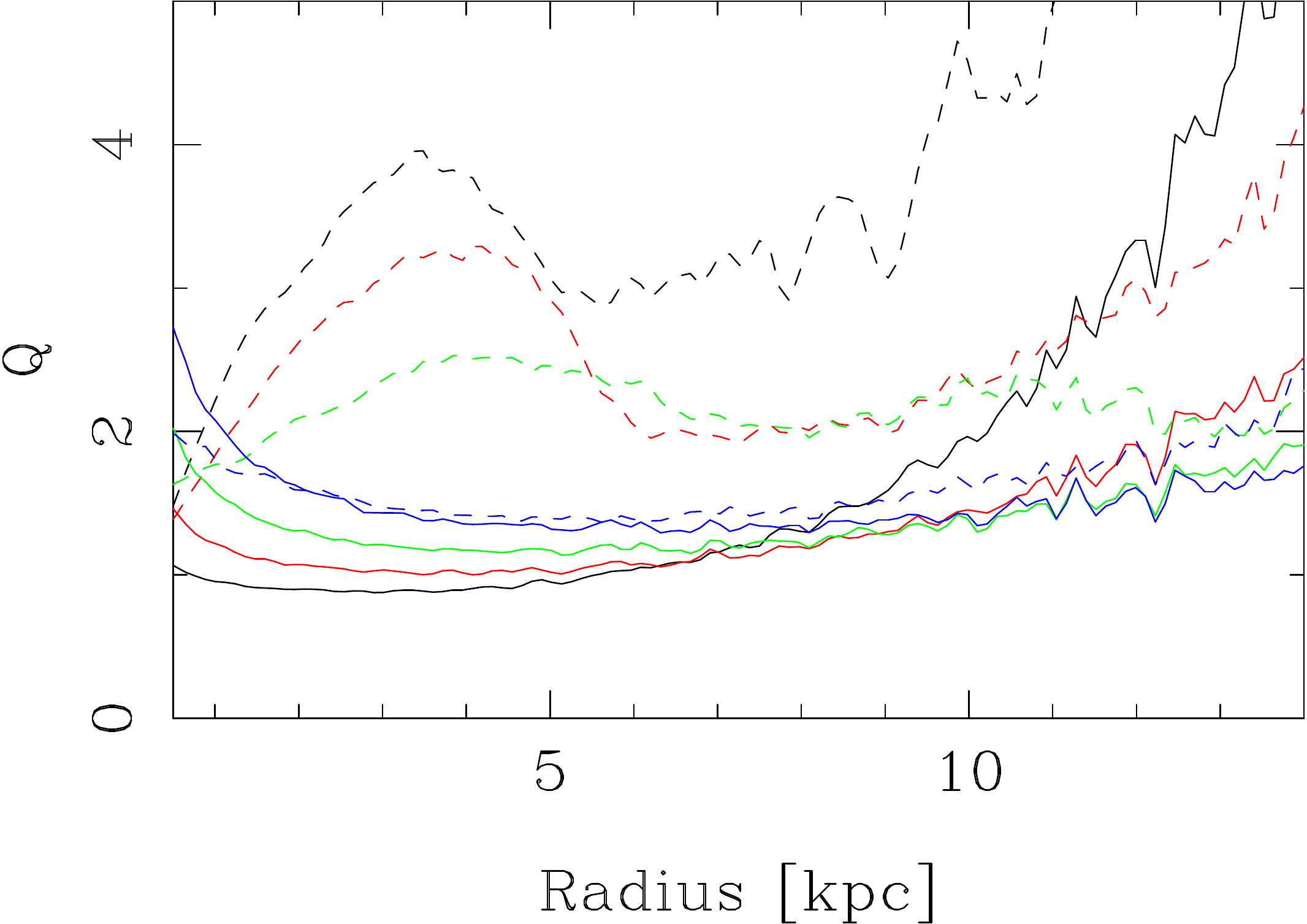}
	\caption{Solid and dashed lines show the initial and
			final $Q$ value for the isolated models.
			Black, red, green, and blue lines represent
			models $A\lambda 03$, $A\lambda 04$,
			$A\lambda 05$, and $A\lambda 06$ 
			respectively.\label{fig:Q}}	
\end{figure}

The initial conditions are generated following the methodology delineated
in SW99 and MO98 as described before.
Our models have been evolved from 0 to 6
Gyr. In the four models we present here, only one initial parameter was modified:
the spin parameter $\lambda\equiv\lambda_h$. Most of
the structural parameters are given in Figure \ref{fig:rotationcurve}. 
It shows models from disk dominate to halo dominate ones.
We remark that these models were generated with the same $M_D/M_H$ ratio,
changing then the disk radial scale length, central surface density
and the Toomre parameter $Q$ as well. The Toomre parameter
$Q$ increases according to the spin parameter in the initial
conditions (Fig. \ref{fig:Q}).

\begin{table}
	\centering
	\caption{\deleted{This table shows the} Initial disk stability parameters for the isolated models.}
	\label{tab1}
	\begin{tabular}{lcccl} 
		\hline
		Models & $\lambda_{d,init}$ & $\lambda_{crit}$ & $\epsilon_m$ & Final Status\\
		\hline
		A$\lambda03$ & 0.03 & 0.052 & 0.76 & Strong bar\\
		A$\lambda04$ & 0.04 & 0.055 & 0.86 & Barred\\
		A$\lambda05$ & 0.05 & 0.056 & 0.95 & Weak bar\\
		\added{A$\lambda05\_M14$} & 0.05 & 0.056 & 0.95 & Weak bar\\
		A$\lambda06$ & 0.06 & 0.057 & 1.03 & Unbarred\\
		\hline
	\end{tabular}
\end{table}

The isolated models have $7\times 10^6$ particles: $2\times 10^6$
to simulate the disk and $5\times 10^6$ to simulate the halo.
We ensured that the mass of the halo particles is not larger
than 8 times the mass of the disk particles. 
\added{Additionally, we repeat the simulation $A\lambda05$
doubling its number of particles to $14\times 10^{6}$: 
$4 \times 10^{6}$ to the disk and $10 \times 10^{6}$ 
to the halo, and also we reduce to half the softening
parameters keeping all the other physical parameters
unchanged. As the analysis show (section \ref{resultsI}), the
main characteristics of the model remain unchanged:
the spin parameter
($\lambda_d$) and the stability experimental parameter 
($\epsilon_m$) time evolution do not depend on these 
numerical parameters.} 

The  simulations were performed with 
the Gadget2 code
\citep{2005MNRAS.364.1105S} with its \replaced{standard units}{default units, where the velocity unit is equal to 1km/s, the length unit is equal to 1kpc and the mass unit 
is equal to $1\times 10^{10}M_{\odot}$}. 
The gravitational forces
were computed with a hierarchical multipole
expansion, which short-range forces are computed
with the `tree' method while long-range forces are
determined with Fourier techniques
with a tolerance parameter $\theta_{tol} = 0.5$.
The softening length for 
the disk particles is $\epsilon=0.01$ and for the halo
ones is $\epsilon= 0.1$. They are
chosen so that 
the maximum inter-particle force shall not exceed the typical 
mean-field strength \citep{2011EPJP..126...55D}.
Thus, we ensure 
that two-body relaxation will not
artificially induce chaotic orbits.
Time integration is
based on a quasi-symplectic
scheme where long-range and short-range forces
can be integrated with different time-steps
given by $\Delta t = \sqrt{2\eta\epsilon/|\vec{a}|}$
where $\eta=0.01$ and $\vec{a}$ is the acceleration of
the last time-step.
With these parameters, we ensure that the energy
conservation was better than $10^{-3}$. We assessed
the numerical robustness by experiment with less
number of particles and bigger softening according to
\cite{2011EPJP..126...55D}, getting similar results.

\subsection{Setting up the encounters}\label{interaction_setpu}

We use the models setted up in \ref{isolated_setup}
to subject them to a perturbation. In \ref{isolated_setup}, 
we modeled four disk galaxies
where we only change the spin parameter 
to get two classical models; two models where the disk
dominates, another one where the halo dominates and 
another one where the disk and the halo have the
same contribution in the inner region of the
rotation curve. 

\begin{table*}
	\centering
	\caption{This table shows the status of the models
	at the beginning and at the end of the simulations, 
	the masses of the target and perturbed model, and
	some measurements at the pericentre; $M_D$ disk
	mass, $M_t$ total mass of the target model, $M_p$
	total mass of the perturbed model, $t_p$ time at
	the pericentre, $r_p$ pericentre distance,
	$\Omega_p$ angular velocity of the perturber the
	time at the pericentre, $\alpha$ angle between 
	the bar and the perturber, and $F_i$ the
	interaction force.\label{tab2}}
	\begin{tabular}{lccccccccccl} 
		\hline \hline
		Group&Models & Status at the Pericenter & $M_D$ & $M_t$ & $M_p$ & $t_p$ & $r_p$ & $\Omega_p$ & $\alpha$ & $F_i$ & Final Status\\
		\hline \hline
		  &$PwA\lambda03$     & Strong bar   & 2.6 & 51.1 &23.8&2.95&18.99&34.03&-8.37 &-19976& Strong bar\\	
		Pw &$PwA\lambda04$     & Weak bar     & 2.6 & 51.1 &23.8&2.95&17.56&37.12&-78.53&-20747& Strong bar\\
		&$PwA\lambda05$        & Unbarred     & 2.6 & 51.1 &23.8&2.95&17.21&37.64&76.42 &-21320& Barred\\
		&$PwA\lambda06$      & Unbarred     & 2.6 & 51.1 &23.8&2.95&17.65&37.01&27.39 &-20642& Barred\\ 
		&\added{$PwA\lambda06$\_M2 }       & Unbarred     & 2.6 & 51.1 &23.8&2.95&17.63&37.35&25.70 &-20519& Barred\\ \hline
		&$PmA\lambda03$        & Strong bar   & 2.6 & 51.1 &47.6&2.92&17.28&34.43&-47.98&-31235& Strong bar\\
		 &$PmA\lambda04$      & Weak bar     & 2.6 & 51.1 &47.6&2.95&17.56&37.12&-76.79&-30963& Strong bar\\
		Pm  &$PmA\lambda05$    & Unbarred     & 2.6 & 51.1 &47.6&2.92&15.78&38.32&49.53 &-32492& Barred\\
		&$PmA\lambda06$        & Unbarred     & 2.6 & 51.1 &47.6&2.92&16.12&37.61&17.47 &-31563& Barred\\ \hline
		&$PsA\lambda03$        & Strong bar   & 2.6 & 51.1 &95.2&2.92&16.94&40.70&-57.19&-46415& Strong bar\\
		 &$PsA\lambda04$      & Weak bar     & 2.6 & 51.1 &95.2&2.92&15.85&44.75&70.91 &-47123& Strong bar\\
		Ps  &$PsA\lambda05$    & Unbarred     & 2.6 & 51.1 &95.2&2.92&15.47&44.88&42.33 &-47484& Strong bar\\
		&$PsA\lambda06$        & Unbarred     & 2.6 & 51.1 &95.2&2.92&15.72&44.93&11.97 &-46217& Barred\\ \hline	
		\hline
	\end{tabular}
\end{table*}

\begin{figure}
	\centering
	\includegraphics[scale=0.47]{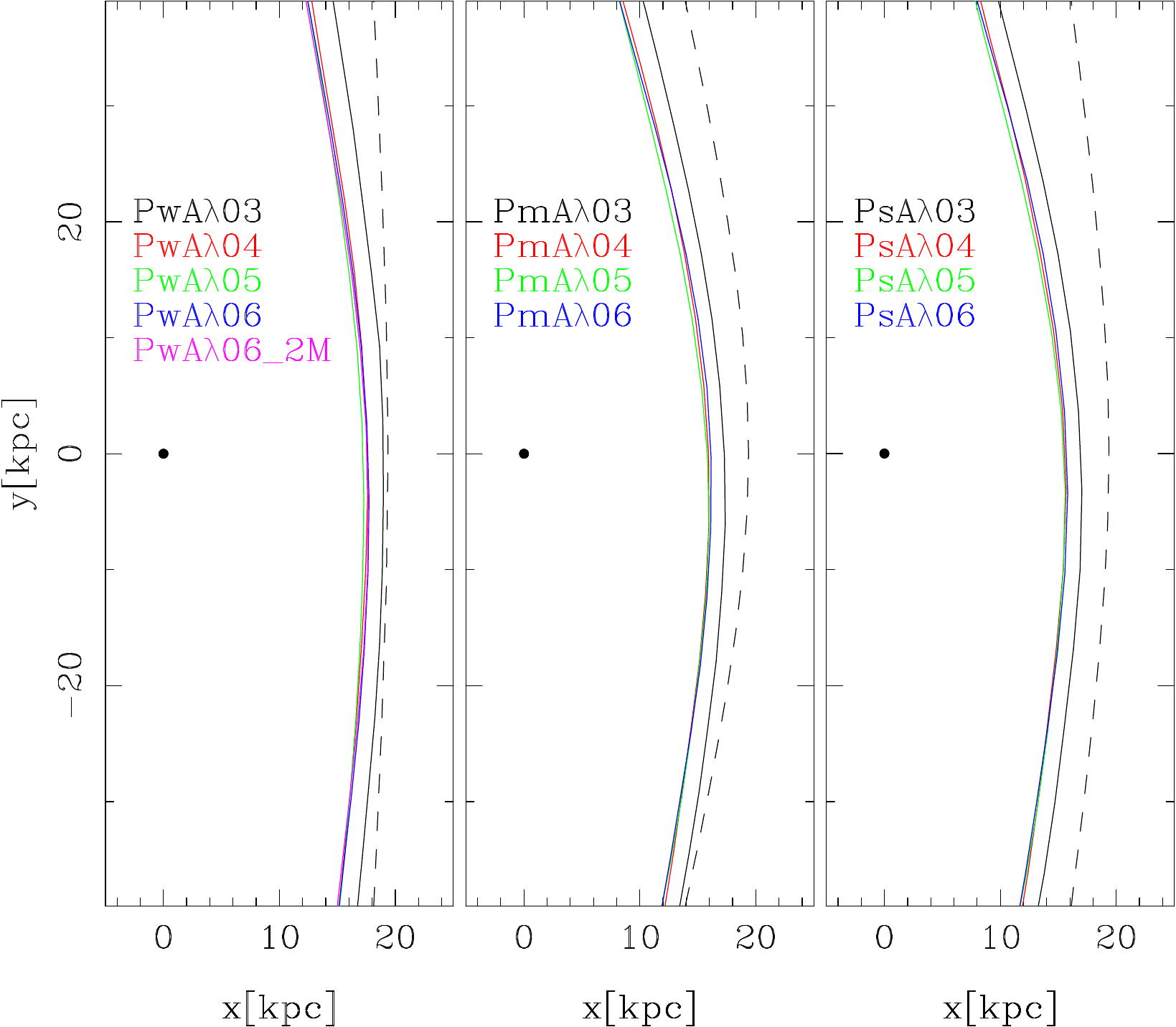}
	\caption{Orbits traced by the perturbation from the
	disk density center (the bar) of the target
	galaxy, the color lines are represented in the
	figure, and the dashed lines show the theoretical
	orbits. We can see that the pericentre is less 
	than the pericentre calculated by the theoretical
	orbit.\label{fig:orbs}}
\end{figure}

We use an Hernquist  profile to built the stellar and 
the halo component of the perturbation using the methodology
established by \cite{1999MNRAS.307..162S} and 
\cite{2005MNRAS.361..776S}. We set up three groups
of encounters where we change the mass of the
perturbed to have different tidal forces
at the pericentre;
table \ref{tab2} enlist such interactions.
The first group (Pw) are interactions where 
the total mass of the perturber is approximately half of
the total mass of the target model, the second group (Pm)
are interactions where the total mass of the perturber
is similar to the total mass of the target model,
and finally, the last one (group Ps) are interactions
where the the total mass of the perturber is almost
two times the total mass of the target model.

Later in the article, we will show that 
isolated models form a bar 
at different times, e.g. in model A$\lambda$03 after the
first gigayear, in model A$\lambda$04 after the
second gigayear, and in model A$\lambda$05  at the
end of the simulation while model A$\lambda$06 does not form
a bar-like structure. We set up the interactions from the
second gigayear in \added{these} isolated models, 
and we calculate the
passage at around the third gigayear to analyze
the tidal perturbation at different stages of the bar
as well as the growth of the bar in the stable model.

The interactions follow a coplanar, prograde
and hyperbolic orbit around the disk
reaching the pericentre at the first gigayear of its
evolution (third gigayear in our interactions). 
We set the pericentre distance for all
interactions approximately to $r_p=3r_{75}=19.4kpc$,
where $r_{75}$ is the radius containing
the 75\% of the disk mass of model A$\lambda$03. 

Due to our experience, we use the mass contained 
up to the radius $r_p$ of target galaxy $M_t(r_p)$ 
and perturbation galaxy $M_p(r_p)$ as a point mass
to calculate an approximation of the orbit 
that the perturbation will trace like two bodies; where
\begin{equation}
M_t(r_p) = M_{Ht}(r_p)+M_{Dt}(r_p),
\end{equation}

and
\begin{equation}
M_p(r_p) = M_{Hp}(r_p)+M_{Sp}(r_p),
\end{equation}
where the subscripts $H$, $D$, and $S$ mean the halo, disk, and
spherical components.

The dark matter mass distribution was modeled with an
\cite{1990ApJ...356..359H} profile (eq. \ref{eq:Hernquist}),
this profile has the same dark matter to NFW profile
\citep{1996ApJ...462..563N, 1997ApJ...490..493N}, i.e.
\begin{equation}
\rho(r)=\rho_{crit}\frac{\delta_0}{(r/r_s)(1+r/r_s)};
\end{equation} 
within the $r_{200}$ radius to be related to cosmological
halos ($r_{200}$ is the radius of which the mean enclosed dark
matter density is 200 times the critical density
$\rho_{crit}$, i.e. it contains the virial mass).
The NFW profile is often given in terms of the 
concentration index $c$, defined as $c=r_{200}/r_s$, 
where $r_s$ is the scale length of the NFW halo,
and the characteristic over-density is
\begin{equation}
\delta_0=\frac{200}{3}\frac{c^3}{\ln(1+c)-c/(1+c)}.
\end{equation}

Then, the mass within radius $r_p$ for the halos $M_{Ht}$ 
and $M_{Hp}$ is:
\begin{equation}
M_H(r_p)=4\pi\rho_{crit}\delta_0r_s^3\left[\frac{1}{1+cx}-1+\ln(1+cx)\right],
\end{equation}
where $x=r_p/r_{200}$.

The stellar component of the perturbation $M_{Sp}$
was \added{also} modeled with an Hernquist profile 
\deleted{too}; therefore the mass contained at $r_p$ radius is
\begin{equation}
M_{Sp}(r_p)=M_{Sp0}\frac{r_p^2}{r_p^2+a},
\end{equation}
where $M_{Sp0}$ is the total mass of the stellar
component for the perturbation galaxy, and the mass
within radius $r_p$ of the exponential disk  is then
\begin{equation}
M_{Dt}(r_p) = M_D\left[1-\left(1+\frac{r_p}{r_d}\right)e^{r_p/r_d}\right].
\end{equation}

Then, we get the position and the initial velocity of the
perturber to get that orbit from the motion equations of two bodies.
Obviously, this theoretical
orbit is slightly different from the simulated orbit.
Figure \ref{fig:orbs} compare the theoretical and
the simulated orbits.

The target models are $A\lambda03$, $A\lambda04$, 
$A\lambda05$ and $A\lambda06$, the perturbation model 
have $1\times 10^6$ particles: $1\times 10^4$ to simulate
the stellar component (it is a Hernquist profile), and 
$9.9\times 10^5$ to simulate the halo. 
The  simulations were performed with 
the Gadget2 code\citep{2005MNRAS.364.1105S} 
\deleted{with the standard units} where the tolerance
parameter is $\theta_{tol}=0.5$,
\replaced{the softening length for the stellar component
of the perturbation is $\epsilon = 0.01$ and for the halo is
$\epsilon = 0.01$.}{and the softening length is 
$\epsilon=0.01$}. In the same form as we mention in section
\ref{isolated_setup},
\added{ we perform 	again the interaction $PwA\lambda06$ but 
	doubling the number of particles in the perturbation
	($2\times 10^6$ particles: $2\times 10^4$ to simulate
	the stellar component, and $1.98\times 10^6$ to simulate the halo,
	$PwA\lambda06$\_2M); also, we reduce to half
	the softening parameters keeping all the other physical
	parameters unchanged. The analysis of all characteristics of the perturbed disk (section \ref{results}), shows that the evolution of the disk is very similar to the case with a perturbation of low N particles. Likewise,}
we assessed the numerical robustness by experiment with less
number of particles and bigger softening according to
\cite{2011EPJP..126...55D}, getting similar results.

\subsection{Measurement of parameters}

We determine the spin parameters $\lambda_d$
using equation \ref{eq:lh} and equation \ref{eq:lm}
from the phase space of the simulation.
The $\lambda_{crit}$ was calculated performing
equation \ref{eq:lcrit}. First, we calculate the potential
as function of radius, then we determine the derivative
of this curve to get the rotation curve $v_c(r)$
for a snapshot, and so we obtain the $V_{max}$.
To calculate $f_c$, we fit the halo profile to the NFW profile
to determine the scale radius of the halo $r_s$ and
the radius $r_{200}$,
then we calculate the halo concentration
using the equation \ref{eq:ah}. Finally, we fit the profile of the disk
to the exponential disk (eq. \ref{eq:pd}) to
obtain the scale length radius of the disk, then
we calculate the integral $f_r$ (eq \ref{eq:fr}).
Also, we use these parameters to calculate
the experimental stability parameter $\epsilon_m$.
It should be mentioned that we calculate all these
parameters using the phase space for each saved snapshot
of the simulations.

As we said before, the bars are characterized by three main 
observational parameters: length, strength and pattern speed.
In order to measure these parameters, we compute Fourier coefficients
for modes from $m=1$ to $m=10$ in the disk particles and monitor 
their amplitudes and phases variation across the disk 
as a function of time.
We use the amplitude of $m=2$ to measure the strength and 
the growth rate of the bar which is being formed in the disk. 
The phase of $m=2$ of the Fourier coefficients also was used to
calculate the instantaneous angular velocity of the bar $\Omega_b$, 
and then, it was used to fix the bar reference frame.
From the bar reference frame,
we calculate the bar axes length by providing
a density threshold to define the bar limits.
In order to measure the size of the bar; first,
we obtain a profile of the bar along with both the
major and minor axes dividing them into cells and then
calculating their surface density; so,
the limits of the bar were calculated using the density
threshold provided. The next step was to change iteratively
the size or number of cells until to get a
convergence of $10^{-2}$ on the length of bar axes.

\section{RESULTS FOR ISOLATED MODELS}\label{resultsI}

In this section, we describe the growth of the bar,
the calculations of the disk instabilities through the time,
and the results from the Fourier Transform analysis. 

\subsection{Stability criterion through the time}

\begin{figure*}
	\includegraphics[scale=1.3,angle=-90]{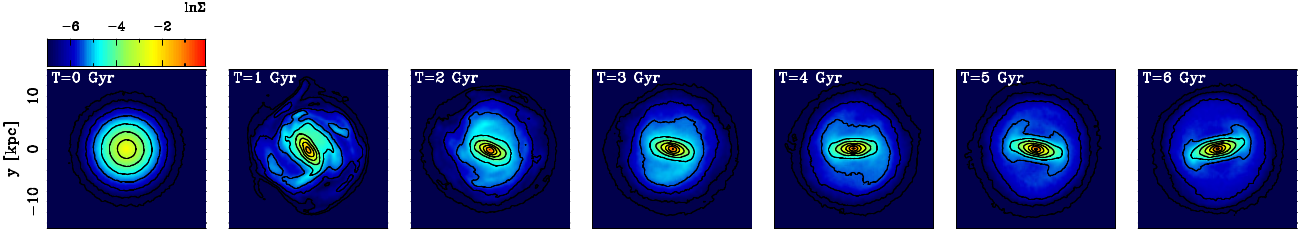}
	\includegraphics[scale=1.3,angle=-90]{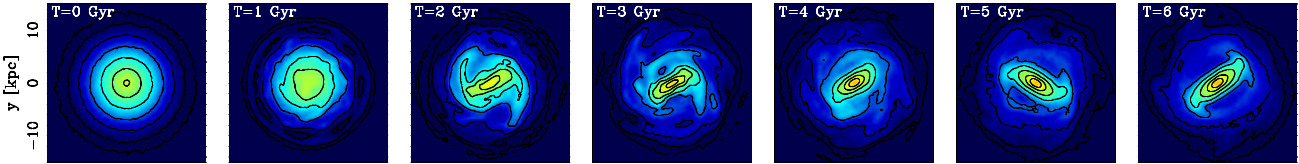}
	\includegraphics[scale=1.3,angle=-90]{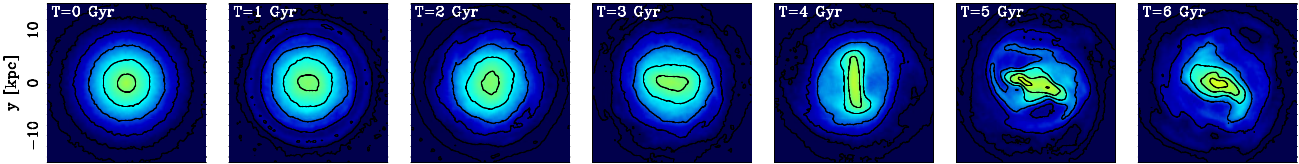}
	\includegraphics[scale=1.3,angle=-90]{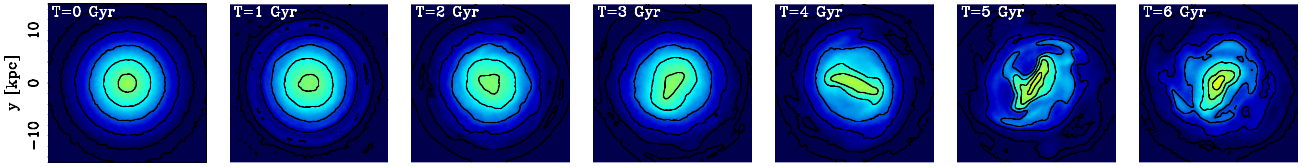}
	\includegraphics[scale=1.3,angle=-90]{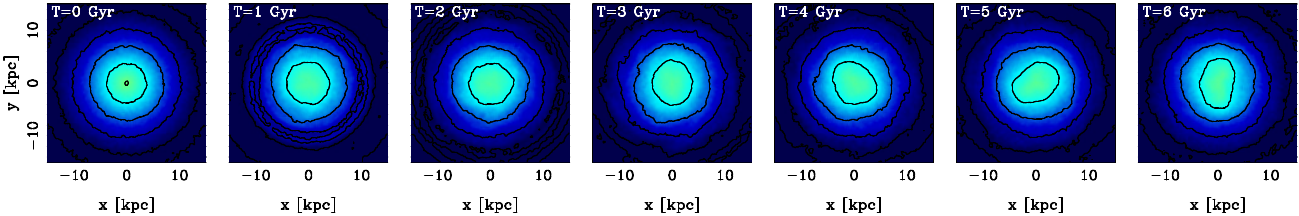}
	\caption{Face-on surface logarithm density maps for our
	\added{isolated}
	models and their evolution. The time increase to the right
	in units of Gyr. Color scale at the top left correspond
	to the logarithm of the surface density. 
	From top to bottom, we present the evolution of models $A\lambda03$,
	$A\lambda04$, $A\lambda05$, \added{$A\lambda05$\_M14} 
	and $A\lambda06$, respectively.
	We can observe how the bar forms and evolves in the three upper models.}
	\label{fig:snaps}
\end{figure*}

\begin{figure}
	\includegraphics[scale=0.34,angle=-90]{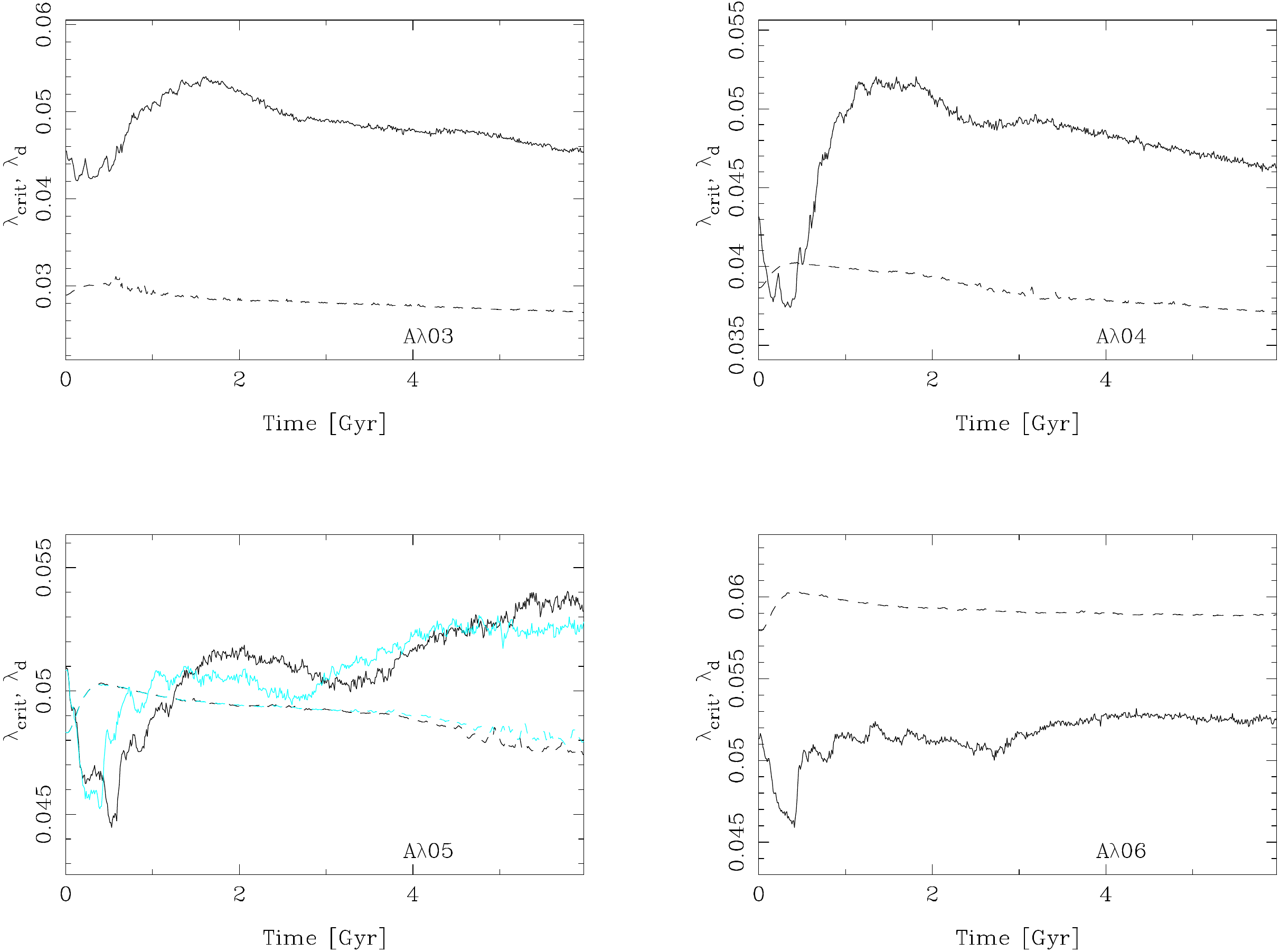}
	\caption{The spin parameter of the disk $\lambda_d$ (equation \ref{eq:lm})
		is depicted with a dashed line, while its critical spin
		parameter $\lambda_{crit}$ (equation \ref{eq:lcrit}) is depicted with a continuous
		line as a function of time. The model name is given at the bottom right corner. 
		\added{The left bottom panel depict the comparison between $A\lambda05$ (black lines) and $A\lambda05$\_M14 (cyan lines) models.}
		We can observe that models forming a bar 
		exhibit $\lambda_d<\lambda_{crit}$, 
		and the stable model against the bar formation 
		exhibits $\lambda_d>\lambda_{crit}$.\label{fig:lambdas}}	
\end{figure}

\begin{figure}
	\includegraphics[scale=0.7,angle=-90]{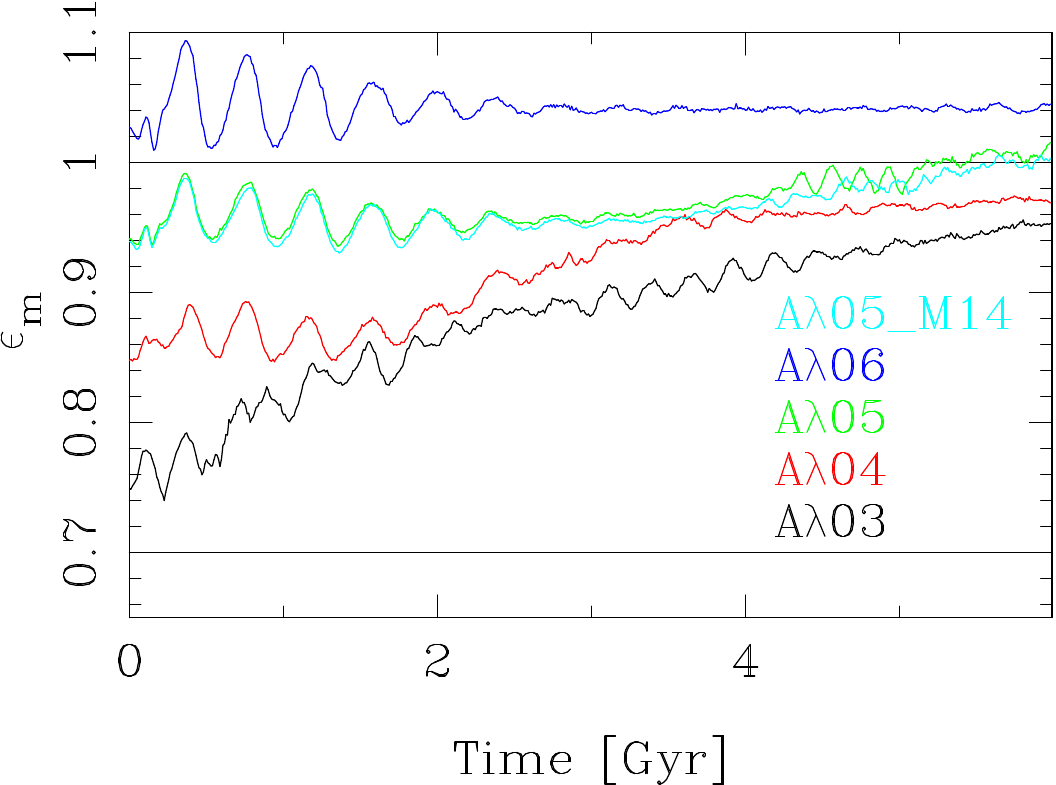}
	\caption{This figure shows the evolution of stability 
	experimental parameter $\epsilon_m$ defined in EF82. We observe how
	the bar models begin in the range $0.7<\epsilon<1$
	and evolve asymptotically toward the unity, while
	the stable model keeps $\epsilon_m$ beyond the unity.
	\added{The $A\lambda05$ and $A\lambda05$\_M14 models
	show a very similar behavior.}}
	\label{fig:em}	
\end{figure}

Figure \ref{fig:snaps} shows the face-on logarithm
surface density maps for all of our models, \added{including
the $A\lambda05$\_M14 model,} at times of
0, 1, 2, 3, 4, 5, 6 Gyr. Model $A\lambda03$ on top row
has the lowest value of $\epsilon_m$ and $\lambda_{d}$
being $\lambda_{d}<\lambda_{crit}$ (see Figure
\ref{fig:lambdas}), this model maintains
$\lambda_{d}<\lambda_{crit}$ for the full evolution.
This model forms a bar very quickly and it is kept
during all the evolution. Model $A\lambda04$ is 
shown in the second row; it also has 
$\lambda_{d} < \lambda_{crit}$ and $0.7<\epsilon_m<1$
(see Figure \ref{fig:em}),
and it forms the bar around two to three Gyr, and
it is also maintained during the entire evolution.
The third row shows the model $A\lambda05$; this model
has $\lambda_{d} \approx \lambda_{crit}$ and
$\epsilon_m$ a little less than unity (see table
\ref{tab1}). It forms a weak bar perturbation around
the fourth Gigayear; \added{besides, the $A\lambda05$\_M14 model has also a very similar nature.
We notice that the numerical resolution affects modestly the angular velocity of the bar structure as we can see in Figure \ref{fig:snaps} where the position of bar is different between these models at the same snapshot; however, we observe that the DSP does not have a high dependence on the resolution like shows the Figures \ref{fig:lambdas} and \ref{fig:em}.} 
The last row present the model
$A\lambda06$; this model has 
$\lambda_{d} > \lambda_{crit}$ and $\epsilon_m>1$.
This model is stable and shows some weak spiral waves.

Figures \ref{fig:lambdas} show the evolution of the DSP
($\lambda_d$, and $\lambda_{crit}$) for our
simulations. Model $A\lambda03$, in upper left panel
of figure \ref{fig:lambdas}, begins with 
$\lambda_d <\lambda_{crit}$ and this configuration 
is maintained in the entire simulation despite 
the formation of the bar. \replaced{This behavior is 
similar in model $A\lambda04$}{The $A\lambda04$ model presents a similar behavior as mentioned above} (see upper right 
panel of Figure \ref{fig:lambdas}). 
\replaced{Model $A\lambda05$, in bottom left panel,
 starts}{The $A\lambda05$ (black lines) and 
 $A\lambda05$\_M14 models (cyan lines)
in bottom left panel, start} with
 $\lambda_{d} \approx \lambda_{crit}$, but
once the bar begins to form, $\lambda_{d}$
becomes smaller than $\lambda_{crit}$
and remains like that to the end of the simulation.
The model $A\lambda06$, in bottom right panel, shows
that the spin parameter is larger than the critical spin
parameter during the whole evolution. 

\begin{figure}
	\includegraphics[scale=0.5]{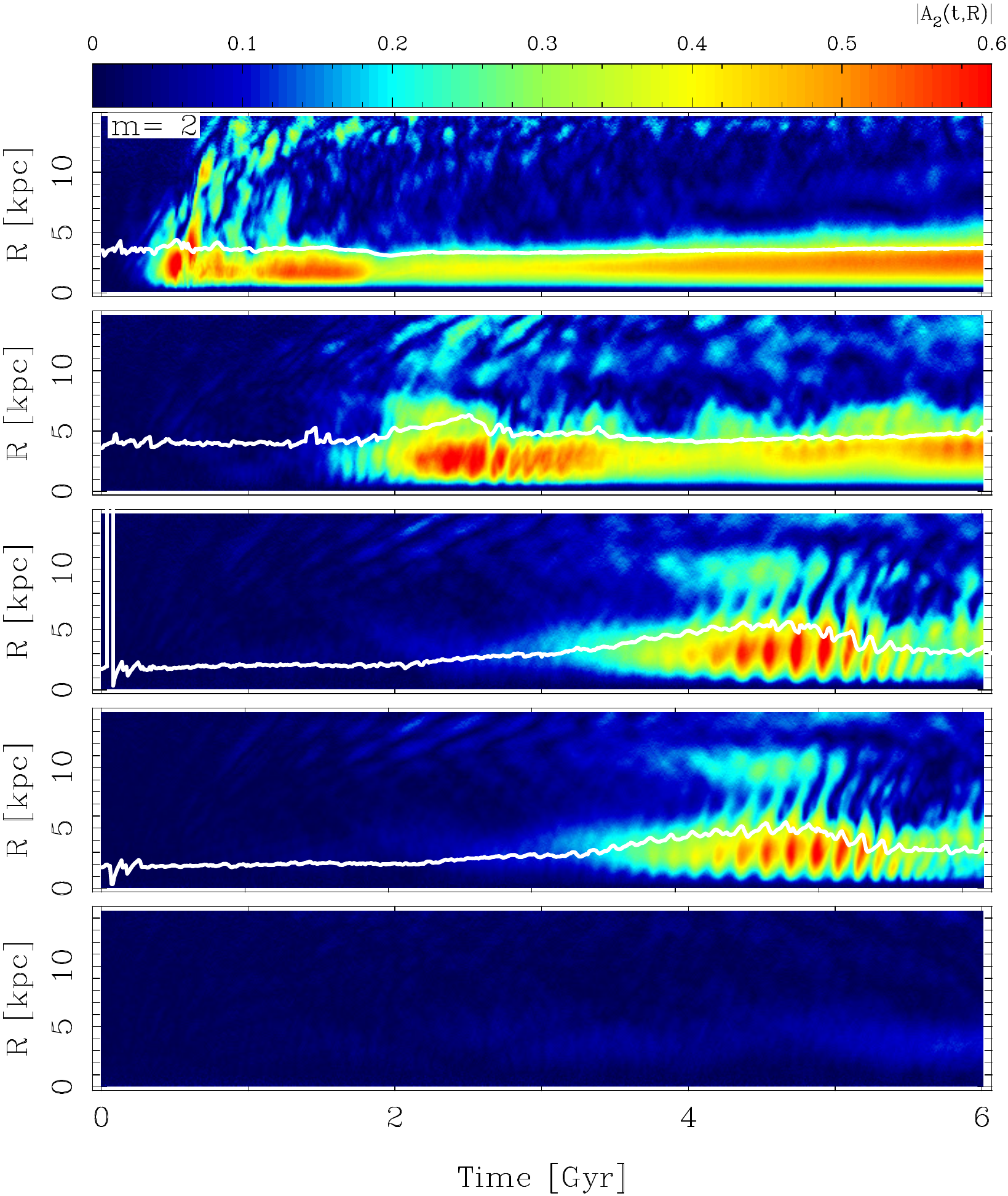}
	\caption{The figure presents the $m=2$ Fourier
		amplitude in all of our \added{isolated} models as a
		function of time and radius. From top to bottom, models
		$A\lambda03$, $A\lambda04$, $A\lambda05$, 
		\added{$A\lambda05$\_M14} and $A\lambda06$. 
		\added{Note that the $A\lambda05$\_M14 
		$A\lambda05$ models present a very similar time
		evolution.}
		We clearly see as the increase of the $\lambda$ value
		delays the formation of the bars.}
	\label{fig:AmFT1D_I}	
\end{figure}

\begin{figure}
	\includegraphics[scale=0.34,angle=-90]{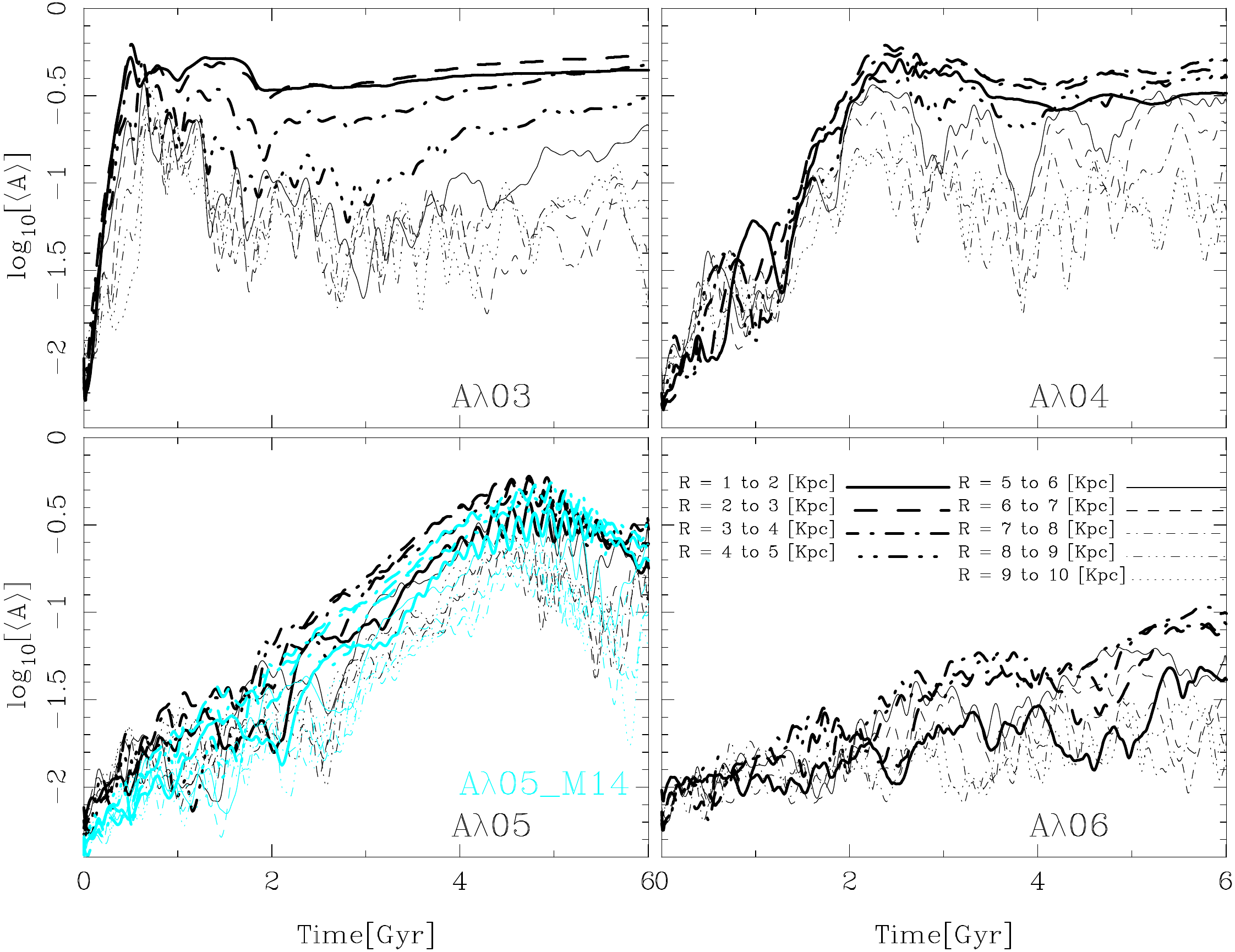}
	\caption{The average of the FT1D amplitude for some radial
		ranges for $m = 2$. In these plots, one can note as the amplitude for inner
		radii (thicker lines) grows and becomes more or less constant
		representing the formation of the bar in the central part of the disk. 
		The amplitude for outer radii (thinner lines) oscillates representing 
		the formation of transient spiral structures.
		\added{The left bottom panel also show the amplitude
		of the $A\lambda05$\_M14 model which is akin to
		the experiment of low N particles.}}
	\label{fig:averaStrong}	
\end{figure}

The experimental stability parameter,  
$\epsilon_m$, is plotted in Figure \ref{fig:em}.
This Figure shows the evolution
of $\epsilon_m$ for all of the models. The black line depicts
the $A\lambda03$ model, red line depicts the $A\lambda04$
model, green line shows the $A\lambda05$ model,
\added{the cyan line depicts the $A\lambda05$\_M14 model} and 
blue line depicts the model $A\lambda06$. 
While the evolution of
$\epsilon_m$ parameter shows values in the range 
from 0.7 to 1 for barred models,
the evolution of $\epsilon_m$ for the $A\lambda06$ model
displays values larger than one. Furthermore, 
the $\epsilon_m$ parameter
for the models $A\lambda03$ and $A\lambda04$ show an increase
during the evolution, but not exceed the unity while 
$\epsilon_m$ for the models $A\lambda05$ 
\added{and $A\lambda05$\_M14} fluctuates 
around the unity at the end of simulation.
\added{This last model evolves very similarly to the
$A\lambda05$ model which means that the DSP and the 
$\epsilon_m$ parameter do not depend on the number of
particles in our models.}

\subsection{Growth of the bar}

\begin{figure*}
	\includegraphics[scale=0.19,angle=-90]{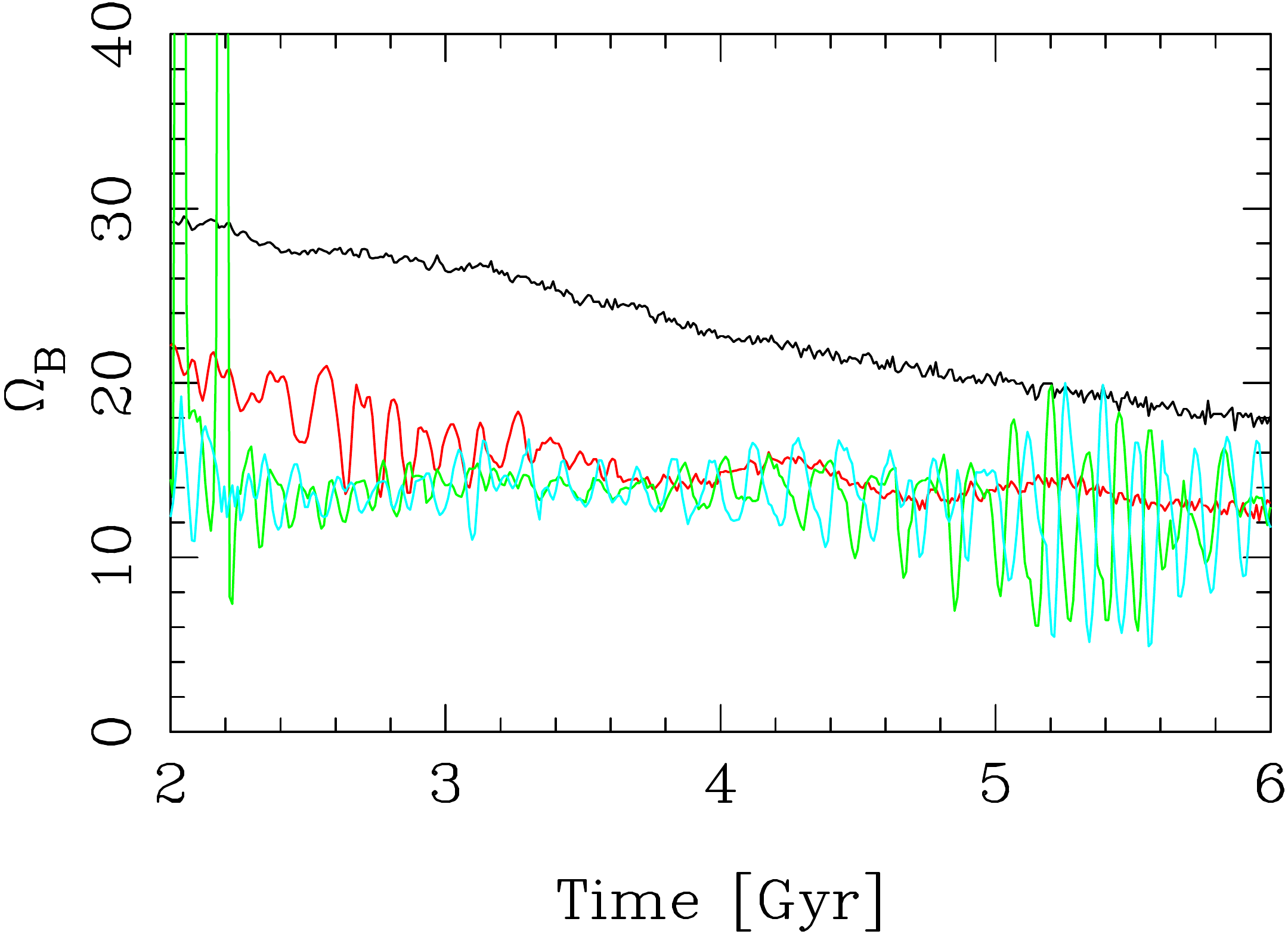}
	\includegraphics[scale=0.19,angle=-90]{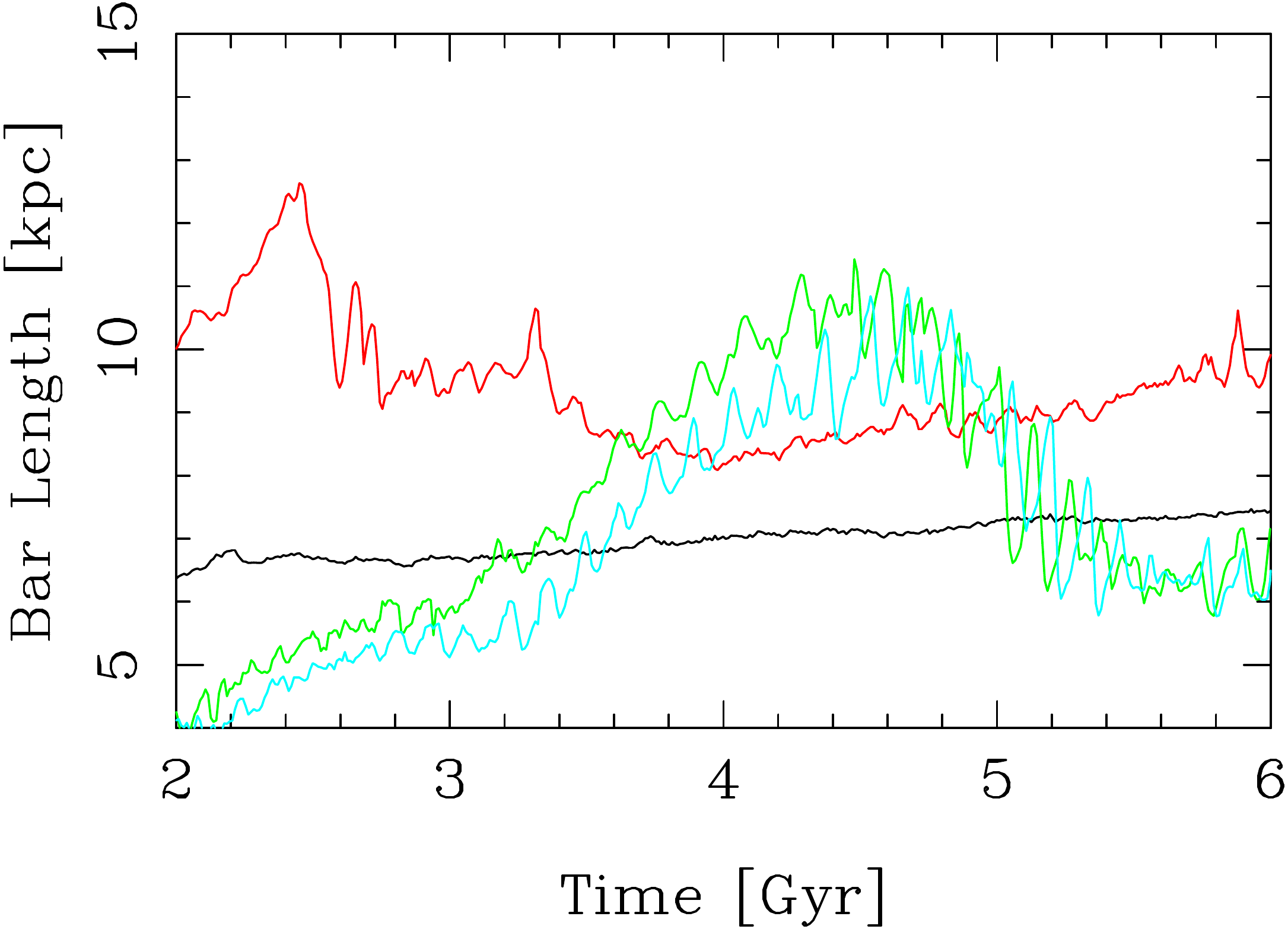}
	\includegraphics[scale=0.19,angle=-90]{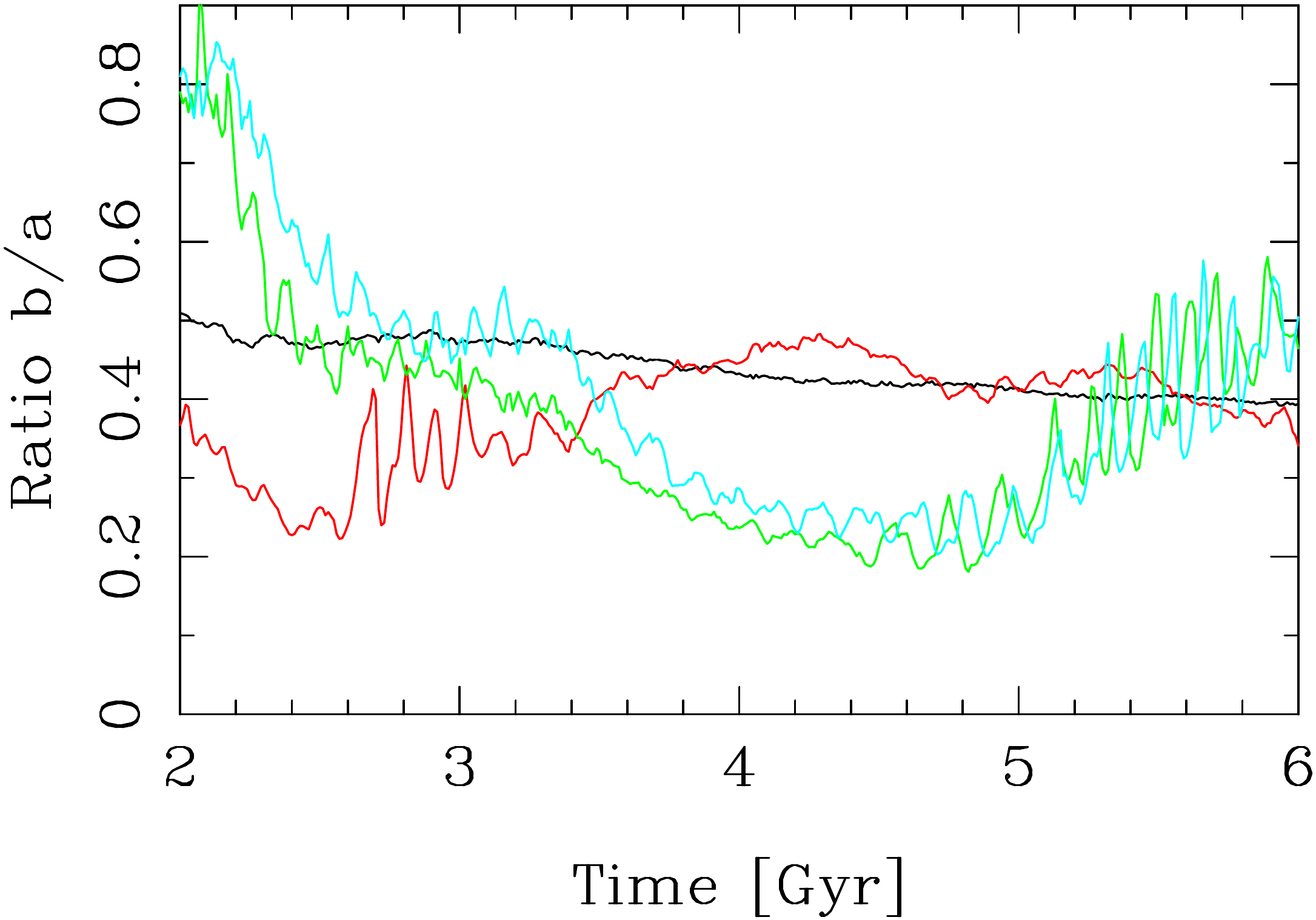}
	\includegraphics[scale=0.19,angle=-90]{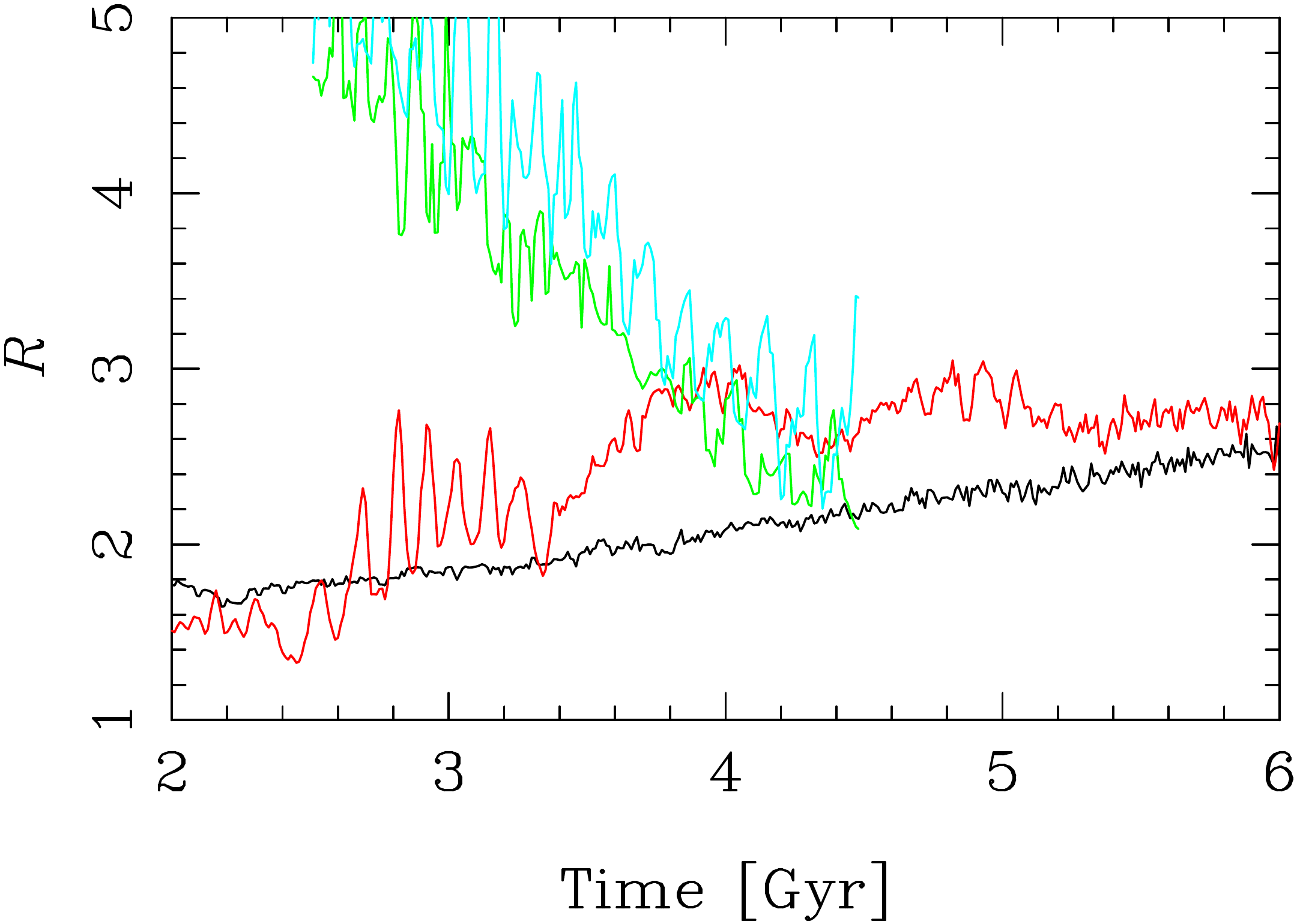}
	\caption{Time evolution of some bar parameters. 
		Black, red and green lines correspond to values for $A\lambda03$,
		$A\lambda04$ and $A\lambda05$, respectively. First panel
		shows the instantaneous bar pattern speed, second panel
		shows the length of the bar, the third panel shows the ratio between the minor and major axes of the
		bar, the last panel shows the $\mathcal{R}=R_{CR}/R_{bar}$ parameter.}
	\label{fig:OmLenR_I}	
\end{figure*}

The evolution of the bar and the spiral structure is
shown in Figure \ref{fig:AmFT1D_I}, in which we plot
the Fourier Amplitude for the $m=2$ mode, $A_2(t,R)$.
For model $A\lambda03$ (top panel),
we observe that the rapid growth of the bar is followed
by spiral waves up to the 
$A_2$ reaches a maximum; after that, the disk rises
its velocity dispersion maintaining the bar perturbation, 
and some weak spiral waves are driven transiently by the bar
\citep{1980A&A....88..184A,2010ApJ...715L..56S}.
The second panel, model $A\lambda04$,
the bar growts at around 2-3 Gyr, and 
the spiral structures generated here are stronger 
than those generated in model $A\lambda03$.
Models $A\lambda05$ (third panel)
\added{and $A\lambda05$\_M14 (fourth panel)} show that the bar
growts at around 4-5 Gyr and the saturation 
at around 6 Gyr; this \replaced{model also}{these
panels} show the
bar growth accompanied by some strong bi-symmetrical structures at
larger radii. 
Finally, model $A\lambda06$ (bottom panel) only shows
weak and transient waves in the $m=2$ mode.
The white line in these panels represent
the radius of the bar, which is half of
the length of the bar (second panel of Figure \ref{fig:OmLenR_I}).
  
\begin{table}
	\centering
	\caption{The table shows the growth rate $\omega$ of the bar/oval for isolated models.\label{tab3}}
	\begin{tabular}{ll} 
		\hline
		Model       &  $\omega\ [km\ s^{-1}\ kpc^{-1}]$ \\\hline
		A$\lambda$03 &  \multicolumn{1}{c}{3.82}\\
		A$\lambda$04 &  \multicolumn{1}{c}{0.84}\\
		A$\lambda$05 &  \multicolumn{1}{c}{0.39}\\
		\added{A$\lambda$05\_M14} &  \multicolumn{1}{c}{0.40}\\
		A$\lambda$06 &  \multicolumn{1}{c}{0.14}\\
		\hline
	\end{tabular}
\end{table}

Figure \ref{fig:averaStrong} shows the amplitude for 
mode $m=2$ Fourier coefficient for different radii 
as a function of time. As we show in
\cite{valenciaenriquezetal17}, these plots can be understood
as growing curves of the structures that are being assembled.
It means they represent the strength of the structures 
that are developing in the disk. In these curves, we can identify 
the three main phases of the bar growth
by the amplitude curves taking from the inner radii; 
the first phase, the growth of the bar, corresponds 
to an exponential rise of its amplitude, from which we can get 
the growth rate $\omega$ of the bar (see table \ref{tab3});
the second phase corresponds to saturation of
its amplitude (around the maximum amplitude);
and the final phase, the secular evolution, corresponds
to flattening of its curve (bar saturation).

\begin{figure}
	\includegraphics[scale=0.35,angle=-90]{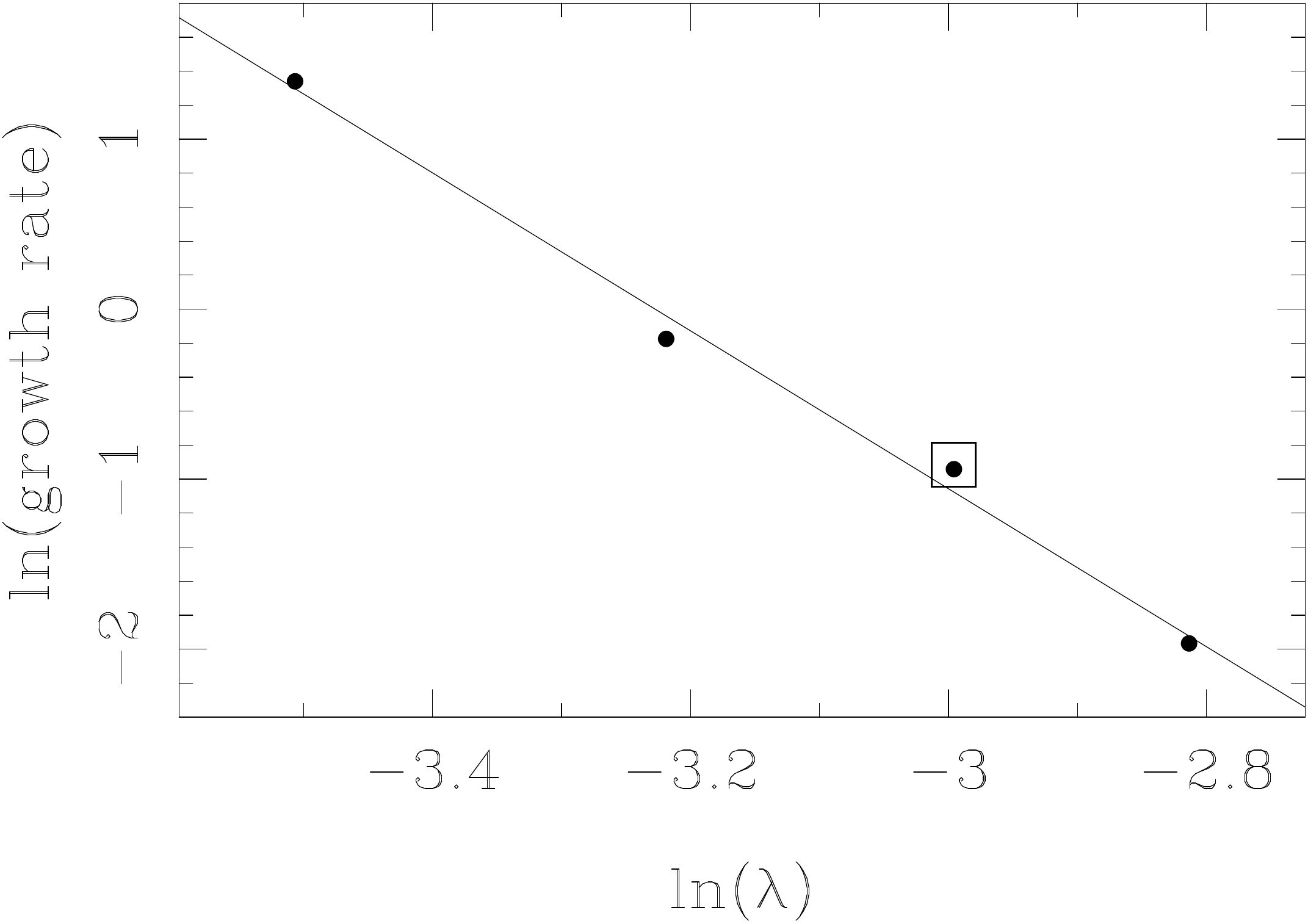}
	\caption{Growth rate in km/sec/kpc versus initial $\lambda$. The fit \added{on points} represents  an slope of $-4.65$. \added{The square presents the growth rate for 
	$A\lambda05$\_M14 model.}}
	\label{fig:growthvsrd}
\end{figure}

In Figure \ref{fig:averaStrong}, model $A\lambda03$,
in the upper left panel, shows the fastest
growth rate of the bar, $\omega=3.82 \ km \ s^{-1} \ kpc^{-1}$,
the maximum amplitude is reached around 0.5 to 1.5 
Gigayear (Gyr), and after
the second Gyr the bar saturates. 
Model $A\lambda04$, in the upper right panel, 
presents a growth rate of
$\omega=0.84 \ km \ s^{-1} \ kpc^{-1}$, 
the second phase is 
around 2 to 3 Gyr, and the bar saturates
after the fourth Gyr. \replaced{Model $A\lambda05$}{The
$A\lambda05$ (black lines) and $A\lambda05$\_M14 (cyan lines)
models}, 
in the bottom left panel, reach 
the maximum amplitude after the fourth 
Gyr, and the growth rate is around of 
$\omega=0.39 \ km \ s^{-1} \ kpc^{-1}$. Finally, model $A\lambda06$,
in the bottom right panel, shows weak bi-symmetrical structures that evolve with a 
very slow growth rate of $\omega=0.14 \ km \ s^{-1} \ kpc^{-1}$. We have found
a relation between the measure growth rate and the initial $\lambda$
in the form $growth\ rate \propto \lambda^{-4.65}$ (figure \ref{fig:growthvsrd}).

Some other measurements were made to characterize the bar.
In Figure \ref{fig:OmLenR_I} we present 
the instantaneous angular velocity
$\Omega_B$, length of the bar $l$, axial ratio $b/a$,
and the ratio $\mathcal{R}=R_{CR}/a_b$ , where $R_{CR}$ and
$a_b$ are the corotation radius and bar length,
respectively. This parametrization permits
a classification of bars into 
``fast'' ($1.0<\mathcal{R}<1.4$) or ``slow''
($\mathcal{R}>1.4$) \citep{2000ApJ...543..704D,
1992MNRAS.259..345A}.
Model $A\lambda03$, depicted with black
line, shows a constant decrease of the angular velocity
approximately from 30 to 20 $km\ s^{-1}\ kpc^{-1}$.
For model $A\lambda04$ (red line), this decreasing
starts from 20 to 12  $km\ s^{-1}\ kpc^{-1}$.
The length of the bars is kept almost constant for both models
as well as the ratio between their axes. Furthermore, 
in their evolution the bars change from almost
fast ($\mathcal{R} \sim 1.4$) to slow ($\mathcal{R} > 1.4$)
as we observe in the rightmost panel 
of Figure \ref{fig:OmLenR_I}. \replaced{The model  $A\lambda05$, depicted with a green line,}
{The $A\lambda05$ and $A\lambda05$\_M14 models,
depicted with green and cyan lines, respectively,} 
show an angular velocity
around of $12\ km\ s^{-1}\ kpc^{-1}$. 
For \replaced{this model}{these models}, the bar grows in size until it
achieves the longest length of 11 kpc \added{approximately}.
Afterwards, the bar shrinks to 7 kpc at 
the maximum amplitude phase (around T=5 Gyr). However,
the bar seems to be destroyed after that,
and the measurements of bar parameters is more difficult and less
precise.

\section{RESULTS FOR INTERACTIONS}\label{results}

\begin{figure*}
	\includegraphics[scale=1.0]{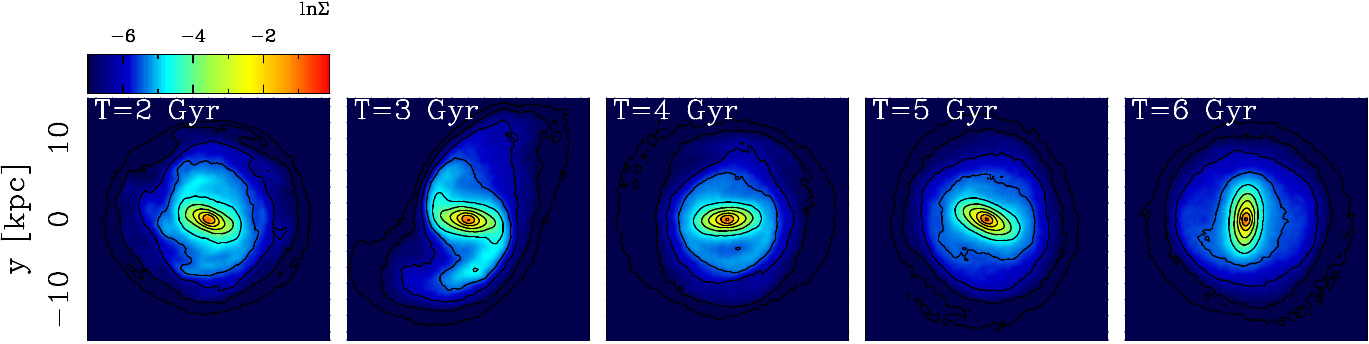}\\
	\includegraphics[scale=1.0]{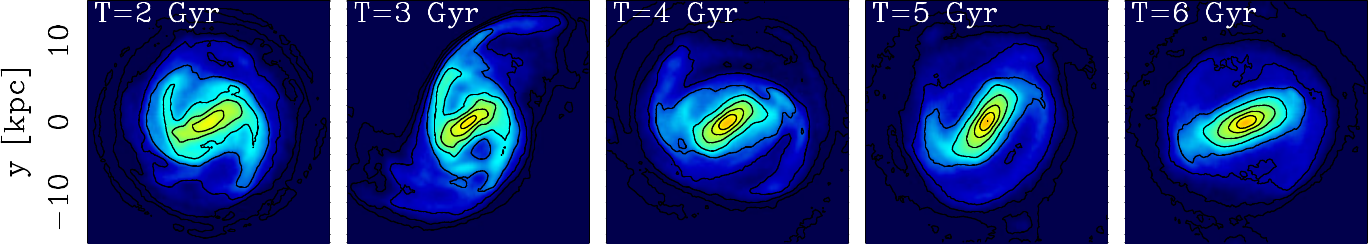}\\
	\includegraphics[scale=1.0]{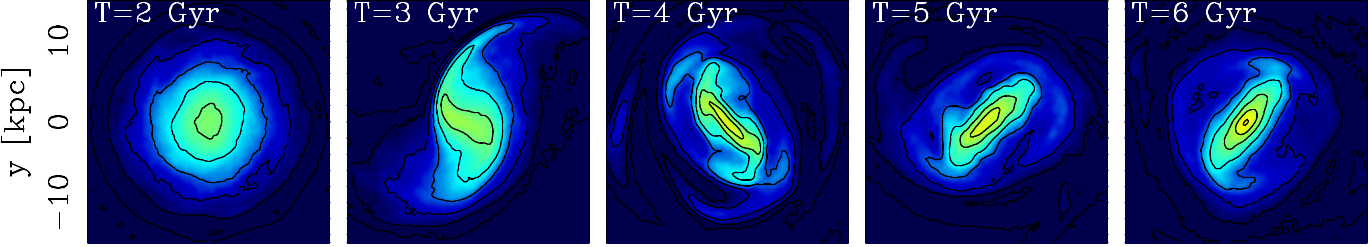}\\
	\includegraphics[scale=1.0]{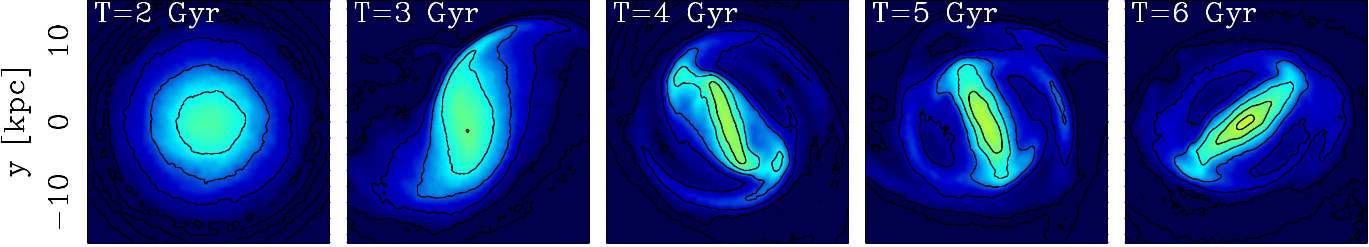}\\
	\includegraphics[scale=1.0]{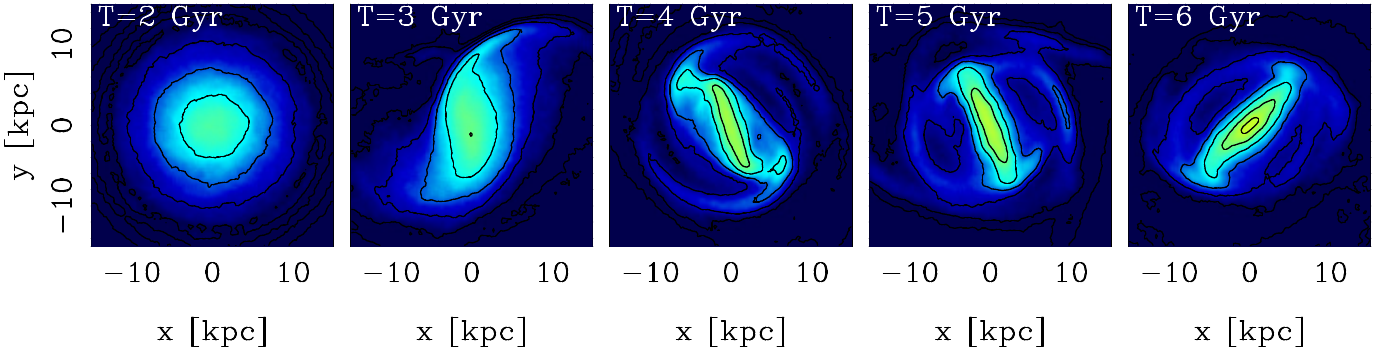}
	\caption{Face-on surface logarithm density maps for 
		interactions of group Pw; the color scale is at the
		top left, the time increase from the left
		to the right in units of Gyr. Top row the 
		snapshots of model $PwA\lambda03$, second row
		shows the snapshots of model $PwA\lambda04$,
		the third row shows the snapshots of $PwA\lambda05$
		and \replaced{the last one shows the snapshots of model
		$PwA\lambda06$.}{the last two ones show the snapshots
		of the $PwA\lambda06$ and $PwA\lambda06$\_M2 models;
		notice that the evolution is very similar.}}
	\label{fig:snpasPwA}
\end{figure*}

\begin{figure*}
	\includegraphics[scale=1.0]{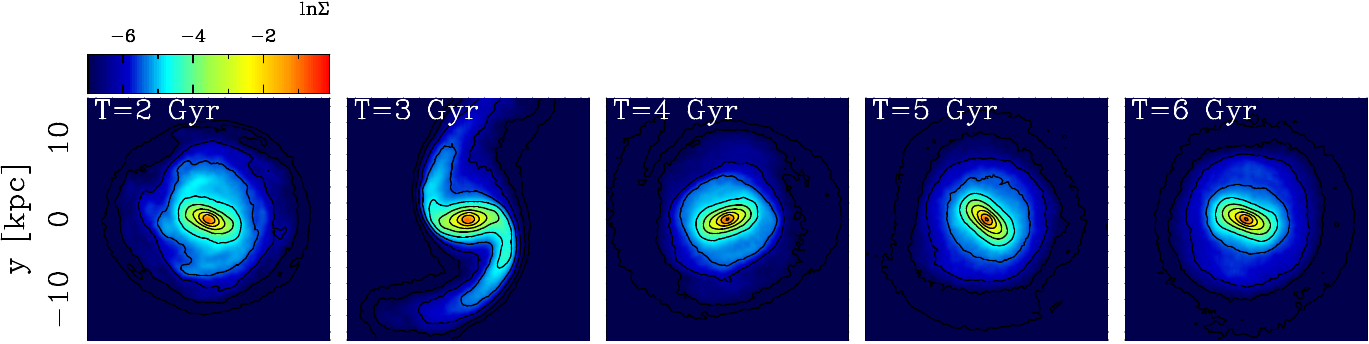}\\
	\includegraphics[scale=1.0]{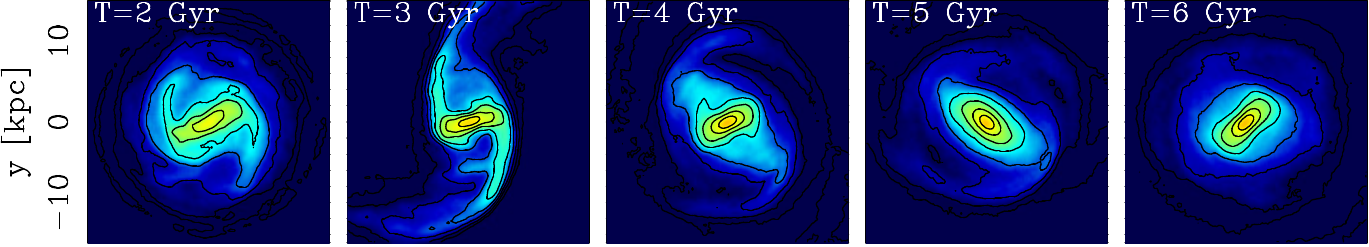}\\
	\includegraphics[scale=1.0]{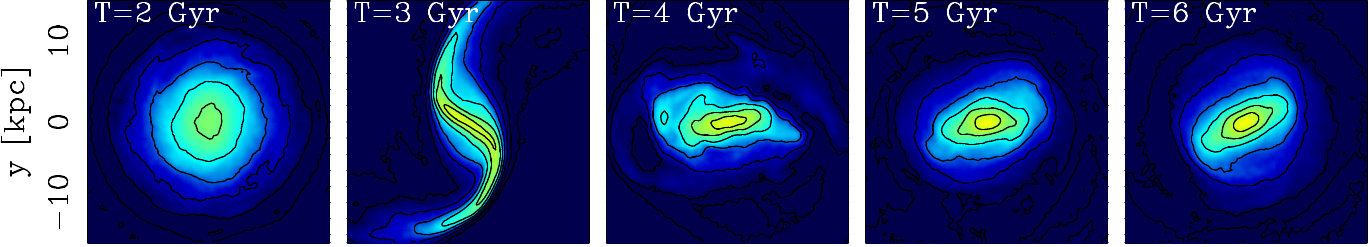}\\
	\includegraphics[scale=1.0]{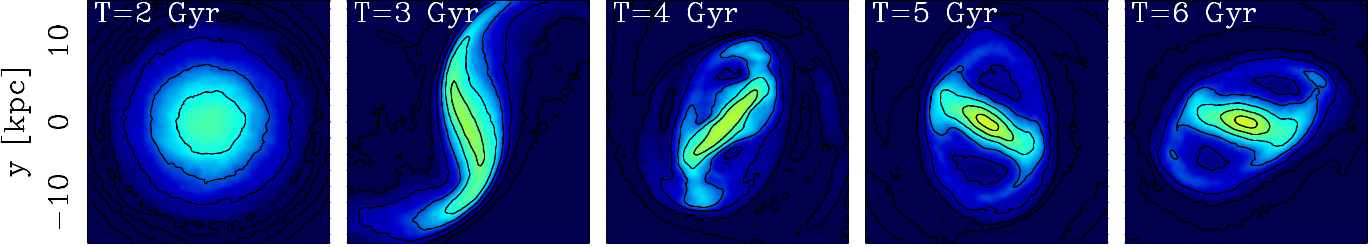}
	\caption{As in figure \ref{fig:snpasPwA}, but for interactions
		of group Pm.}
	\label{fig:snpasPmA}
\end{figure*}

\begin{figure*}
	\includegraphics[scale=1.0]{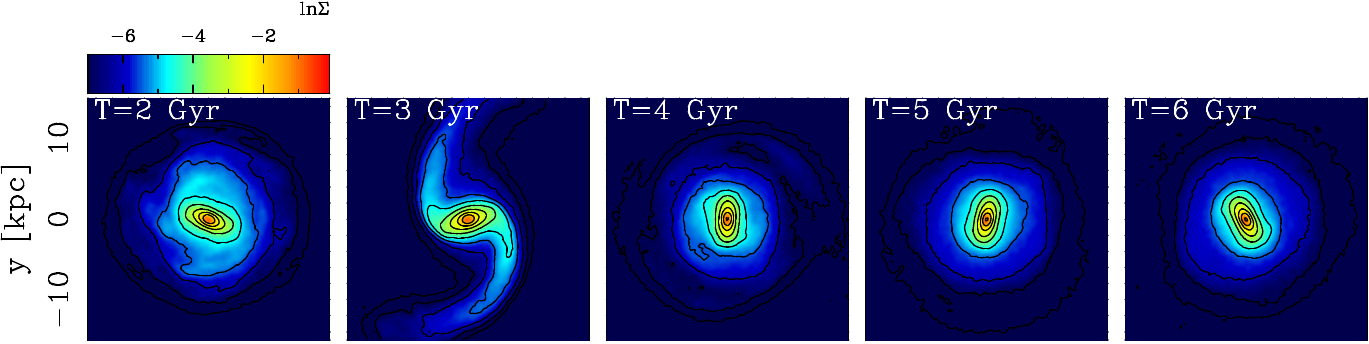}\\
	\includegraphics[scale=1.0]{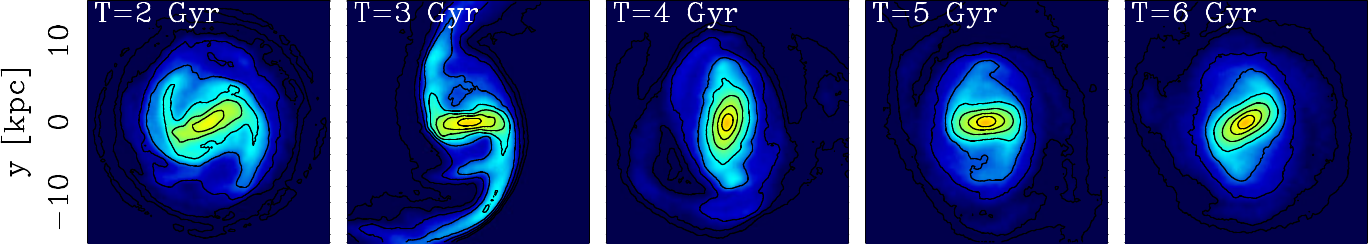}\\
	\includegraphics[scale=1.0]{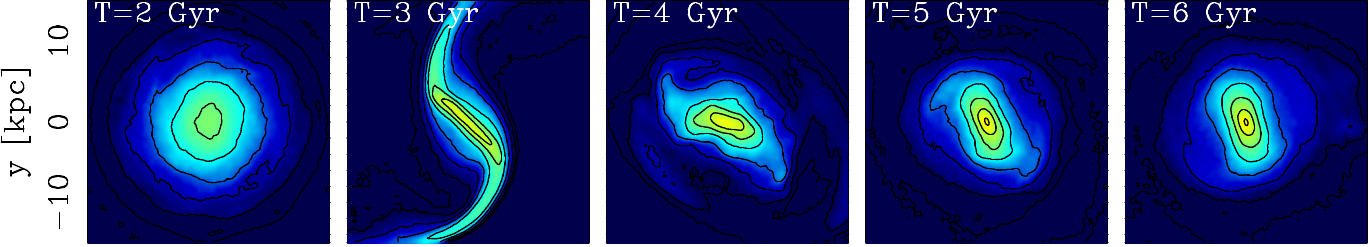}\\
	\includegraphics[scale=1.0]{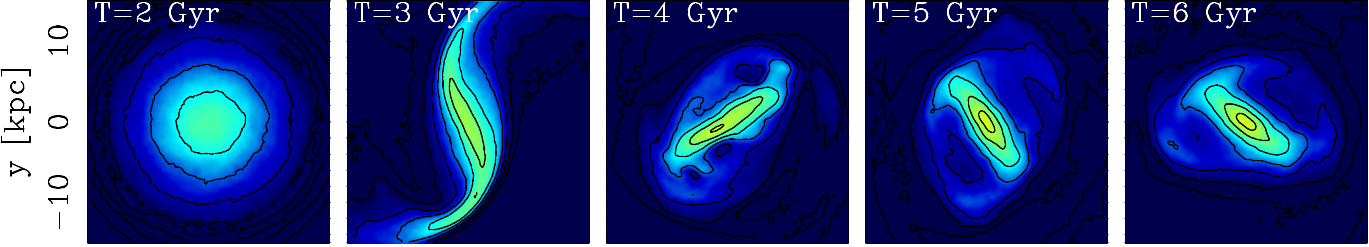}
	\caption{As in figure \ref{fig:snpasPwA}, but for interactions
		of group Ps.}
	\label{fig:snpasPsA}
\end{figure*}

Figures \ref{fig:snpasPwA}, \ref{fig:snpasPmA}, 
\ref{fig:snpasPsA} show the face-on 
logarithm surface density maps for all of 
our interacting models at times of 2, 3, 4, 5, and 6 Gyr;
The simulations start from the second gigayear;
therefore the pericentre passages in all encounters
are around the third gigayear
(second column of these figures);
and we have to bear in mind that
all encounters pass around of
$40km/s/kpc$, therefore the duration of
interaction is similar in all encounters.
We can observe that those tidal interactions
produce  well defined strong spiral arms
and extended tidal features, such as bridge and tail, 
that are all transient, but distinct in nature
\citep{1972ApJ...178..623T,2008ApJ...683...94O}.
The models in which the bar is already formed show 
strong spiral structure, but the bar seems not 
to be affected; while models where the bar
is not created yet show a thin and durable oval 
triggered by the strong tidal pull.
Figure  \ref{fig:snpasPwA} shows the interactions
of group Pw; the tidal pull is the lightest, hence,
these interactions generate wide spirals in all models,
and a wide oval in simulations where the bar
is not formed yet.
\added{We add in the last row the $PwA\lambda06$\_M2 model to compare it to the $PwA\lambda06$ model. We notice that they are apparently very similar in nature which means that the perturbation resolution does not affect the model evolution as long as those have the same mass distribution.}
Figure  \ref{fig:snpasPmA} shows the interactions 
of group Pm; the tidal pull is relatively strong; not only
these interactions generate narrow spirals in
all models, but also they generate a thin central oval
in the disk, where the bar is not form yet.
Finally, Figure  \ref{fig:snpasPsA} 
shows the interactions of group Ps; the tidal
pull is the strongest, generating the thinnest
spirals and oval of all interactions. Observing these
interactions, we can notice while stronger the interaction
thinner the spiral structure in the target model.

\subsection{Disk Instabilities through the time}

\begin{figure*}
\gridline{\figg{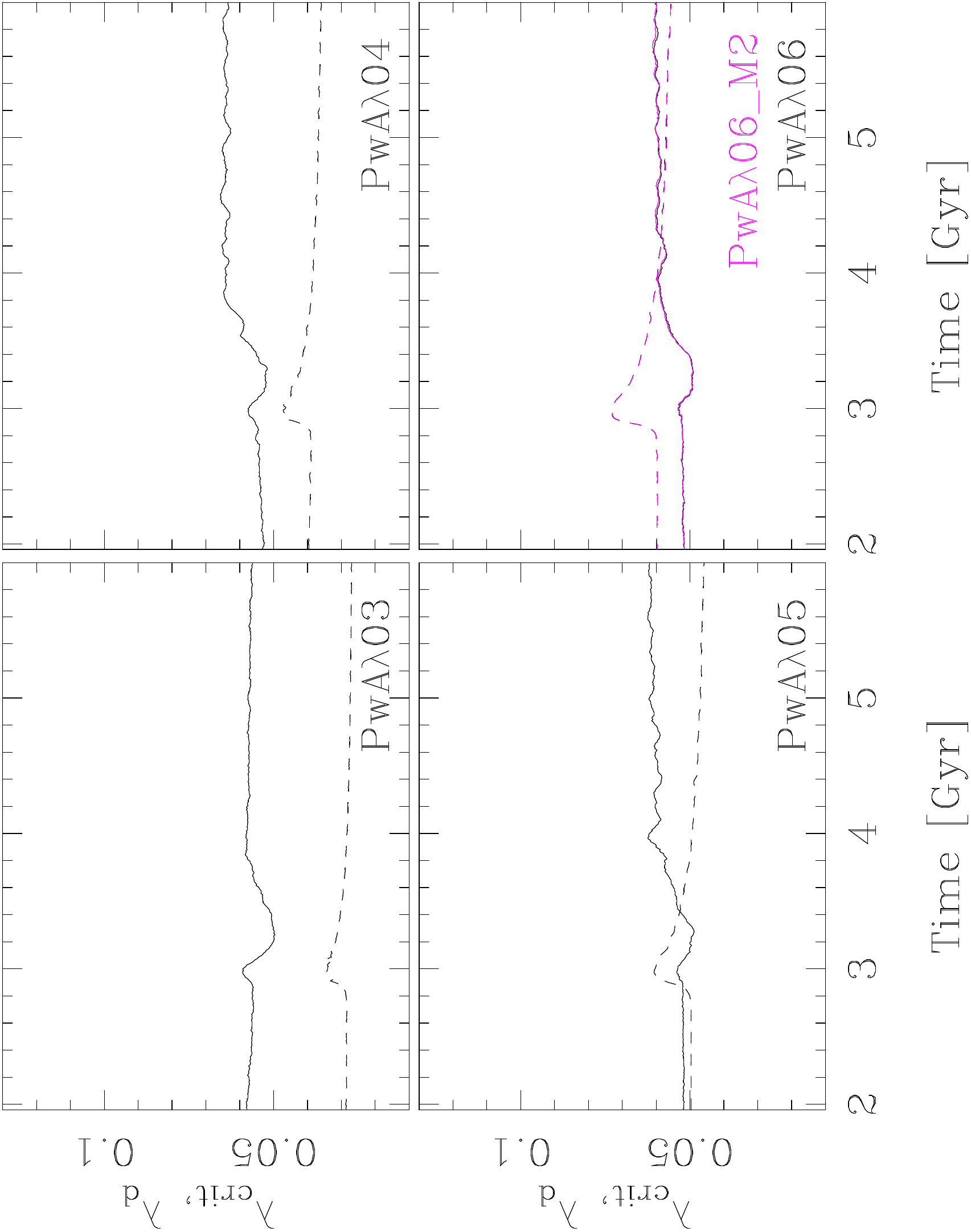}{0.3\textwidth}{(a) Pw set of simulations.}{-90}{0.25}
          \figg{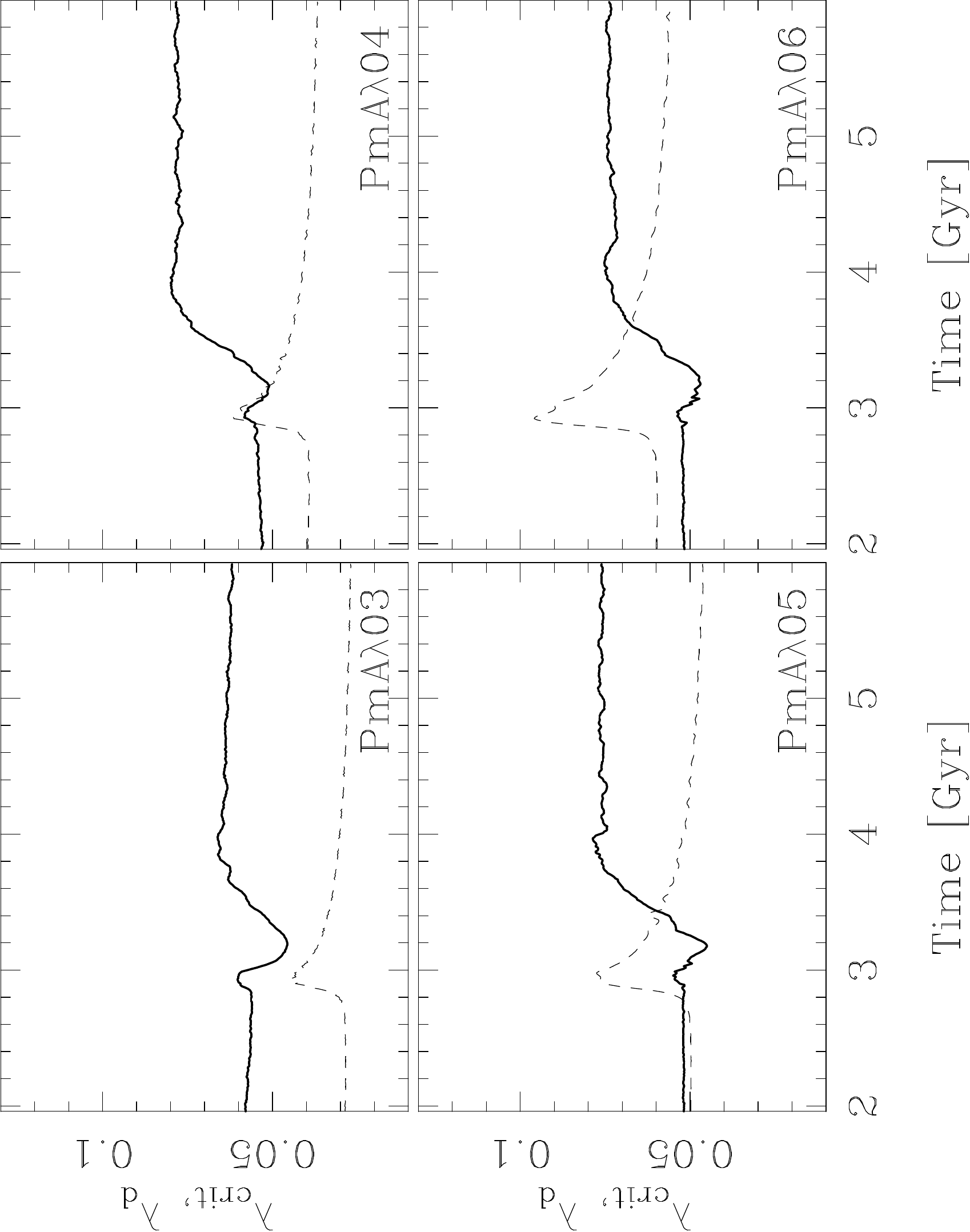}{0.3\textwidth}{(b) Pm set of simulations.}{-90}{0.25}
          \figg{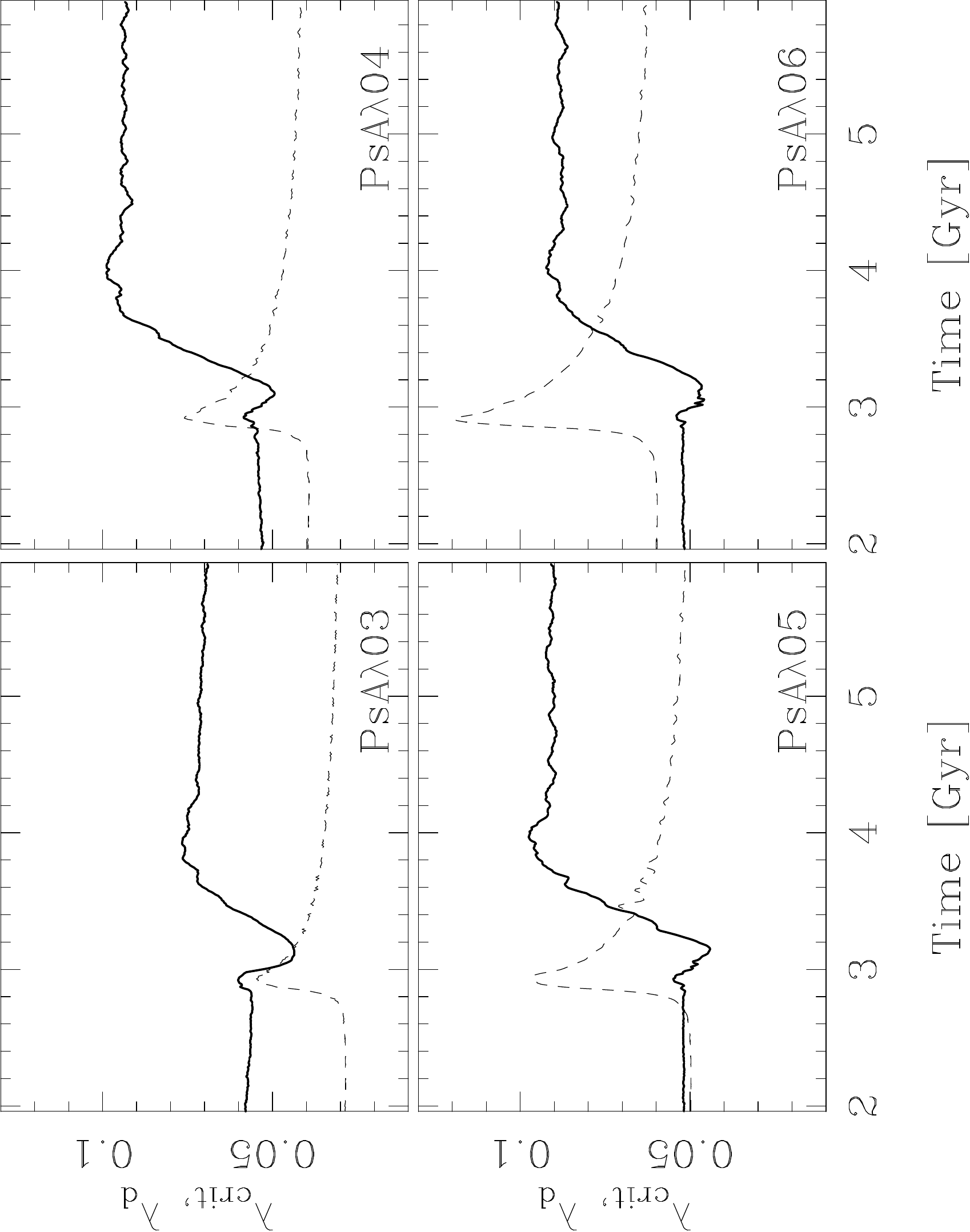}{0.3\textwidth}{(c) Ps set of simulations.}{-90}{0.25}
          }
\gridline{\figg{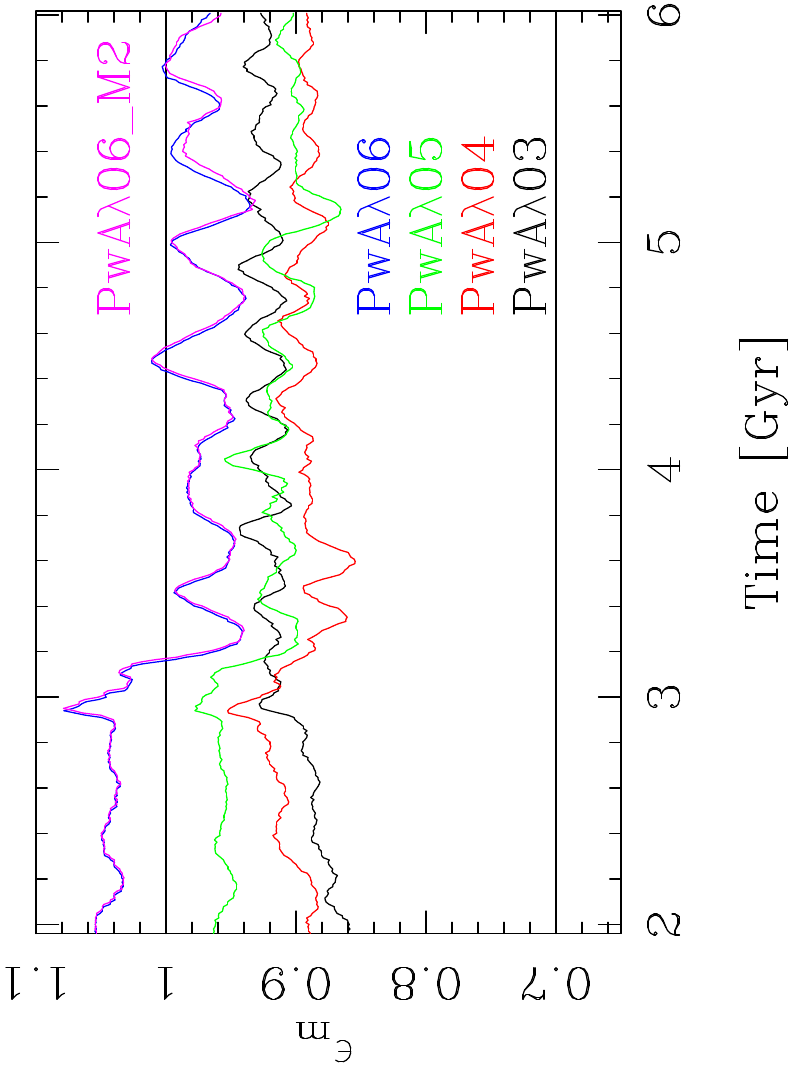}{0.3\textwidth}{(d) Pw set of simulations.}{-90}{0.5}
          \figg{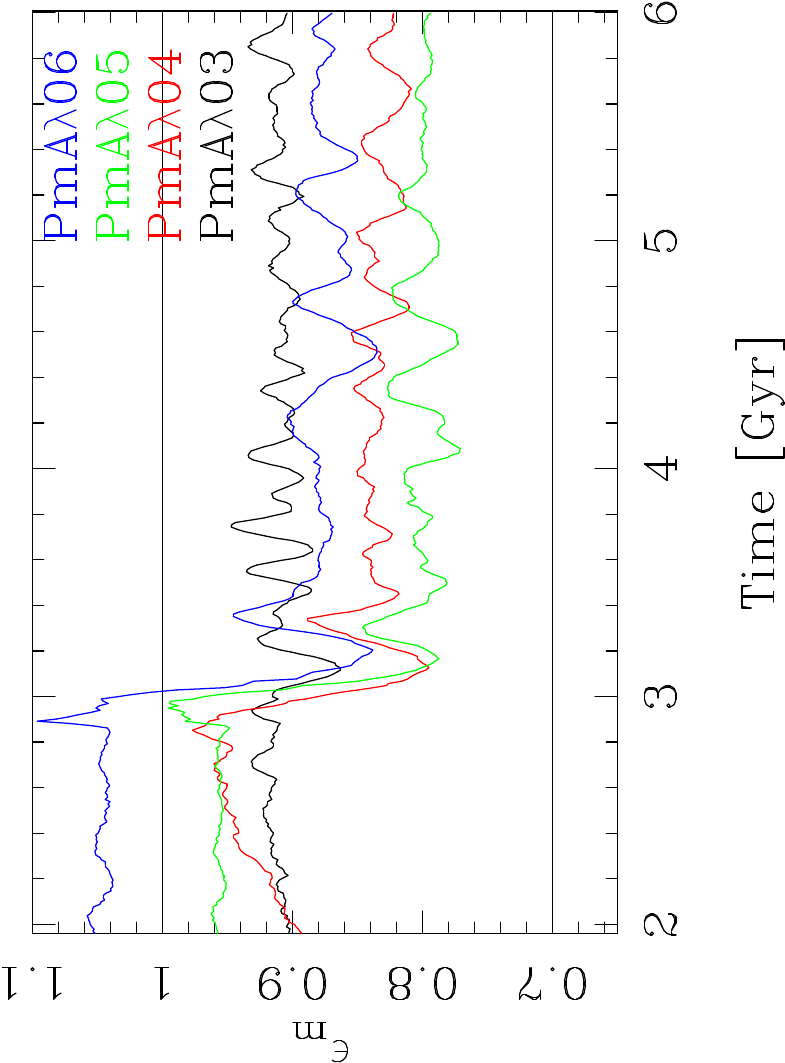}{0.3\textwidth}{(e) Pm set of simulations.}{-90}{0.5}
          \figg{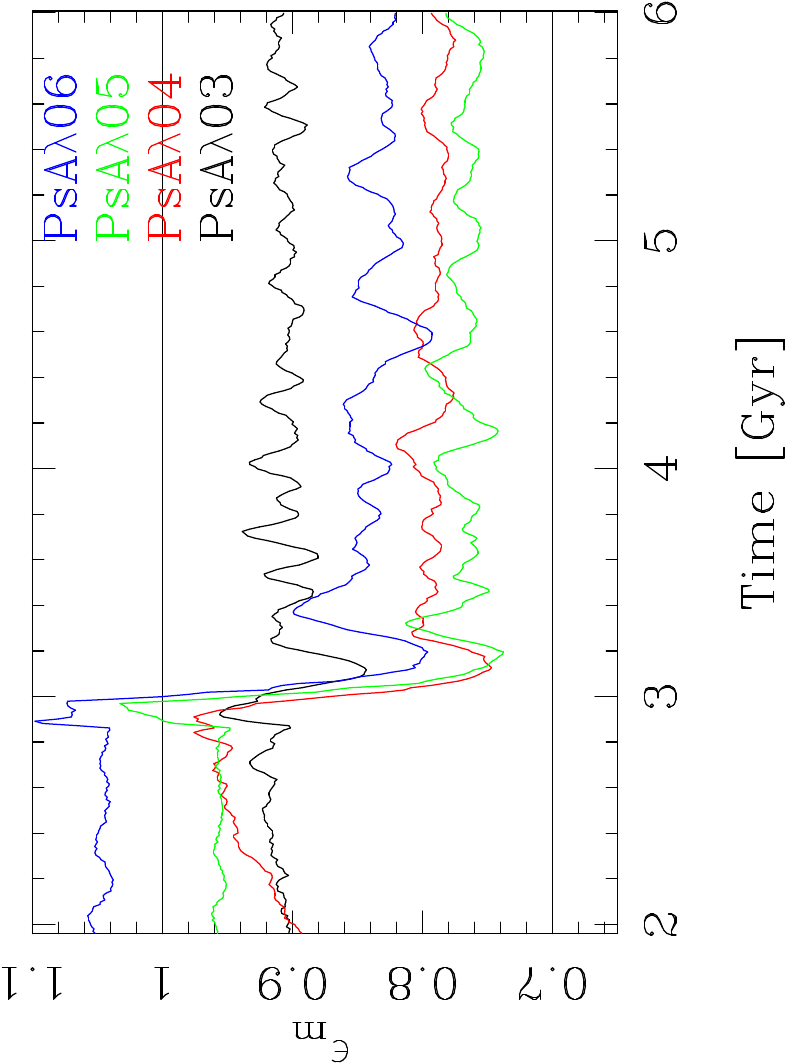}{0.3\textwidth}{(f) Ps set of simulations.}{-90}{0.5}
          }
\caption{Disk Stability Parameters (DSP). Figures
from \reffiglambdasPwA \ to \reffiglambdasPsA \
	show the measurements of the spin parameter $\lambda_{crit}$ (solid line) and
	$\lambda_{d}$ (dashed line); while Figures from
	\reffigemPwA \ to \reffigemPsA \
	the measurements of the stability parameter $\epsilon_m$.
	\added{We observe that the purple line, which
	represents the experiment $PwA\lambda06$\_M2, is
	almost identical to the case with low N particles.}
	\label{fig:GSP}}
\end{figure*}

In Figure \ref{fig:GSP}, 
first column, plots \ref{fig:GSP}\reffiglambdasPwA \
and \ref{fig:GSP}\reffigemPwA, depicts the interactions with the
lightest companion (Pw); 
\added{including the $PwA\lambda06$\_M2 model (purple line)}, the second column depicts the interactions 
with the companion that has a similar mass to the target
galaxy (Pm) and the last column depicts the
interactions with the heaviest perturbation (Ps).
The measurements of the Disk Stability Parameters (DSP),
Figures from \ref{fig:GSP}\reffiglambdasPwA \ to
\ref{fig:GSP}\reffiglambdasPsA,
show similar behavior in all interactions. In these
figures, the spin parameter $\lambda_{crit}$ are drawn
as solid lines, while the $\lambda_{d}$ is
depicted as dashed lines.
On the other hand, the experimental stability parameter
$\epsilon_m$ is displayed in Figures from
\ref{fig:GSP}\reffigemPwA \ to \ref{fig:GSP}\reffigemPsA. 

In general, at the time of encounter,
the impulse on particles of the target galaxy given
by the perturbation makes that the spin
parameter $\lambda_d$ and the critical spin
parameter $\lambda_{crit}$ increases and
decreases, respectively, and the stability parameter
$\epsilon_m$ decrease as well. After the perturbation
passes and is far from the studied galaxy,  $\lambda_d$
tends to come back to values it had before the flyby while
$\lambda_{crit}$ tends to increase to higher values than
$\lambda_d$ causing the disk becomes unstable.
The $\epsilon_m$ parameter is set down almost
constant by the rest of the simulation.

The change of DSP depends on the mass, distance, and velocity around the
pericentre which means that the growth of spirals and bar
properties evolve different
\citep{1987MNRAS.228..635N,1990A&A...230...37G,
	1991A&A...245L...5S,1998ApJ...499..149M,
	2008ApJ...683...94O,2017MNRAS.464.1502M,
	2017arXiv170306002M}.
For our purpose, we only experiment with the mass
of the perturbation setting down almost constant 
the pericentre distance and the angular velocity 
of the perturbation.

\subsection{Evolution of bar parameters} 

\begin{figure*}
\includegraphics[scale=0.35]{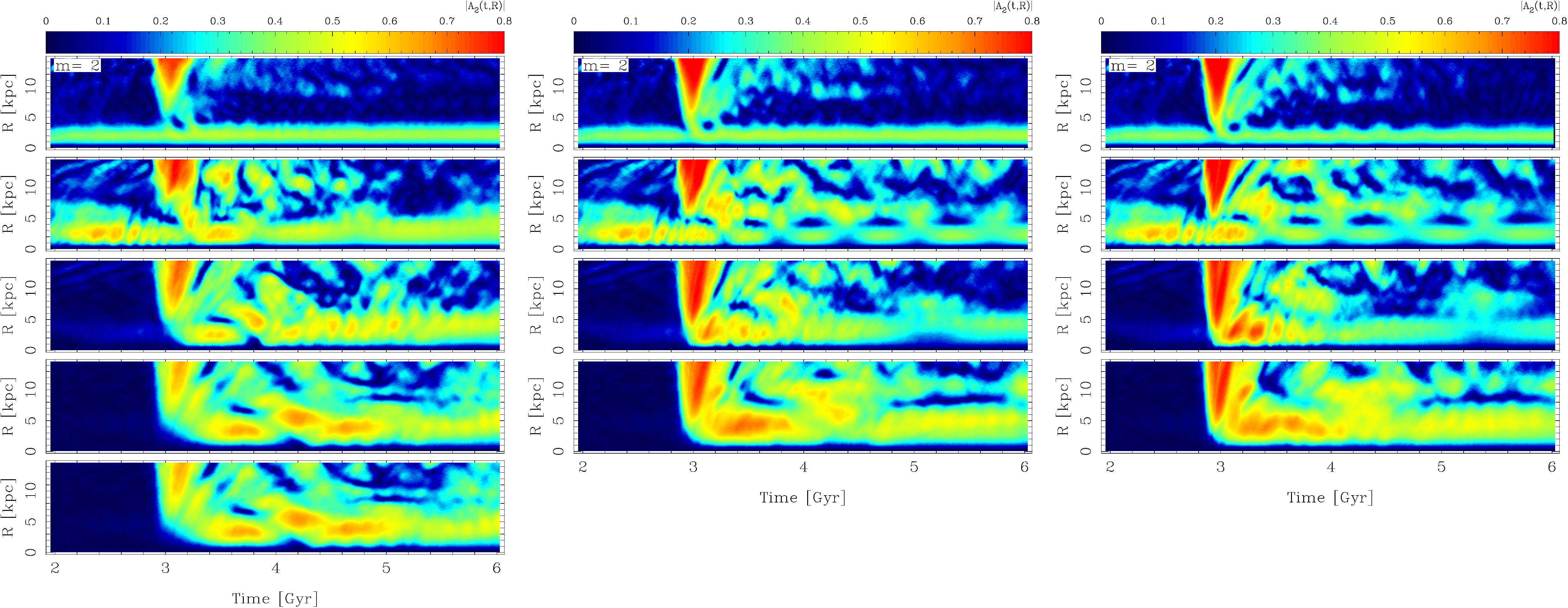}
\gridline{
	{\qquad \qquad (a) Pw set of simulations.\qquad      \qquad \qquad \qquad}
    {(b) Pm set of simulations.\qquad \qquad \qquad \qquad }
    {(c) Ps set of simulations.\qquad \qquad \qquad}
        }
\caption{The figures from \reffigAmFTPwA \ to \reffigAmFTPsA \
present the $m=2$ Fourier amplitude in all 
of our models as function of time and radius. From top to the bottom, models $*A\lambda03$, $*A\lambda04$, 
$*A\lambda05$, $*A\lambda06$ 
\added{and we add in the bottom most panel of figure \reffigAmFTPwA, the results of $PwA\lambda06\_M2$ simulation, which is quite similar to the results for $PwA\lambda06$.}
We clearly see the passing of the perturbation generates
strong features in the amplitude. 
\label{fig:AmFT1D}.
}
\end{figure*}

\begin{figure}
\gridline{\figg{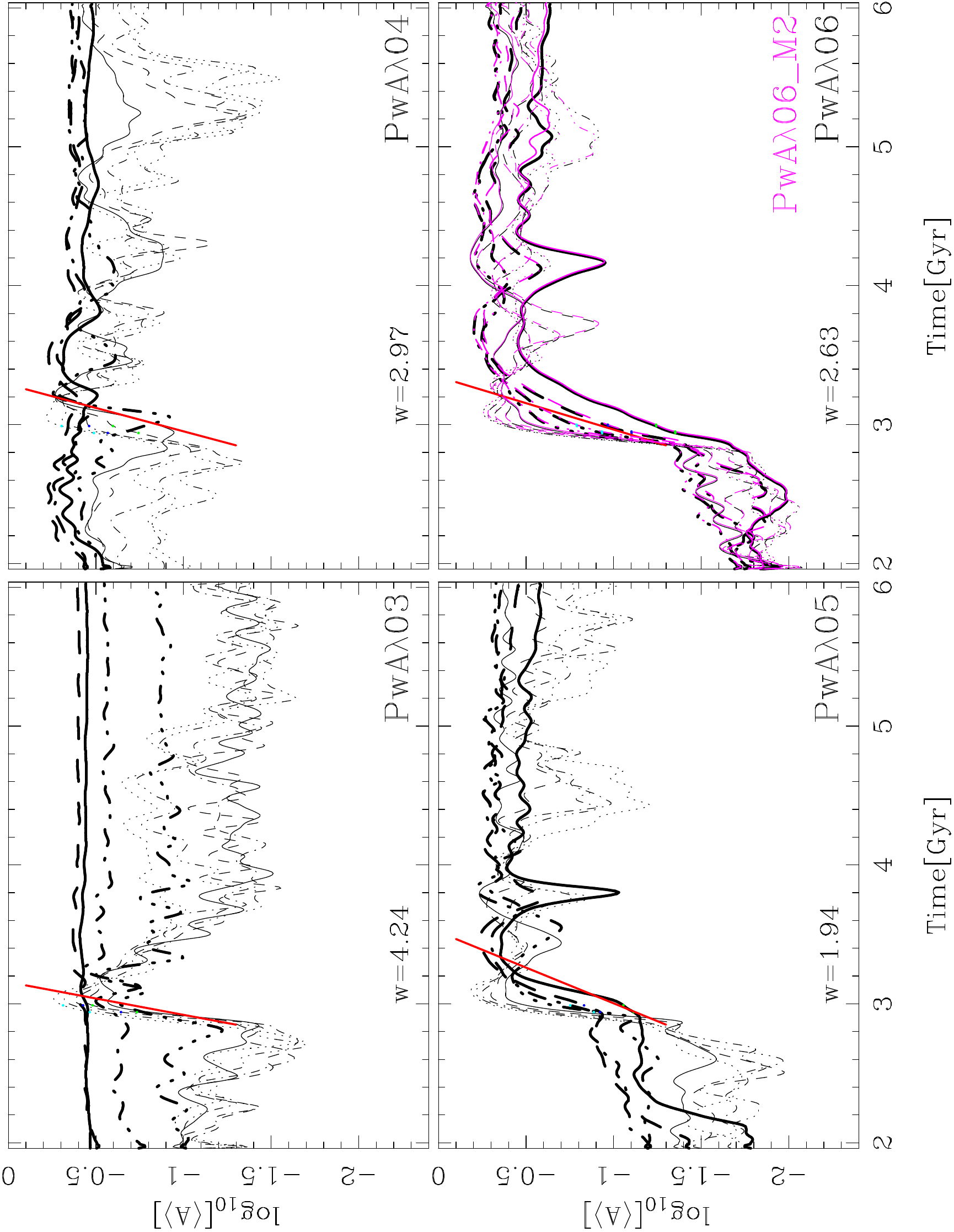}{0.5\textwidth}{(a) Pw set of simulations.}{-90}{0.31} }
\gridline{\figg{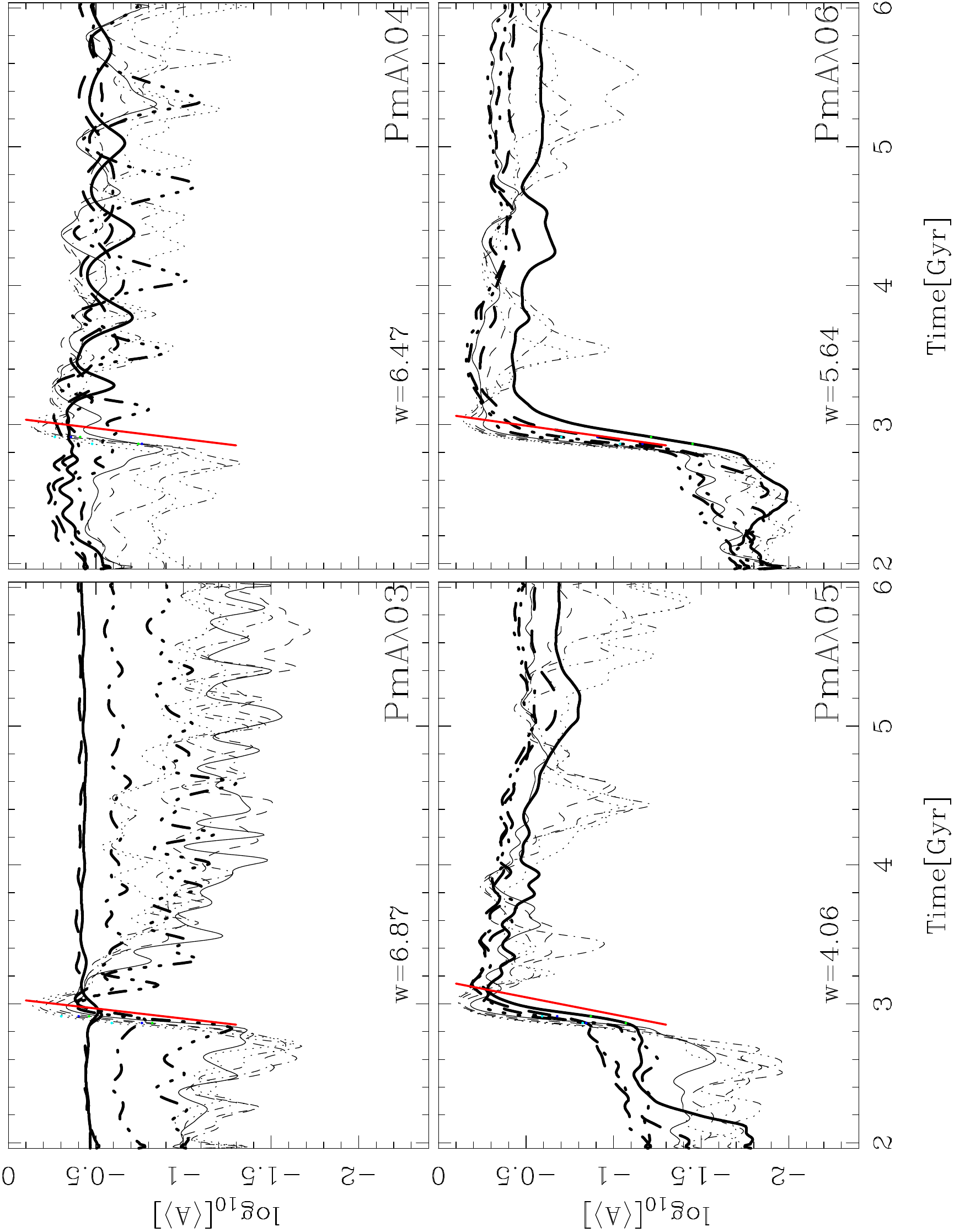}{0.5\textwidth}{(b) Pm set of simulations.}{-90}{0.31} }
\gridline{\figg{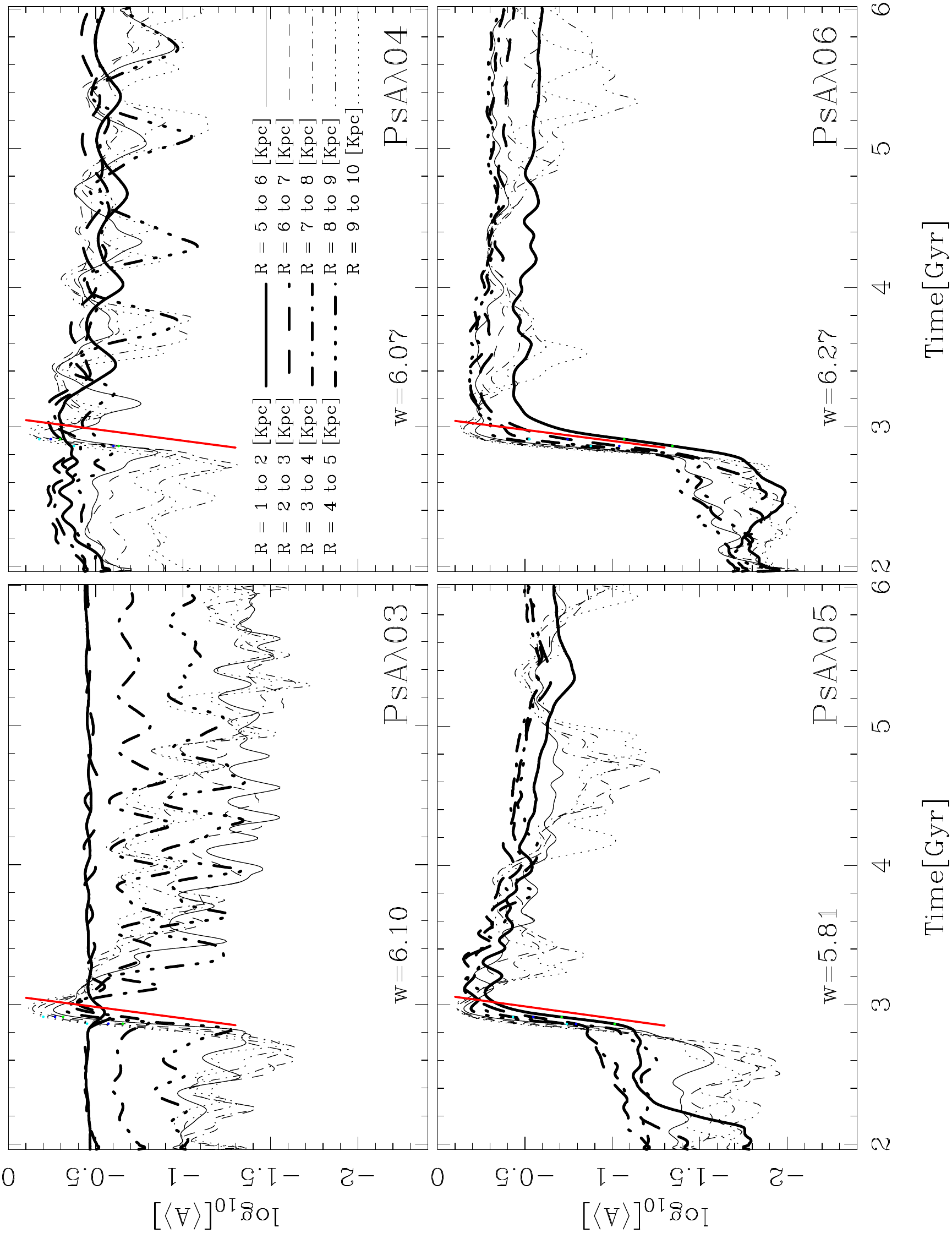}{0.5\textwidth}{(c) Ps set of simulations.}{-90}{0.31} }
	\caption{The figures from \reffigAmFTwPwA \ to
	\reffigAmFTwPsA \ show the average of Fourier 
	amplitude for some radial ranges (see Figure \reffigAmFTwPsA). In models with a bar 
	($*A\lambda03$ and $*A\lambda04$), one can note as the amplitude of the inner radii (thicker lines) 
	keep almost constant, while the amplitude for outer radii (thinner lines) increase rapidly when the 
	perturbation pass. In contrast, models without a bar formed yet ($*A\lambda05$ and $*A\lambda06$), 
	we observe as the amplitude for inner radii grows and becomes more or less constant representing 
	the bar formation which was triggered by the interaction; besides, the amplitude for outer radii 
	stay longer with high amplitudes.
	\added{Also, in the right bottom panel of figure 	\reffigAmFTwPwA, we compare the models which use $1\times10^6$ and $2\times10^6$ particles for the perturbation. The red solid line in all panels represents the growth rate.}
	\label{fig:AmTF_P}}
\end{figure}

Since the bar and tidal pull produce
strong features in the mode $m=2$ of the Fourier
component, we measure such amplitude and display it
in Figure \ref{fig:AmFT1D} for all our interactions. 
Figures \ref{fig:AmFT1D}\reffigAmFTPwA \ to
\ref{fig:AmFT1D}\reffigAmFTPsA \ show the 
amplitude in color scale, and Figures
\ref{fig:AmTF_P}\reffigAmFTwPwA \ to
\ref{fig:AmTF_P}\reffigAmFTwPsA \
show curves of that amplitude for different radii.
The passing of the perturbation causes that the amplitude
increase strongly and transiently from the outer
to the middle region of the disk.
From there, substantial amplitudes move towards
the inner part of the disk in all interactions.

Models $PwA\lambda03$, $PmA\lambda03$ and $PsA\lambda03$
($*A\lambda03$\footnote{Models 
	$*A\lambda03$; the asterisk refers to 	models
	$PwA\lambda03$, $PmA\lambda03$ and $PsA\lambda03$.
	We write the asterisk when refers to models
	with the same initial parameter $\lambda$ from PI}) 
represented in the top panels of Figures \ref{fig:AmFT1D}\reffigAmFTPwA,
\ref{fig:AmFT1D}\reffigAmFTPmA \ and
\ref{fig:AmFT1D}\reffigAmFTPsA,
respectively, show the evolution of bar in the
inner region of disk, which is almost constant through
the time, and it appears not to be affected by the interaction
(see also upper left panels of Figures \ref{fig:AmTF_P}\reffigAmFTwPwA,
\ref{fig:AmTF_P}\reffigAmFTwPmA, and \ref{fig:AmTF_P}\reffigAmFTwPsA).
Besides, in the outer part of the disk, the ``growing curves''
(Figures \ref{fig:AmTF_P}\reffigAmFTwPwA, \ref{fig:AmTF_P}\reffigAmFTwPmA, and
\ref{fig:AmTF_P}\reffigAmFTwPsA) show that
the amplitude of transient tidal spirals
is higher than the one in the bar at the interaction time.
Then, after the perturbation
is far from the target galaxy, some spiral can survive
transiently, while other weaker spiral wave may
be driven temporarily by the bar.

In Models $PwA\lambda04$, $PmA\lambda04$ and
$PsA\lambda04$ ($*A\lambda04$), the perturbation
passes when the bar gets a maximum amplitude for
mode $m=2$ (MA phase) which causes the MA phase
finish quickly saturating the inner part of the disk
(see second panels of Figures: \ref{fig:AmFT1D}\reffigAmFTPwA,
\ref{fig:AmFT1D}\reffigAmFTPmA \ and \ref{fig:AmFT1D}\reffigAmFTPsA,
and upper right panels of Figures
\ref{fig:AmTF_P}\reffigAmFTwPwA, \ref{fig:AmTF_P}\reffigAmFTwPmA \
and \ref{fig:AmTF_P}\reffigAmFTwPsA),
and strong tidal spirals can survive 
for more time than those of models $*A\lambda03$.

Before the perturbation passes,
models $PwA\lambda05$, $PmA\lambda05$ and
$PsA\lambda05$ ($*A\lambda05$) in both
third panels of Figures \ref{fig:AmFT1D}\reffigAmFTPwA,
\ref{fig:AmFT1D}\reffigAmFTPmA \ and \ref{fig:AmFT1D}\reffigAmFTPsA,
and bottom left panels of Figures
\ref{fig:AmTF_P}\reffigAmFTwPwA, \ref{fig:AmTF_P}\reffigAmFTwPmA \
and \ref{fig:AmTF_P}\reffigAmFTwPsA, respectively,
do not form a bar yet.
Thus, the interaction
happens before the MA phase, causing
high amplitudes of mode $m=2$ which grow first in the outer
region of the disk, and then less strong amplitudes
grow in the inner region of the disk, then
accelerating the formation of a larger narrow bar.
Tidal spirals can survive for more time
and it seems to be connected to the bar
transiently. In particular, interaction $PsA\lambda05$,
which has the heaviest perturbation, shows
this behavior more conspicuous than the others.

Models $PwA\lambda06$, $PmA\lambda06$ and
$PsA\lambda06$ ($*A\lambda06$), shown in both
last panels of Figures \ref{fig:AmFT1D}\reffigAmFTPwA,
\ref{fig:AmFT1D}\reffigAmFTPmA \ and \ref{fig:AmFT1D}\reffigAmFTPsA,
and bottom right panels of Figures
\ref{fig:AmTF_P}\reffigAmFTwPwA, \ref{fig:AmTF_P}\reffigAmFTwPmA \
and \ref{fig:AmTF_P}\reffigAmFTwPsA, respectively,
show similar behavior to models
$*A\lambda05$. As we show before, model $A\lambda06$
is stable against bar formation; however,
when it is subjected to a perturbation, the
DSP parameters fall below the stability limits
(see Figures \ref{fig:GSP}\reffiglambdasPwA, \ref{fig:GSP}\reffiglambdasPmA \
and \ref{fig:GSP}\reffiglambdasPsA) causing
the bar formation (see the snapshots and 
measurements of the Fourier Transform).
\added{Likewise, the $PwA\lambda06$\_M2 model, which is
shown in the last panel of figures
\ref{fig:AmFT1D}\reffigAmFTwPwAd \ and \ref{fig:AmTF_P}\reffigAmFTwPwA, presents an akin nature with 
respect to its fiducial model,}

Additionally, from the ``growing curves'' we calculated
the growth rate of spirals and bar, which is
depicted with a straight red line in panels of Figure
\ref{fig:AmTF_P}\reffigAmFTwPwA, \ref{fig:AmTF_P}\reffigAmFTwPmA \ and
\ref{fig:AmTF_P}\reffigAmFTwPsA. For models
$*A\lambda03$ and $*A\lambda04$
the bar is already formed, thus we 
calculated the growth rate of strong spirals that was triggered
by the interaction. For the other models, we
calculated the growth rate of the bar.
We summarize in table \ref{tab4} the growth rate of
tidal spirals (bold numbers) and the growth rate
of the bar.

\begin{figure*}
	\includegraphics[scale=0.19,angle=-90]{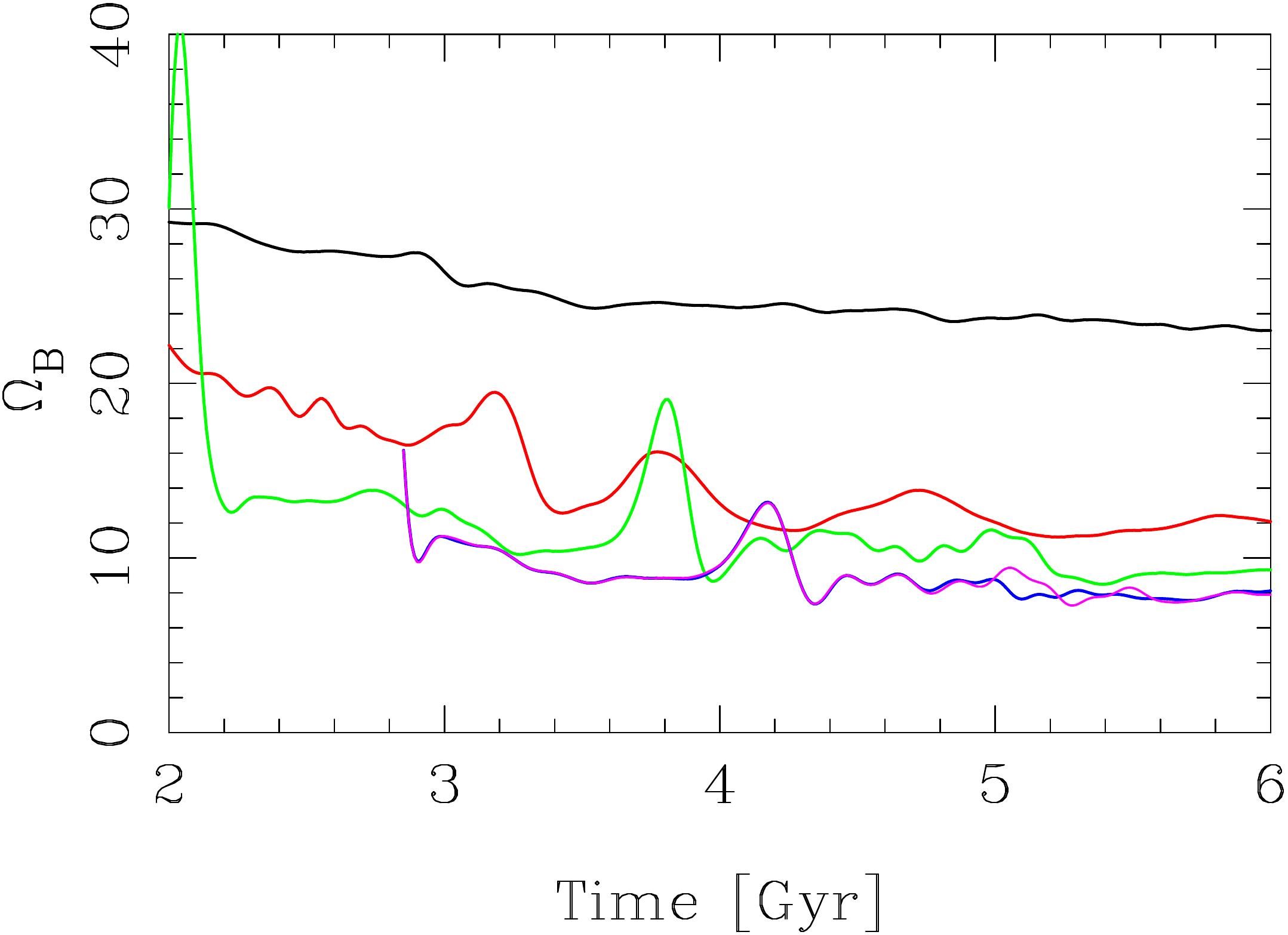}
	\includegraphics[scale=0.19,angle=-90]{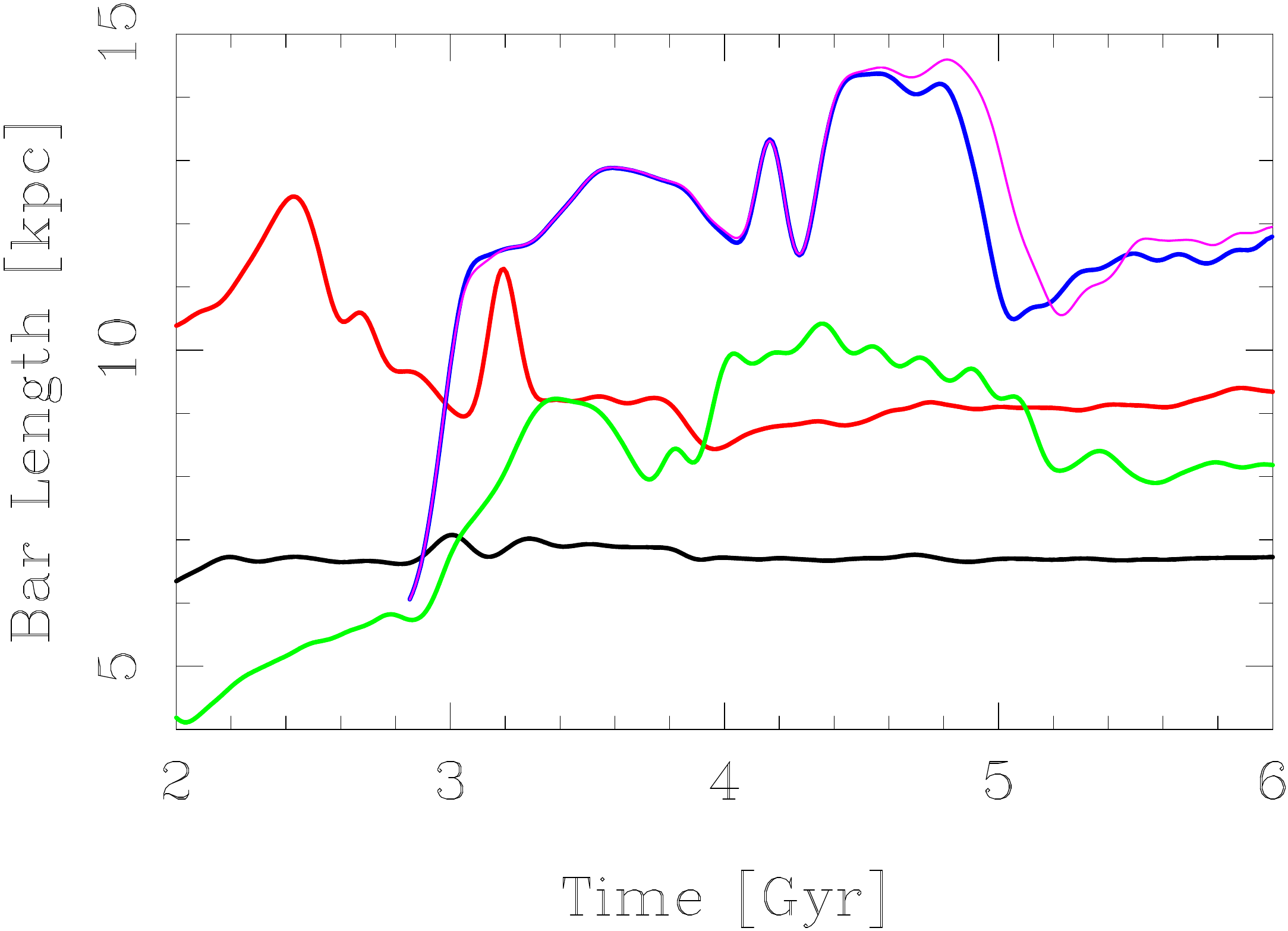}
	\includegraphics[scale=0.19,angle=-90]{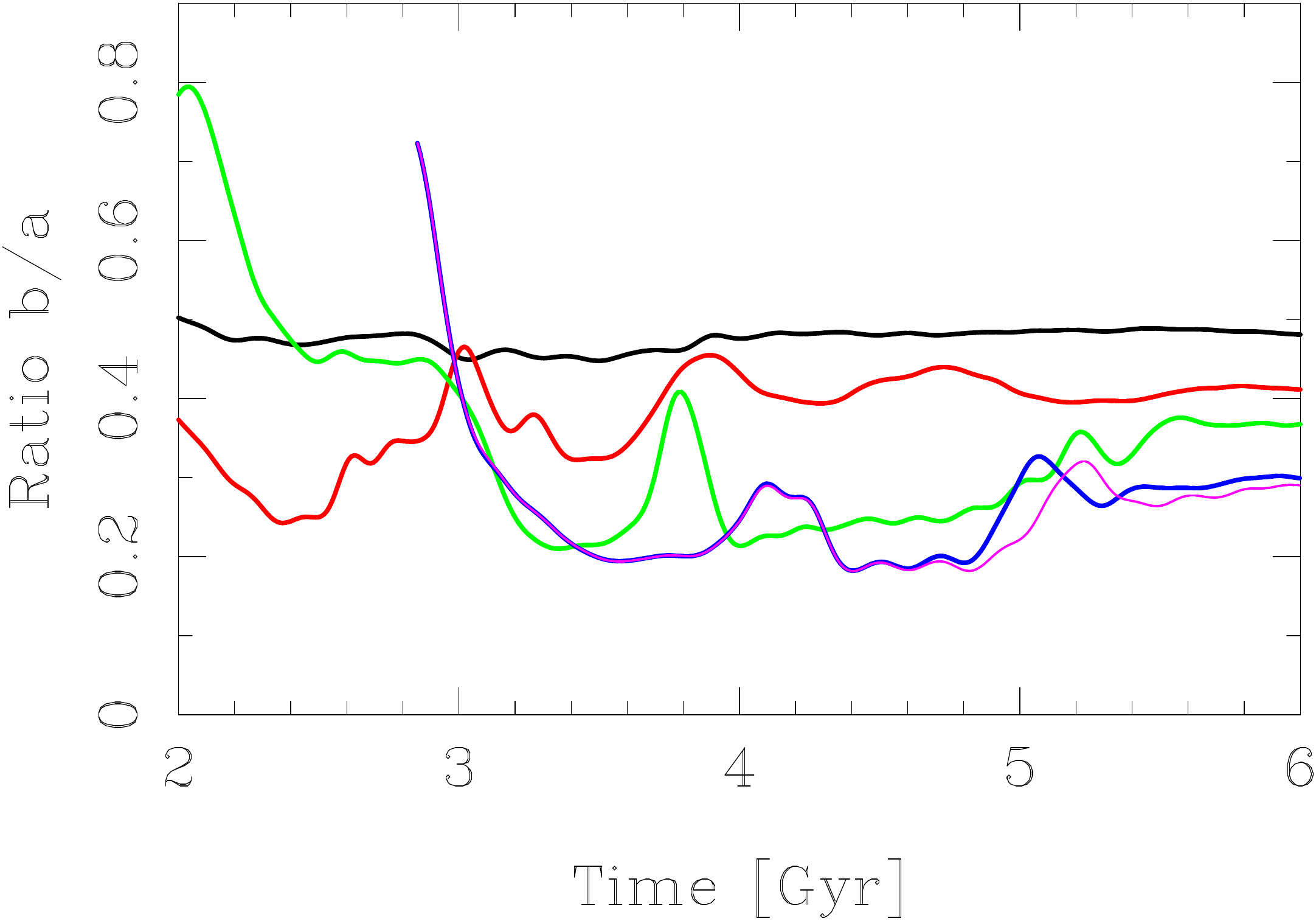}
	\includegraphics[scale=0.19,angle=-90]{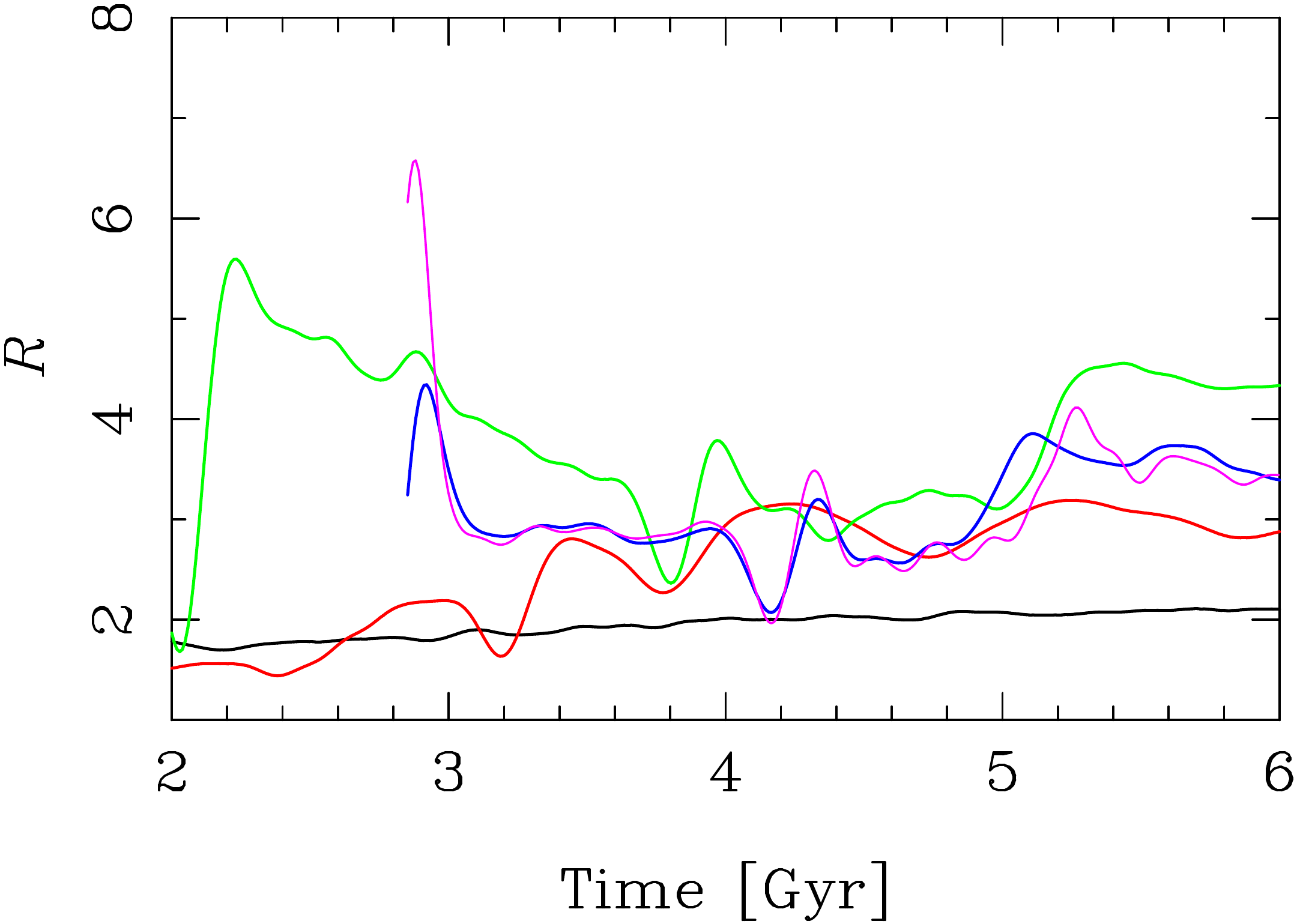}
	\\
	\includegraphics[scale=0.19,angle=-90]{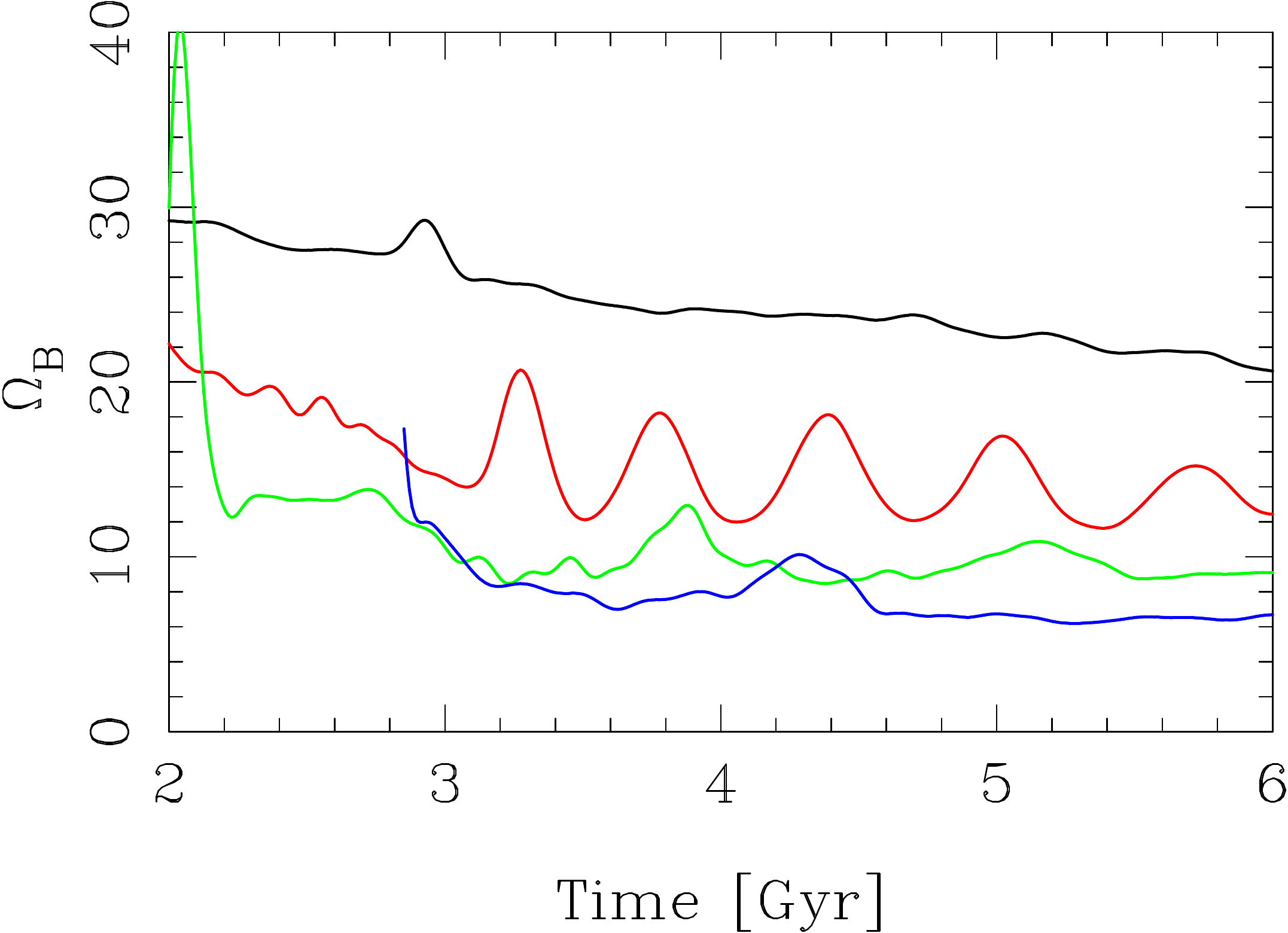}
	\includegraphics[scale=0.19,angle=-90]{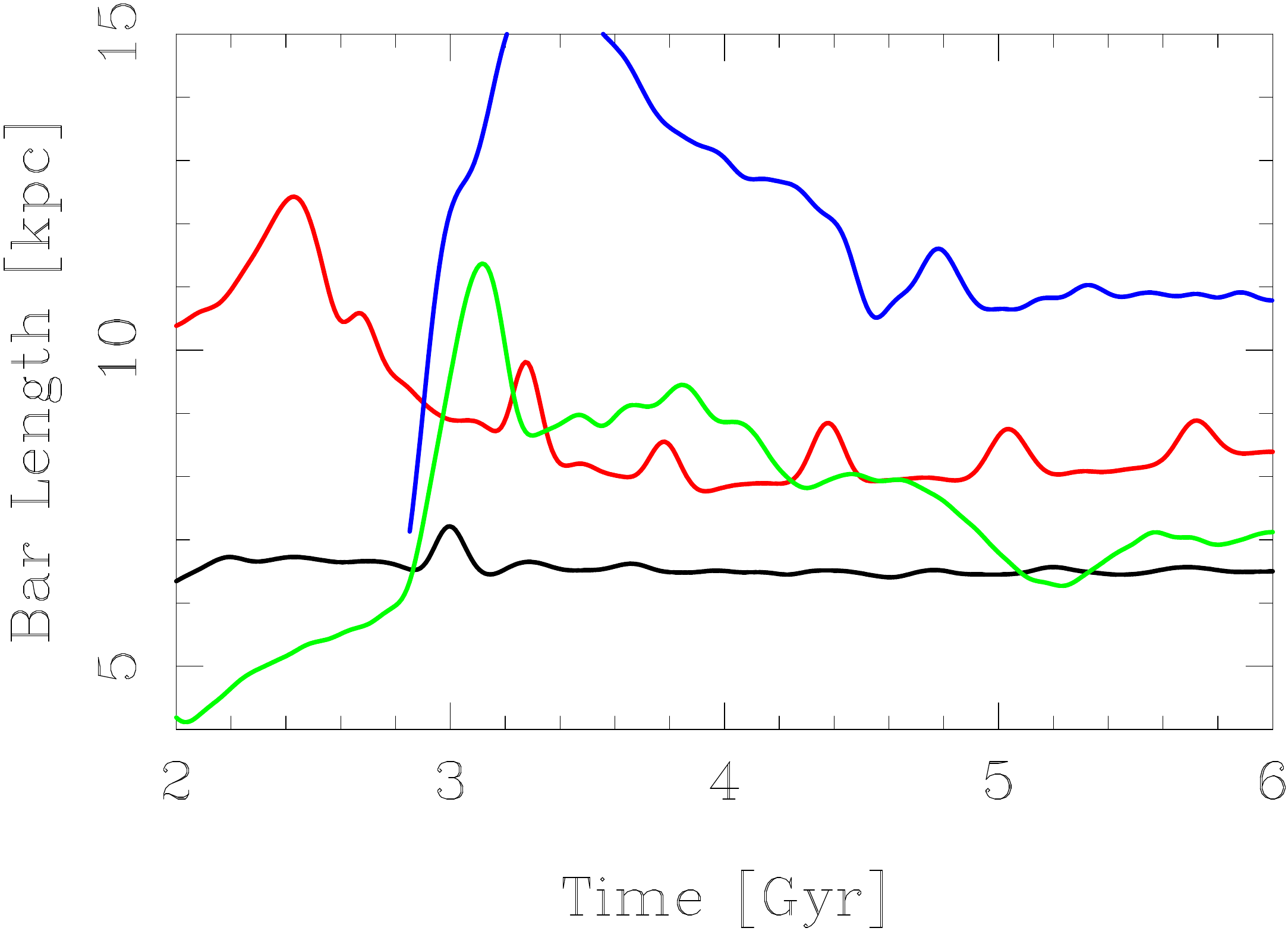}
	\includegraphics[scale=0.19,angle=-90]{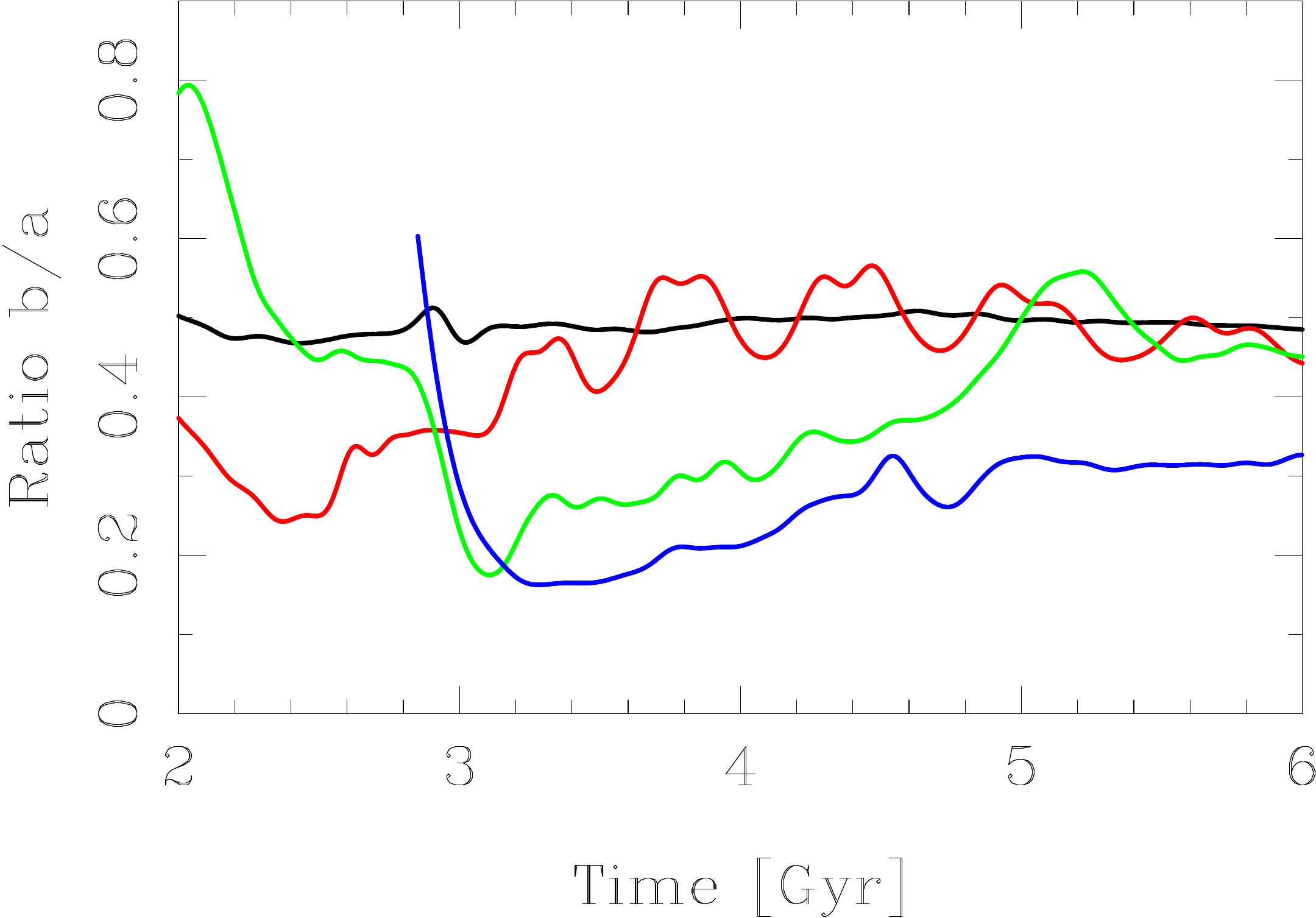}
	\includegraphics[scale=0.19,angle=-90]{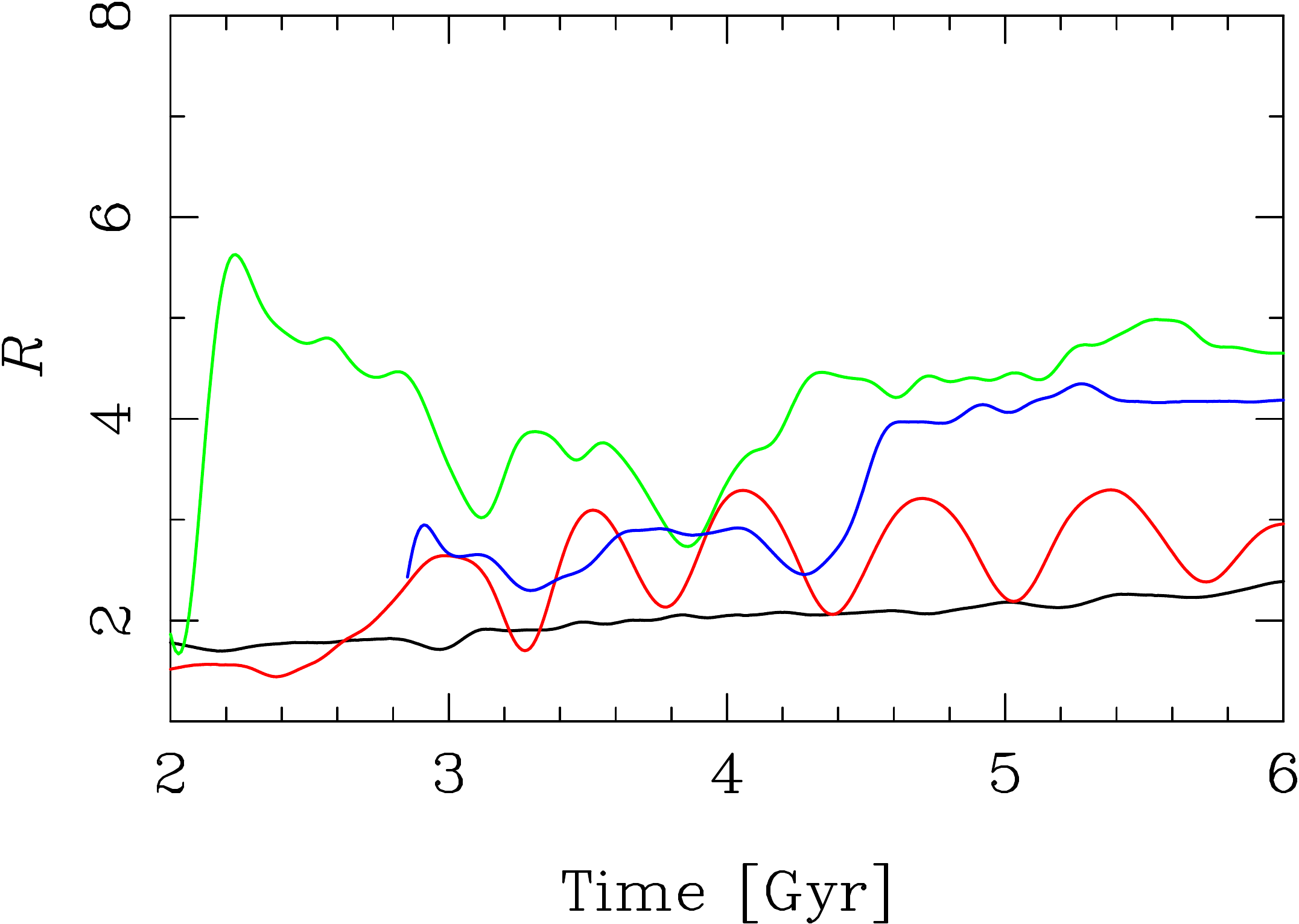}
	\\
	\includegraphics[scale=0.19,angle=-90]{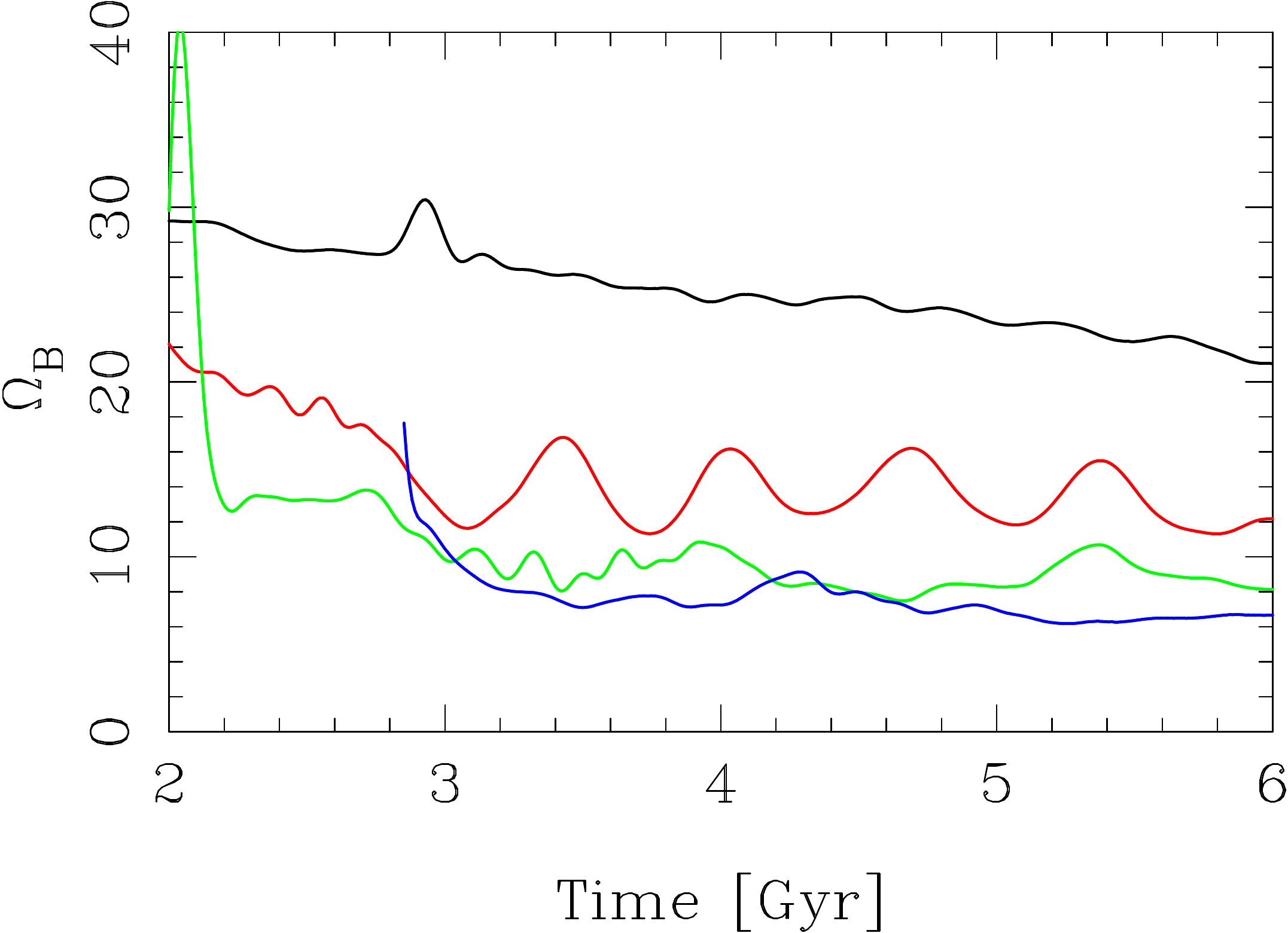}
	\includegraphics[scale=0.19,angle=-90]{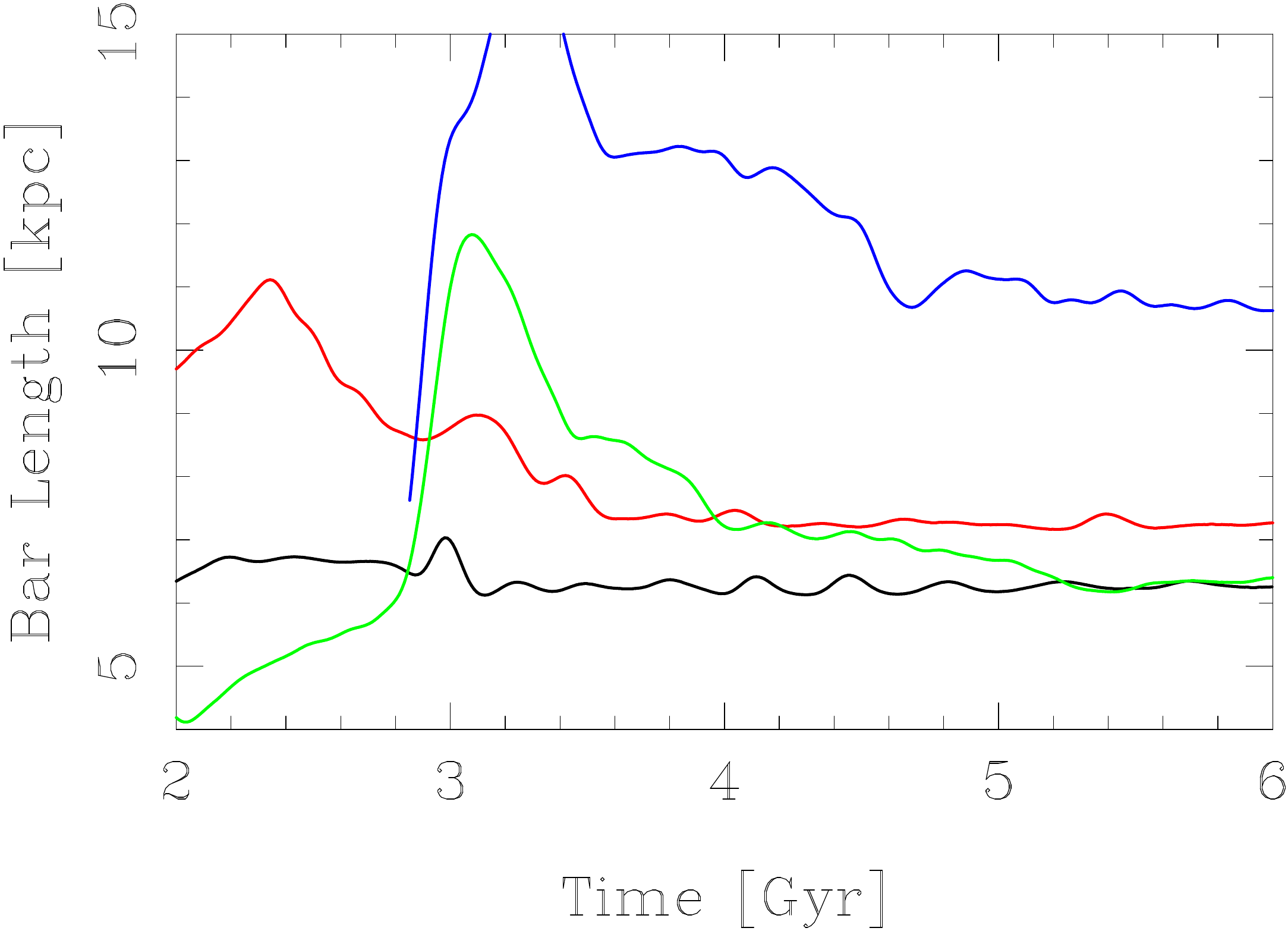}
	\includegraphics[scale=0.19,angle=-90]{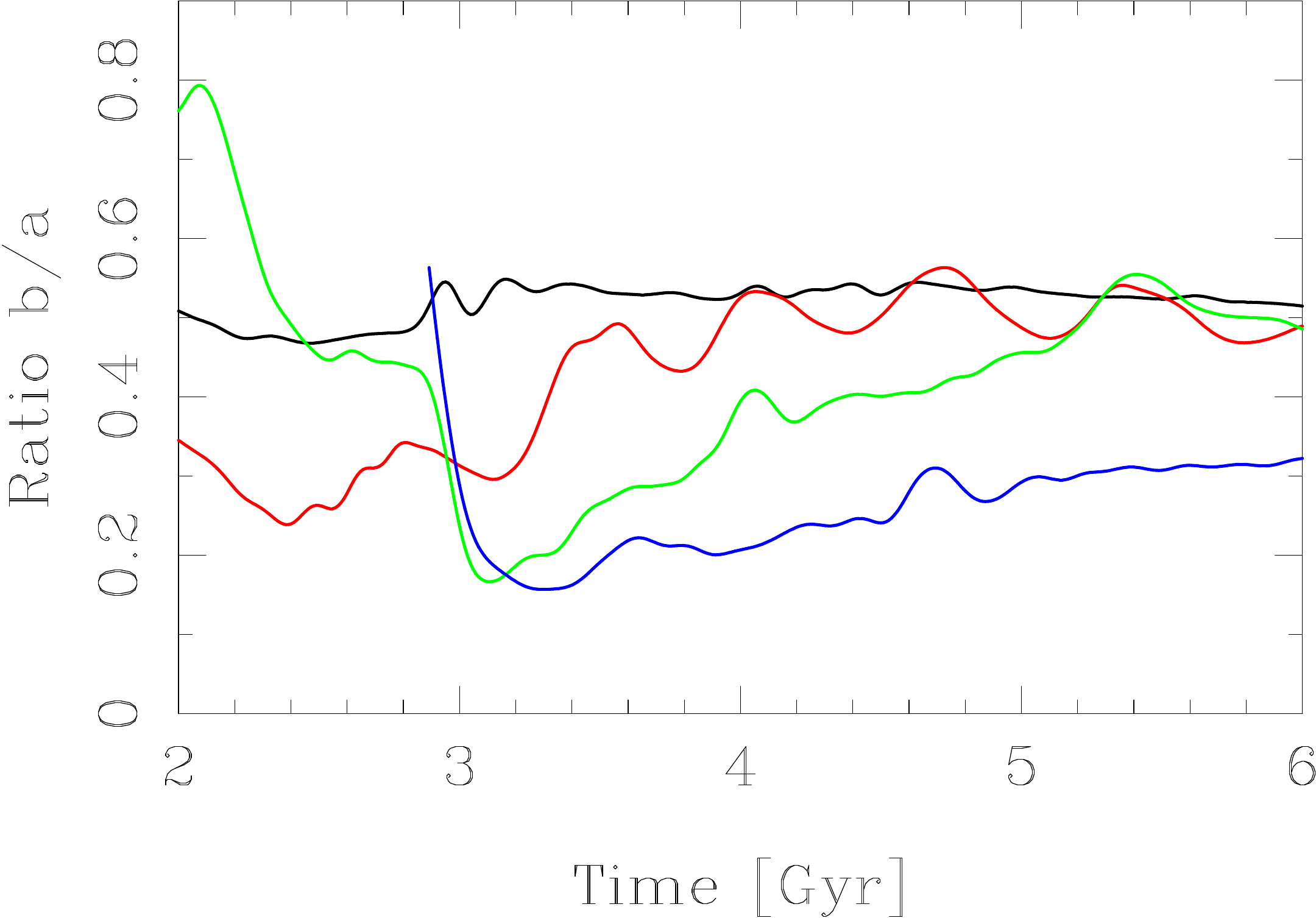}
	\includegraphics[scale=0.19,angle=-90]{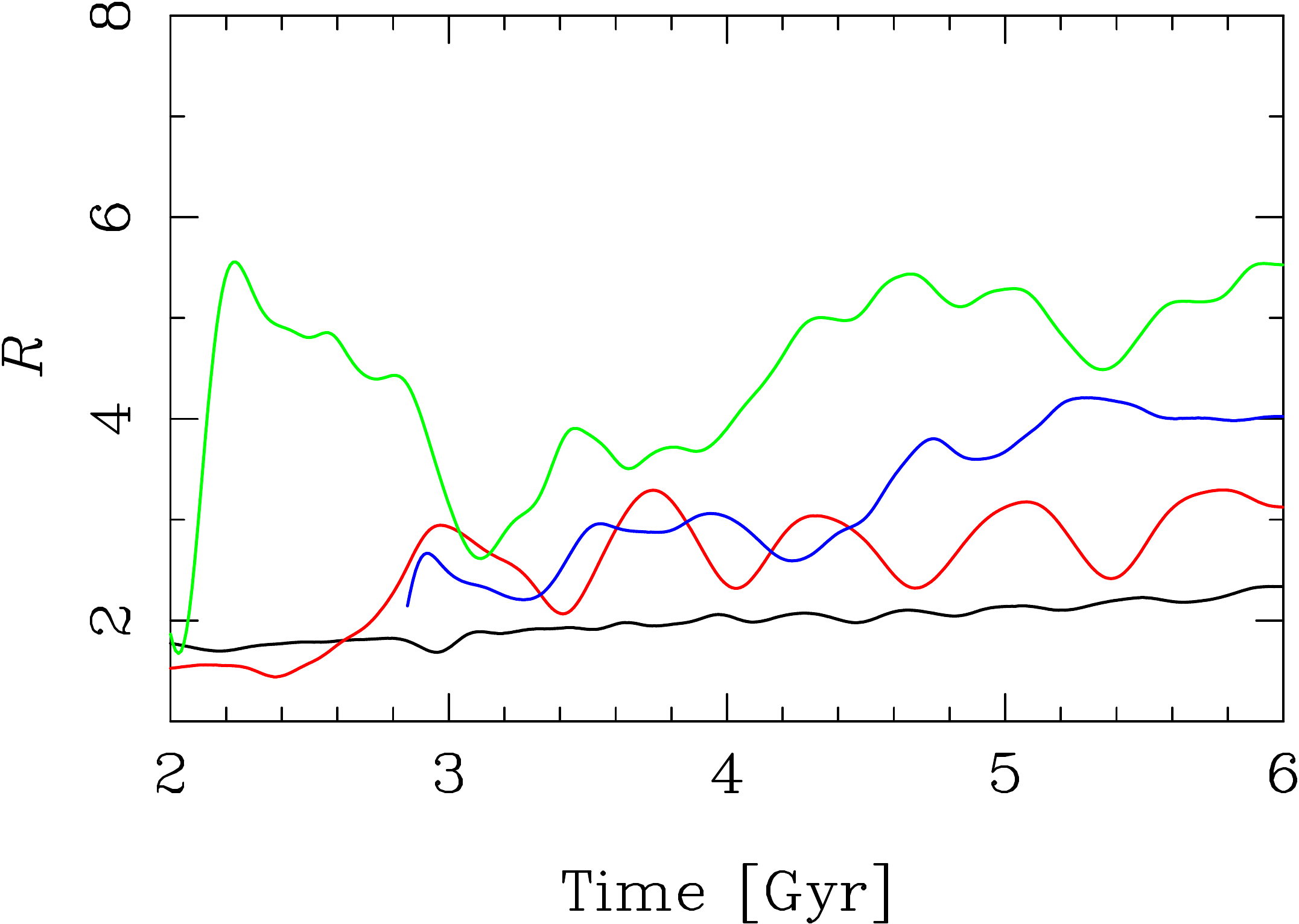}
	\caption{Time evolution of observational bar parameters. 
		Black, red, green and blue lines represent $*A\lambda03$,
		$*A\lambda04$, $*A\lambda05$ $*A\lambda06$, respectively.
		We display on first row the models of group Pw,
		in second row the models of group Pm, and in third row
		the models of Ps group.
		For all simulations, first column shows the
		instantaneous bar pattern speed,
		second column shows the length of the bar, the ratio
		between the axes of the bar is shown in third column,
		and the last one shows the $\mathcal{R}$ parameter.
		\added{Besides, we present with purple line the results of the $PwA\lambda06$\_M2 model to compare with the $PwA\lambda06$ model; observing that they are very similar.}
		 \label{fig:OmLenR}}	
\end{figure*}

Figure \ref{fig:OmLenR} shows the growth of the bar
which is characterized by their observational parameters.
From the \replaced{second}{top} row to the last one, we display the
evolution of the bar for models of groups Pw, Pm, and Ps,
respectively. We can observe in all encounters
that at the time of interaction the measurements
of $\Omega_B$ and $l$ show a bump due
to the impact given by the perturbation.

The evolution of bar for models $*A\lambda03$ 
(black lines), and $*A\lambda04$ (red lines) are
not much affected by the
perturbation. They evolve similarly to their isolated counterpart
(see Figure \ref{fig:OmLenR_I}).
However, there is a slight difference in the
bar axis length, e.g., while model $A\lambda03$
shows an increase around of one $kpc$  
during the interval
from two to six , perturbed models $*A\lambda03$
do not present such increment and after the
perturbation passes, the bar seems to maintain
the same size throughout the evolution. On the other hand,
the slowdown of the bar for model $A\lambda04$ 
falls at a constant rate from 21 to 12 $km s^{-1} kpc^{-1}$
while such slowdown seems to stop after the perturbation
overfly, decreasing it and keeping it around of 12 
$km s^{-1} kpc^{-1}$. Particularly,
the $\Omega_B$ for model $PsA\lambda04$
still oscillating between 10 and 16 
$km s^{-1} kpc^{-1}$ during the rest of simulation.

The interactions $*A\lambda05$
(green lines), the third row of Figure
\ref{fig:OmLenR}, cause that the bar formation starts
earlier than in the isolated model $A\lambda05$.  
While the bar angular velocity of
$A\lambda05$ model is around of 12 $km s^{-1} kpc^{-1}$,
this, for perturbed models, changes from
12 $km s^{-1} kpc^{-1}$ at the beginning to
9 $km s^{-1} kpc^{-1}$ when the perturbation is far away.
Moreover, the bar reaches its maximum length
at the MA phase: e.g. for model $A\lambda05$ is around 11 $kpc$
at 4-5 Gyrs, for model $PwA\lambda05$ is around 10 $kpc$
at 3-5 Gyrs, for models $PmA\lambda05$ and 
$PsA\lambda05$ are around 11 $kpc$ at 3 Gyrs. 
These variations cause that the
$\mathcal{R}$ parameter also gets large changes: e.g.
the bar of models $A\lambda05$ and $*A\lambda05$ appears as
slow rotators and tends to become fast rotators.

We can observe clearly the bar formation of models
$*A\lambda06$ (blue lines)
\added{and $PwA\lambda06$\_M2 (purple lines)} 
which was triggered by the
perturbation. After the perturbation passes, the bar
appears with the lowest angular velocity, which is around
of 9 $km s^{-1} kpc^{-1}$ and it stays with such
speed throughout the evolution. Besides,
the bar length in these models reach the largest
radius in both the MA phase and after
that event as well, and the bar in these models are the most
narrow (see the ratio between the axes of the bar, blue lines
in the third column of Figure \ref{fig:OmLenR}).
Particularly, the bar in  \replaced{model $PwA\lambda 06$}
{the $PwA\lambda 06$ and $PwA\lambda 06$\_M2 models}, which
is the lightest interaction, appears as
slow rotator and evolves toward fast rotator; then
when the perturbation is far away, the bar evolves from
fast to slow rotator. In contrast, heavier interactions
($PmA\lambda 06$, and $PsA\lambda 06$) the bar appears
as fast and evolves towards slow rotator.

\begin{table}
	\centering
	\caption{The table shows the growth rate $\omega$
		of the bar/oval, and bold numbers represent the growth rate of
		the spirals triggered by the perturbation.\label{tab4}}
	\begin{tabular}{lllll} 
		\hline
		Model          &  A       & Pw       & Pm         & Ps    \\\hline
		$\lambda03$    & 3.82     &{\bf 4.24}&{\bf 6.87} & {\bf 6.10}  \\
		$\lambda04$    & 0.84     &{\bf 2.97}&{\bf 6.47} & {\bf 6.07}  \\
		$\lambda05$    & 0.39     & 1.94     & 4.06       & 5.81  \\
		$\lambda06$    &{\bf 0.14}& 2.63     & 5.64       & 6.27  \\
		\added{$\lambda06$\_M2}    & -- &  2.65    & --  & --\\
		\hline
	\end{tabular}
\end{table}

\section{DISCUSSION}\label{discussion}

\subsection{Discussion of Isolated models}

In the set-up of the isolated models, we only changed 
the spin parameter $\lambda$ to generate disk dominate 
and halo dominate models and we study the DSP and the properties of the bar
through the time.

Disc dominated models form a bar
relatively quickly while a halo dominated ones
do not form such structure.
It has shown that the growth of a bar in a disk galaxy
is more efficient when the rotation curve is dominated
by the disk due to the exchange of angular momentum between 
e.g. halo and disk \citep{2013seg..book..305A}.  Thus, the rate
at which bar parameters change depends on the properties of the
model,
as well as the initial $Q$ parameter.
The local stability, $Q$ Toomre parameter, keep almost
constant in model $A\lambda06$, however this parameter
increase with the growth of the bar 
in the other models (see Figure \ref{fig:Q}).
Although this parameter is a good indicator to 
know if a model is susceptible to the bar formation at
the beginning of its evolution, the grow of the bar makes that
this parameter increase leaving thus unknown what is the stability
limits. However, MO98 and SW99 showed that disk stability
possesses a lower limit on its spin parameter to get
a stable disk, but it remains unclear what
is this limit and how these disk stabilities
behave during the evolution of a disk. 
In other words, are the DSP maintained  
($\lambda_d>\lambda_{crit}$ or $\lambda_d<\lambda_{crit}$)?,
or do they change like to the $Q$ parameter?

We get a stable disk against
bar formation when the model begins with the spin parameter
greater than its critical spin parameter;
the model $A\lambda06$ start 
with that configuration 
($\lambda_d > \lambda_{crit}$)
which is kept during the whole simulation
and the model does not form a bar 
(see bottom right panel of Figure \ref{fig:lambdas}).
On the other hand, we obtain an unstable disk to 
bar formation when its spin parameter is less than
its critical spin parameter; 
models $A\lambda03$ and $A\lambda04$ start with
that configuration 
($\lambda_d < \lambda_{crit}$); although
the models change $\lambda_d$ and $\lambda_{crit}$
during the evolution of the simulation, they conserve
their configuration (see upper panels of Figure
\ref{fig:lambdas}). 

\added{Additionally, we show that the time evolution
of the DSP do not depend on number of particles. The
$A\lambda05$ and $A\lambda05$\_M14 models, which have
different number of particles but they are equals 
in nature, start with stability parameters very
close to lower stability limit. They begin to form an
oval structure from 3 to 4 Gyr when the critical 
spin parameter $\lambda_{crit}$ rises up larger 
than the spin parameter $\lambda_d$. These
models conserves that configuration until the end of
simulation.}\deleted{The model $A\lambda05$, which starts with
stability parameters very close to lower stability
limit, forms a bar only after 4 Gyrs when its critical 
spin parameter $\lambda_{crit}$ rises up larger 
than its spin parameter $\lambda_d$, and it conserves 
that configuration until the end of the simulation.}

The growth of the bar for all barred
models shows some differences due to the central
properties of the model.
For $A\lambda03$ model, the growth rate of the bar
is  higher, its spirals are weaker 
(figures \ref{fig:AmFT1D_I} and \ref{fig:averaStrong}), 
the instantaneous angular velocity 
of the bar is higher, and its length is shorter (see figure
\ref{fig:OmLenR_I}) than those parameters for model $A\lambda04$. 
We also noticed that the $\mathcal{R}$ 
parameter evolves from slow to fast
for both models $A\lambda03$ and $A\lambda04$,
(see last panel of figure \ref{fig:OmLenR_I})
similar to classical bars showed in different works.
The observational parameters
of the bar, for model $A\lambda05$, are very diffuse because
the bar is just forming at the end of the simulation.
In fact, At the beginnings this model seems to be stable;
but after the fourth Gyr, the model starts to form a
bar structure when $\lambda_{crit}$ becomes higher than
$\lambda_d$ reaching the second phase of its growth 
at the end of the simulation.

Unlike \cite{2013MNRAS.434.1287S} and \cite{2008MNRAS.390L..69A},
the parameter $\epsilon_m$ save the conditions established
by EF82 in our models. The model $A\lambda06$ present $\epsilon_m>1$,
which is stable against bar formation, while
the other models show $0.7<\epsilon_m<1$ which are
unstable to bar formation. This parameter seems to
work well in our models, which has a larger halo with a NFW profile. 

\subsection{Discussion of the interactions}\label{discusison_interact}

We  measured the evolution of the Disk Stability Parameters 
on isolated models to characterize the properties 
of a galactic disk to be stable or unstable to bar formation.
We showed that the DSP configuration of an initial disk
susceptible to bar formation keeps such configuration 
below the stability limits through the entire evolution;
in contrast, a stable disk holds such DSP above the stability
limits. The growth rate of the bar depends of 
how close the DSP are from the stability limits showing that
if the DSP are below and far from the stability limits,
the growth rate of the bar is higher. 

In this work, we subjected the isolated models
to different perturbations in coplanar
hyperbolic orbit to examine the evolution of
their DSP and how this
affects the growth of bar in models stable
and unstable to the bar
formation; we do not take
care of the evolution of the perturbation.
The perturbations were modeled by an extended live spherical
halo and stellar components with an
Hernquist profile, respectively,
so that the interactions are more realistic.
Using a spherical galaxy as a perturbation permits
that the pull given by this one is smoother
than if the perturbation were a point of mass.

We explore the interactions only changing the
mass of the perturbation where the total
mass of the perturbation is half (Pw), similar
(Pm), and two times (Ps) the total mass
of the target galaxy. Therefore, the
interaction force at the pericentre
is stronger with heavier perturbations
(see table \ref{tab2}). The different perturbations
affect similarly the DSP and the bar parameters; 
however, stronger are the interactions, more noticeable
are the changes in the measurements of the parameters. 

The flyby of the perturbation causes a bump in
the spin parameter  $\lambda_d$ and then when the
perturbation is far from the target galaxy, it tends
to returns to similar values. Conversely, the critical
spin parameter $\lambda_{crit}$ decreases slightly
at the time of interaction and then it increases
overtaking during the simulation the
previous values that it had before the
interaction. The increment on $\lambda_{crit}$
is because the scale radius of the halo $r_s$
shrink a larger percentage of distance than the radius
$r_{200}$; thus the halo becomes more concentrate.
Likewise, the $\epsilon_m$ parameter decrease
abruptly. This fall is due to that
the pass of the perturbation causes a vigorous
exchange of angular momentum between resonances
and components (halo and disk), and then
the disk also changes the radial scale $r_d$.
Nevertheless, the small decrease of $r_d$
does not affect much the parameter $\lambda_{crit}$.
We notice that the DSP of perturbed models 
have the same behavior to that of isolated models,
as well as the experimental stability parameter.
For example, the parameter $\lambda_{crit}$
for interactions $*A\lambda06$, which is stable to
bar formation in isolation, overtakes the spin parameter $\lambda_d$
after the moment of interaction, as well as 
the $\epsilon_m$ parameter is set to the range
$0.7 < \epsilon_m < 1$; then it makes that the model
is identified as unstable to bar formation.

Although the $\epsilon_m$ parameter is a simple
comparison of the rotation curve and the circular
velocity of an hypothetical particle subject to a
point-mass potential which has a mass equal to 
that of the disk seems to work well in our 
interacting models. Therefore, it could be
a good indicator to assess the stability of a disk
at least approximately, and also it could be used
in real galaxies to assess and restrict some parameter
of a disk galaxy.

\cite{1972ApJ...178..623T} demonstrated that tidal
perturbations distort extended portions of a disk
to produce elongated and narrow features, phenomenologically
called bridge and tail. The bridge is built
on the near side of the disk toward the perturber, while
the tidal tail, or counter stream, forms on the far side 
\citep{2008ApJ...683...94O}. Together, these two
features generate high amplitudes in the Fourier
Transform of mode $m=2$. Such amplitude illustrates
the dynamical responses of disks to a tidal perturbation.
Figures \ref{fig:AmFT1D} show that the tidal pull evolves
from the outer to the inner region of the disk generating
a vigorous spiral wave which excites the epicycle
orbits of individual particles. Therefore, this
triggers the bar formation rapidly in the disk without
a bar. However, the bar already developed
in a disk is marginally affected; e.g. we observe that the bar
becomes slightly oval, but the axes ratio is almost constant
during the evolution of the simulation (see Figure
\ref{fig:OmLenR}). After the perturbation is far,
the tidal tail and bridge dissipate quickly,
but some spirals growth transiently. These
spirals are stronger and longer in models without
bar before the perturbation passes. 

On the other hand, the perturbation does not affect
much the observational parameters in barred models.
In contrast, when the
perturbation trigger the bar, the observational
parameters appear with less angular velocity,
longer and shorter major and minor axis of the
bar, respectively, and when the perturbation
is heavy (e.g., Pm and Ps groups) the bar rotates from 
fast to slow, but when the perturbation is
light, the bar grows in two phases: first appears
as slow rotator and evolve toward fast rotator, then
it evolves from fast toward slow rotator (see models
$PwA\lambda 06$ \added{and $PwA\lambda 06$\_M2}).

Finally, different to other results, e.g.,
\cite{2017MNRAS.464.1502M}, we find that
the bar growth, which is triggered by a light interaction,
develops into two phases; the bar grows
from slow to  fast rotator; however, when the value
of $\mathcal{R}$ is close to one, the growth
changes from fast to slow rotator.

\section{CONCLUSIONS}\label{conclusions}

In this work, we have followed for the first time
in N-body simulations the critical spin parameter
$\lambda_{crit}$ and the experimental stability
parameter $\epsilon_m$ to characterize the stability
of a disk galaxy model. We get an unstable 
isolated model to bar formation setting up 
$\lambda_d < \lambda_{crit}$ (its rotation curve
is dominated by the disk), while a stable
model against bar formation is achieved setting up
$\lambda_d > \lambda_{crit}$ 
(its rotation curve is dominated by the 
halo). Moreover, we show that the configuration of
stability e.g $\lambda_d > \lambda_{crit}$ or
$\lambda_d < \lambda_{crit}$ is saved for
long time no matter what structure is forming in the
disk. Following the same line of research, we perturbed
the isolated models to understand the nature of the
formation and evolution of a bar in disk galaxies.
There, we illustrate how the DSP are affected when the 
models are subjected to a perturbation. We reported
one of the most important conclusion of the whole work,
which in general is:``the bar in our disk galaxy models
are formed below the stability limits in both isolated
and perturbed disk, and this depends on how close are
the parameters of their critical values''.

The growth rate of the bar in our models
depends on the configuration 
of the DSP ($\lambda_d$, $\lambda_{crit}$).
We have found that the growth rate is high if the
spin parameter of the disk $\lambda_d$ is lower
than the stability limit $\lambda_{crit}$, as well
as $\epsilon_m$ is far from the unity.
The model $A\lambda05$ \added{($A\lambda05$\_M14) as well}
shows that the bar starts
to evolve when the stability limit $\lambda_{crit}$
overtake the spin parameter $\lambda_d$, i.e., even if the initial
configuration sets a model as stable, its own evolution with
exchanges of angular momentum between the disk and halo can place
it in the unstable regime.
The stability parameter $Q$ 
increases during the evolution; in contrast, we have found that
the models maintain their initial configuration on DSP
(unstable or stable regime), except when the initial Disk Stabilities Parameters
are close to the critical stability limits, e.g. model $A\lambda05$.


With respect to observational parameters of the bar,
we show that the evolution of those parameters mainly 
depend on the central properties of the model. A more
concentrate mass distribution in a disk generates
a shorter bar that rotates faster than a more extended 
mass distribution model in both isolated and perturbed models.


\section*{Acknowledgements}

D.V.E and I.P. thanks the Mexican Foundation Conacyt for grants that
support this research. I.R. thanks the Brazilian agency CNPq (Project 311920/2015-2).
Part of the numerical work was developed using the Hipercubo Cluster 
(FINEP 01.10.0661-00, FAPESP 2011/13250-0 and FAPESP 2013/17247-9) at IP\&D--UNIVAP.

\listofchanges

\end{document}